\documentclass[a4paper,onecolumn,11pt,unpublished]{quantumarticle}
\pdfoutput=1
\usepackage[utf8]{inputenc}
\usepackage[english]{babel}
\usepackage[T1]{fontenc}
\usepackage{amsmath}
\usepackage{amssymb}
\usepackage{amsthm}
\usepackage{hyperref}

\usepackage{tikz}
\usepackage{lipsum}

\usepackage{multicol}
\usepackage{bm}
\usepackage{mleftright}

\newtheorem{theorem}{Theorem}

\newtheorem{observation}{Observation}

\usepackage{chngcntr}
\newcommand{\id}{\mathbbm{1}}

\usepackage{graphics, setspace}
\usepackage{physics}

\newcommand{\pud}{P_{\, \textnormal{UD}}}
\newcommand{\pq}{P_{\, \textnormal{Q}}}
\newcommand{\pt}{P_{\, \textnormal{T}}}

\newcommand{\pme}{P_{\, \textnormal{ME}}}

\newcommand{\lmin}{\lambda_{\textnormal{min}}}
\newcommand{\umin}{u_{\textnormal{min}}}
\newcommand{\vecr}{\vec{r}}
\newcommand{\conv}{\textnormal{conv}}

\newcommand{\m}{\mathcal{M}}

\newcommand{\qcrit}{Q_{\textnormal{crit}}}

\newcommand{\tcrit}{T_{\textnormal{crit}}}

\newcommand{\ecrit}{\varepsilon_{\textnormal{crit}}}

\newcommand{\sgeq}{\succcurlyeq}

\newcommand{\perr}{P_{\textnormal{err}}}

\newcommand{\psin}{P_{\textnormal{single}}}
\newcommand{\pj}{P_{\textnormal{joint}}}

\newcommand{\ve}{\varepsilon}
\newcommand{\aep}{A_\varepsilon}

\newcommand{\qmax}{Q_\textnormal{max}}

\usepackage{graphics, setspace}

\newcommand{\tmax}{T_{\textnormal{max}}}

\usepackage{tikz}
\usepackage{mathtools}
\usepackage{bm}
\usetikzlibrary{arrows.meta}
\usetikzlibrary{decorations.markings,quantikz2}
\usetikzlibrary{fit,shapes.arrows}

\usepackage[numbers,sort&compress]{natbib}

\usepackage{multicol}
\usepackage{subcaption}

\usepackage{pgfplots}
\pgfplotsset{compat=1.18}

\usepackage{float}

\usepackage{changes}

\usepackage{multicol}

\usepackage{multirow,array}

\newcommand{\qber}{\textnormal{QBER}}

\begin{document}

\title{Quantum randomness under unambiguous attacks}

\author{Fionnuala Curran}
\affiliation{ICFO-Institut de Ci\`encies Fot\`oniques, The Barcelona Institute of Science and Technology,
Av. Carl Friedrich Gauss 3, 08860 Castelldefels (Barcelona), Spain}
\orcid{0000-0001-9036-8113}
\email{fionnuala.curran@icfo.eu}
\maketitle
\begin{abstract}
Intrinsic randomness is generated when a quantum state is measured in any basis in which it is not diagonal. In an adversarial scenario, we quantify this randomness by the probability that a correlated eavesdropper could correctly guess the measurement outcomes. What if the eavesdropper is never wrong, but can sometimes return an inconclusive outcome? Inspired by analogous concepts in quantum state discrimination, we consider the randomness generated by a measurement setup when an eavesdropper guesses unambiguously, and, relaxing the assumption of perfect accuracy, with a fixed rate of inconclusive outcomes. We solve the \emph{maximal} randomness of any quantum state under an unambiguous or FRIO attack, optimized over all projective measurements. We also solve both problems for any state and projective measurement in dimension two. If a noisy setup is used to choose random bases for a prepare-and-measure quantum key distribution protocol, an unambiguous attack may allow an eavesdropper to gain knowledge about the secret key, without causing any disturbance. Moreover, while experimental noise is typically ascribed to a single device, we find, given a state measured in an unbiased basis of any dimension, that an eavesdropper correlated only to a noisy state is outperformed by an eavesdropper with joint correlations to both a noisy state \emph{and} a noisy measurement. In fact, we identify a critical noise parameter beyond which the joint-noise eavesdropper achieves a \emph{perfect} guessing probability, ending all hopes of private randomness. 
 \end{abstract}

\section{Introduction}
Quantum physics is inherently random. This quality is baked into the very pillars of quantum theory, via the superposition principle and the Born rule: the result of measuring a pure state in any basis in which it is not diagonal is unpredictable. Complications arise, however, if we replace the pure state with a mixed, or \emph{noisy} one. Mixed states possess a second form of randomness arising from ignorance, since they can be physically realized as one of infinitely many mixtures of pure states. In an adversarial scenario, we must assume that this extra \emph{side-information}, though unknown to the experimentalist, is known by a malevolent eavesdropper, who uses it to guess the measurement outcomes. What is the extent of the eavesdropper's knowledge? This question, motivated by foundational interest, as well as the recent proliferation of quantum random number generators (QRNGs) \cite{Herrero_Collantes_2017,Mannalatha_2023}, is the focus of the present work. 

A standard way to quantify randomness is to consider the probability that the eavesdropper guesses the outcomes correctly, optimized over all possible decompositions of the mixed state \cite{Law_2014}. This quantity has operational significance, as it can be directly related to the conditional min-entropy \cite{Konig_2009}, which lower bounds the number of (statistically \emph{and} private) random bits that can be extracted from the measurement outcomes \cite{renner2006thesis,Konig_2011}. Recently, the maximal intrinsic randomness of any quantum state has been solved by maximizing the conditional min-entropy over all projective \cite{Meng_2024} or general \cite{anco_2024secure} measurements. Typically, it is supposed that the eavesdropper makes a guess on \emph{every} measurement round, incurring, inevitably, some fraction of incorrect guesses, which she aims to minimize. But what if she wanted a higher, or even perfect, level of accuracy?  

Applying concepts from quantum state discrimination \cite{Ivanovic_1987,Dieks_1988,Peres_1988,Jaeger_1995}, we consider the randomness generated by a quantum state-measurement pair under an unambiguous attack, as quantified by the optimal guessing probability of an eavesdropper who makes no errors, but who is allowed to output an additional, inconclusive outcome. The \emph{maximal} randomness of the state is then found by minimizing the guessing probability over all measurement bases. In this work, we solve this problem for any state, providing a sufficient condition for the optimal measurement.
Unambiguous state discrimination has already been considered in the context of quantum key distribution (QKD), performed both by the receiving party \cite{Bennett1992, Ekert_1994} and the eavesdropper \cite{Yuen_1996, Dusek_2000, Curty2007, Branciard_2007}, and in the generation of semi-device-independent randomness \cite{Brask_2017}; here, by contrast, it is used by an adversary to \emph{attack} a QRNG. We show that, when a noisy QRNG is used to generate random bases in a prepare-and-measure quantum key distribution (QKD) protocol (e.g., BB84 \cite{BB84}), an unambiguous attack allows an eavesdropper to gain knowledge about the secret key \emph{without being detected}.

Further, again adapting ideas from quantum state discrimination \cite{Chefles_inter_1998,Zhang_inter_1999, Nakahiri_2012}, we present an intermediate scenario wherein the eavesdropper maximizes her guessing probability subject to a fixed rate of inconclusive outcomes (FRIO), a term coined in \cite{Bagan_2012}. For a fixed parameter $Q$, the randomness generated by a quantum state-measurement pair, subject to a FRIO attack, is given by the optimal guessing probability of an eavesdropper who returns an inconclusive outcome with probability $Q$. In the QKD scenario described above, the eavesdropper could choose the parameter $Q$ such that her disturbance of the intercepted states coincides with the maximum error rate allowed by the communicating parties. We solve the maximal randomness of any state under a FRIO attack, again optimized over all projective measurements. We also solve the eavesdropper's unambiguous \emph{or} FRIO guessing probability given any qubit state and projective measurement, matching previous results in the non-adversarial setting \cite{Jaeger_1995, Bagan_2012, Jiménez_2021, Nakahiri_2012}.

We devote special attention to the setting in which a noisy state, defined as a pure state affected by isotropic noise, is measured in a basis that is unbiased to its own diagonal basis. This could be related, for example, to randomness generation using single photons distributed across several paths  \cite{Gräfe_2014,Gao_2023}. Though experimental noise is commonly attributed to a single device, we consider the more realistic picture where the state \emph{and} the unbiased measurement are noisy. The eavesdropper, now correlated to both, can perform a joint decomposition of the state and the measurement \cite{frauchiger2013,Dai_2023,Senno_2023}. We show that such a joint-noise eavesdropper always outperforms her single-noise counterpart given the same amount of total noise. This holds true given any upper bound imposed on the probability of error (i.e. any FRIO parameter). In fact, beyond a critical noise value that is no greater than a half, the joint-noise eavesdropper achieves a perfect guessing probability, with no inconclusive outcome needed, ending all hopes of private randomness. 

There is now a wealth of results on quantum state discrimination, but most of it has not been translated to an adversarial setting. We address this gap here by considering less conventional eavesdropping strategies in a randomness generation scenario. Our joint-noise case study warns that, even in a wholly device-dependent setting, with perfect statistical randomness and stringent constraints on the guessing accuracy, an eavesdropper might still gain some (or even total) information about the outcomes, due to noise. Our work shows, too, that noisy QRNGs challenge the assumption, crucial to quantum cryptography, that information gain from quantum systems implies disturbance.  
\\
\\
The paper is laid out as follows. In Section \ref{sec: setup}, we set up our two scenarios, presenting the optimization problems given unambiguous and FRIO attacks. In Section \ref{sec: results}, we give our results. Concretely, in Section \ref{ssec: any state}, we find analytic solutions for the maximal randomness of any state under an unambiguous or FRIO attack. The former translates to a lower bound on the probability of unambiguous discrimination for any ensemble of linearly independent pure states. In Section \ref{ssec: qubits}, we find the unambiguous and FRIO guessing probability given any state and projective measurement in dimension two, which we use in Section \ref{ssec: random bases} to describe an eavesdropper's attack on a noisy QRNG which is used to select random bases for quantum key distribution.
In Section \ref{ssec: noisy}, we lower bound the guessing probability of a joint-noise eavesdropper and prove that it always exceeds that of a single-noise eavesdropper, for any unbiased state-measurement pair affected by isotropic noise. In Section \ref{ssec: conc}, we present our conclusions and ideas for future work. The majority of proofs and derivations are relegated to the Appendix.

\section{Set-up}\label{sec: setup}
In both of our randomness scenarios, an experimentalist (who we call Alice) performs a measurement on a system, represented by a mixed state $\rho$, which is positive semidefinite ($\rho \sgeq 0$) and has trace one. We represent her measurement by a Positive Operator-Valued Measure (POVM), $\m=\{M_j\}_j$, where the POVM elements $M_j$ are positive semidefinite ($M_j \sgeq 0$) and sum to the identity ($\sum_{j}M_j=\id$). Throughout most of this work, we consider rank-one projective measurements, which have the form $M_j=\ketbra{m_j}{m_j}$, where $\{\ket{m_j}\}$ is an orthonormal basis in finite dimension $d$. 

We allow the eavesdropper, Eve, to hold a purification of the state $\rho$, i.e. she holds the state $\sigma = \tr_S \ketbra{\Psi}{\Psi}_{SE}$, for some joint state $\ket{\Psi}_{SE}$ such that $\rho = \tr_E \ketbra{\Psi}{\Psi}_{SE}$, where the subscripts $S$ and $E$ represent Alice and Eve's respective Hilbert spaces. Eve chooses the optimal measurement on her local system to try to infer Alice's measurement outcomes. From the Schr\"{o}dinger-Gisin-Hughston-Jozsa-Wootters theorem \cite{Schrödinger_1935, gisin_1989, Hughston_1993}, we know that a distant party who performs local measurements on the purification of a state can steer any ensemble of states to the other party. It is equivalent, then, to consider a `classical' eavesdropper, who simply chooses a decomposition $\{p_{\mu}, \, \rho_{\mu}\}$ for the state, where $\sum_\mu p_\mu=1$ and $\sum_{\mu}p_{\mu} \rho_\mu= \rho$. In each measurement round, Eve (unlike Alice) knows which state, say $\rho_j$, is actually used, and guesses the most probable outcome, say $j$, accordingly. As a final step to reconcile the `classical' and `quantum' eavesdropping scenarios, we assume that the states in the decomposition $\{p_\mu, \, \rho_\mu\}$ are independent and identically distributed (i.i.d).

Since we demand that Eve guess Alice's outcomes either unambiguously or with a FRIO parameter $Q$, she must include an `inconclusive' state $\rho_0$ in the decomposition with corresponding probability $p_0$. Whenever the state $\rho_0$ is sent, Eve declines to guess. The decomposition then has the form 
\begin{equation}
    \rho = p_0 \rho_0 + \sum_{j=1}^{d} p_j \rho_j\,.
\end{equation}
In the unambiguous case, we need $\rho_j= \ketbra{m_j}{m_j}$. Eve's guessing probability $\pud \mleft( \rho, \, \m \mright)$ (named for `unambiguous discrimination') is then given by the optimization problem
\begin{equation}\label{eqn: mt UD rand}
\begin{aligned}
  \pud \mleft( \rho, \, \m \mright) = \;\;     \underset{\{ p_j \} }{\text{maximize}}  \, \quad & \sum_{j=1}^{d} p_j  && {}
        \\
         \text{subject to} \quad & p_j \sgeq 0 \quad \textnormal{for all} \;\; j \in \{1, \, \hdots, \, d\} &&   
         \\
          \quad & \rho \sgeq \sum_{j=1}^{d} p_j \ketbra{m_j}{m_j} \,. \quad &&  {}
    \end{aligned}
\end{equation}
 Since the unambiguous discrimination of an ensemble of pure states is only possible when they are linearly independent \cite{Chefles_1998}, given a rank-one projective measurement, an eavesdropper's unambiguous guessing probability is nonzero only for mixed states of full rank.\footnote{More precisely, given a rank-one projective measurement $\m$ in dimension $d$ and a quantum state with rank $n < d$, the post-measurement states steered to the eavesdropper cannot be linearly independent in a space of dimension $d$. They may, however, be linearly independent in a subspace of dimension $n$, if the projection of $\m$ onto the support of $\rho$ results in a rank-one projective measurement in dimension $n$. In this case, we may as well work with this measurement instead and take $\rho$ to be full rank.} The maximal randomness of a state under an unambiguous attack, given by $\pud^{*} \mleft( \rho \mright)$, is then found by minimizing $\pud \mleft( \rho, \, \m \mright)$ over all measurement bases.

When Eve has a fixed probability $Q$ of returning an inconclusive outcome, her guessing probability $\pq \mleft( \rho, \, \m \mright)$ is given by the optimization problem
\begin{equation}\label{eqn: mt opt Q rand}
\begin{aligned}
  \pq \mleft( \rho, \, \m \mright) = \;\;     \underset{ \{p_\mu, \, \rho_\mu \} }{\text{maximize}}  \, \quad & \sum_{j=1}^{d} p_j  \langle m_j | \rho_j | m_j \rangle  && {}
        \\
         \text{subject to} \quad & \rho_\mu \sgeq 0 \quad \textnormal{for all} \;\; \mu \in \{0, \, 1, \, \hdots, \, d\}  &&  
         \\
          \quad &  \sum_{\mu=0}^{d} p_\mu \rho_\mu = \rho \,\quad &&  {}
          \\
          \quad &  p_0 = Q\,. \quad &&  {}
    \end{aligned}
\end{equation}
Once solved, $\pq \mleft( \rho, \, \m \mright)$ can be inverted to find the optimal guessing probability given a fixed rate of error \cite{Herzog_2012}, a scenario with its own counterpart in (non-adversarial) state discrimination \cite{Touzel_2007,Hayashi_2008}. More details on the derivation of \eqref{eqn: mt UD rand} and \eqref{eqn: mt opt Q rand} are given in Appendices \ref{ssec: unamb fom} and \ref{sssec: frio fom}. As before, the maximal randomness of a state under a FRIO attack, given by $\pud^{*} \mleft( \rho \mright)$, is found by minimizing $\pud \mleft( \rho, \, \m \mright)$ over all measurement bases. 

Following \cite{Meng_2024}, we consider only projective measurements in this work, since we allow Alice no access to additional quantum states or other sources of randomness. We show in Appendices \ref{ssec: UD max rand} and \ref{ssec: max rand FRIO} that coarse-graining a measurement does not decrease the eavesdropper's unambiguous or FRIO guessing probability, so, in solving the maximal randomness in each case, we are free to restrict to rank-one projective measurements. We can now proceed to the results.

\section{Results}\label{sec: results}

\subsection{Unambiguous attacks on any state}\label{ssec: any state}
\begin{theorem}\label{thm: any state}
The maximal randomness of any state $\rho$ under an unambiguous attack is given by the guessing probability 
\begin{equation}\label{eqn: mt opt UD any rho}
 \pud^{*} \mleft( \rho\mright) = d \lmin \mleft( \rho \mright)\,. \end{equation}    \end{theorem}
We can prove the lower bound by considering the decomposition
 \begin{equation}
     p_0 \rho_0 = \rho - \lmin \mleft( \rho \mright) \id\,, \qquad p_j \rho_j = \lmin \mleft( \rho\mright) \ketbra{m_j}{m_j} \quad \textnormal{for all} \;\; j \in \{1, \, \hdots, \, d\}\,.
 \end{equation}
 We show in Appendix \ref{ssec: any state UD} that the bound \eqref{eqn: mt opt UD any rho} can be achieved by measuring in a basis $\{\ket{m_j}\}$ for which $\ket{\umin}$, the eigenvector of the state $\rho$ corresponding to its minimum eigenvalue, satisfies 
 \begin{equation}\label{eqn: suff cond}
     \abs{\langle \umin | m_j \rangle }^2 = \frac{1}{d} \qquad \textnormal{for all} \;\; j \in \{1, \, \hdots, \, d\}\,.
 \end{equation}
 This includes bases that are unbiased with respect to the eigenbasis $\{\ket{u_j}\}$ of $\rho$. We also show by counterexample that this condition, while sufficient, is not necessary to saturate the bound. 

 The Schr\"{o}dinger-Gisin-Hughston-Jozsa-Wootters theorem \cite{Schrödinger_1935, gisin_1989, Hughston_1993} allows us to relate a randomness scenario, where Alice holds $\rho$, receives the decomposition $\{p_\mu, \, \rho_\mu\}$ and measures in $\{\ket{m_j}\}$, to a state discrimination scenario, where Eve holds $\rho$, receives the ensemble $\{\eta_j, \, \ket{\phi_j}\}$ and performs the measurement $\{E_\mu\}_\mu$, where
 \begin{equation}
    E_\mu = p_\mu \rho^{- \frac{1}{2}} \rho_\mu \rho^{- \frac{1}{2}}\,, \qquad \sqrt{\eta_j} \ket{\phi_j} = \sqrt{\rho} \ket{m_j}\,.
\end{equation}
(See Appendix \ref{app: rand and SD} for more detail on this equivalence.)
 Consider the set of $d$ linearly independent states $\{\eta_j, \, \ket{\phi_j}\}$, and define $\ket{\xi_j}= \sqrt{\eta_j} \rho^{- \frac{1}{2}} \ket{\phi_j}$.
 Since the $d$ operators $\ketbra{\xi_j}{\xi_j}$ have rank one, are positive semidefinite and sum to the identity in dimension $d$, the vectors $\{\ket{\xi_j}\}$ must define an orthonormal basis, so we find 
 \begin{equation}
     \langle \xi_i | \xi_j \rangle = \sqrt{\eta_i \eta_j}\, \langle \phi_i | \rho^{-1} | \phi_j \rangle = \delta_{i, \, j} \qquad \textnormal{for all} \;\; i, \, j \in \{1, \, \hdots, \, d\}\,.
 \end{equation}
 We can then define the measurement $\{E_\mu\}_\mu$, with
 \begin{equation}
     E_0 = \id - \lmin \mleft( \rho \mright) \rho^{-1}\,, \qquad E_j = \eta_j \lmin \mleft( \rho \mright) \rho^{-1} \ketbra{\phi_j}{\phi_j} \rho^{-1}\,,
 \end{equation}
 which is suitable for unambiguous discrimination because
 \begin{equation}
     \tr \mleft( E_i \ketbra{\phi_j}{\phi_j} \mright) = \frac{\lmin \mleft( \rho \mright)}{\eta_j}\,  \delta_{i , \, j} \qquad \textnormal{for all} \;\; i, \, j \in \{1, \, \hdots, \, d\}\,.
 \end{equation}
 Since $\eta_j \tr \mleft( E_j \ketbra{\phi_j}{\phi_j} \mright)$ is equal for all $j \in \{1, \, \hdots, \, d\}$, 
 $\{E_{\mu}\}_\mu$ is called an equal-probability measurement \cite{Chefles_and_Barnett_1998, Eldar_inter_2003}.
 It gives the bound 
 \begin{equation}
 \pud \mleft( \{\eta_j, \, \ket{\phi_j}\} \mright) \geq \sum_{j=1}^{d} \eta_j \tr \mleft( E_j \ketbra{\phi_j}{\phi_j}\mright) 
     = d \,\lmin \mleft( \rho\mright)\,. 
 \end{equation}
A sufficient condition for which this measurement is optimal is
 \begin{equation}
     \eta_j \abs{\langle \umin | \phi_j \rangle }^2 = \frac{\lmin \mleft( \rho \mright)}{d} \qquad \textnormal{for all} \;\; j \in \{1, \, \hdots, \, d\}\,.
 \end{equation}
We summarize the above in the following theorem.
\begin{theorem}\label{thm: SD}
    Let $\{\eta_j, \,\ket{\phi_j}\}$ be an ensemble of $d$ linearly independent vectors $\ket{\phi_j}$ with prior probabilities $\eta_j$, such that $\sum_{j=1}^{d} \eta_j \ketbra{\phi_j}{\phi_j}=\rho$. Then the probability for unambiguous state discrimination is lower bounded by
    \begin{equation}
 \pud \mleft( \{\eta_j, \, \ket{\phi_j}\} \mright) \geq  d \,\lmin \mleft( \rho\mright)\,. 
 \end{equation}
\end{theorem}
We now consider FRIO attacks for some parameter $0 < Q < 1 - d \lmin \mleft( \rho\mright)$. 
\begin{theorem}\label{thm: any state FRIO}
Let $\{\lambda_k\}$ be the eigenvalues of the state $\rho$, arranged in non-increasing order.
The maximal randomness of $\rho$ under a FRIO attack is given by the guessing probability 
\begin{equation}\label{eqn: mt opt FRIO any rho}
 \pq^{*} \mleft( \rho\mright) = \frac{1}{d}\left( \sqrt{t \left( \, \sum_{k=1}^{t}\lambda_k - Q\right)} + \sum_{l=t+1}^{d} \sqrt{\lambda_l} \right)^2 \,,
 \end{equation} 
 where $t \in \{1, \, \hdots, \, d-1\}$ is chosen such that 
 \begin{equation}\label{eqn: r cond}
    \sum_{j=1}^{t-1} \left( \lambda_{j} - \lambda_{t} \right) < Q \leq \sum_{j=1}^{t} \left( \lambda_{j} - \lambda_{t+1} \right)\,.
\end{equation}
 \end{theorem}
 First, note that the quantity $\sum_{j=1}^{t} \left( \lambda_{j} - \lambda_{t} \right)$ is non-negative and doesn't decrease with increasing $t$. For $t=d-1$, the right-hand side of \eqref{eqn: r cond} is
 \begin{equation}
     \sum_{j=1}^{d-1} \left( \lambda_j - \lambda_{d}\right) = 1 - d \lambda_{d} = 1 - d \lmin \mleft( \rho \mright)\,,
 \end{equation}
 while the left-hand side is 0 for $t=1$,
 so for any $\rho$ there exists some $t \in \{1, \, \hdots, \, d\}$ that satisfies \eqref{eqn: r cond}, given any $Q$ in the range $0 < Q < 1 - d \lmin \mleft( \rho \mright)$. We prove the theorem in Appendix \ref{ssec: any state FRIO}, using a decomposition derived from \cite[Theorem 2]{Nakahiri_2012}. The following is a necessary and sufficient condition for the optimal measurement basis $\{\ket{m_j}\}$,
\begin{equation}\label{eqn: opt cond FRIO}
  \langle m_j | \Lambda \sqrt{ \rho} | m_j \rangle = \frac{\tr \mleft( \Lambda \sqrt{\rho} \mright)}{d} \qquad \textnormal{for all} \;\; j \in \{1, \, \hdots, \, d\}\,,  
\end{equation} 
 where 
 \begin{equation}
     \Lambda = \sum_{j=1}^{d} \sqrt{1 - \chi_j} \ketbra{u_j}{u_j}\,,
 \end{equation}
 where $\{\ket{u_j}\}$ is the eigenbasis of $\rho$ and
 \begin{equation}
    \chi_{j} = \begin{cases}
        1 - \frac{1}{t \, \lambda_{j}} \left( \sum_{k=1}^{t}\lambda_{k} - Q \right)\,, \qquad & j \in \{1, \, \hdots, \, t \}\,,
        \\[4 pt]
        0\,, \qquad & j \in \{t+1, \, \hdots, \, d\}\,.
    \end{cases}
\end{equation}
 The condition \eqref{eqn: opt cond FRIO} is met, for instance, when the measurement basis $\{\ket{m_j}\}$ is unbiased to $\{\ket{u_j}\}$.

\subsection{Qubits}\label{ssec: qubits}
\subsubsection{Notation}
In dimension two, we can represent the state $\rho$ and the measurement elements $M_j$ by the Bloch vectors $\vecr$ and $\pm \vec{m}$, where
\begin{equation}
    \rho = \frac{1}{2} \left( \id + \vecr \cdot \vec{\sigma} \right)\,, \qquad  M_j = \frac{1}{2} \left( \id + \left( -1\right)^{j+1} \vec{m} \cdot \vec{\sigma} \right)\,\qquad \textnormal{for all} \;\; j \in \{1, \, 2\}\,, 
\end{equation}
where $\vec{\sigma}$ is a vector of Pauli matrices, $\vec{\sigma} = \left( \sigma_{x}, \, \sigma_{y}, \, \sigma_{z}\right)$. Since $\m$ is a rank-one measurement, $\abs{\vec{m}}^2=1$, and since $\rho$ is mixed, $\abs{\vecr \,}^2<1$. 
Without loss of generality, we introduce a `normal' vector $\vec{n}$ that is orthogonal to $\vec{m}$, i.e. $\vec{m} \cdot \vec{n}=0$, and write $\vecr$ as
\begin{equation}\label{eqn: qub form}
    \vecr = m \vec{m} + p \vec{n}\,, \qquad m^2 + p^2 < 1\,,
\end{equation}
where we further assume that $m$ and $p$ are non-negative. (As a slight abuse of notation, we will sometimes refer to $\vecr$ as the state and to $\{\pm \vec{m}\}$ as the measurement.) 
We can write Eve's decomposition in terms of Bloch vectors $\{\vecr_\mu\}$ as
\begin{equation}
    \vecr = p_0 \vecr_0 + p_1 \vecr_1 + p_2 \vecr_2\,.
\end{equation}
In the unambiguous case, we must have $\vecr_1 = + \vec{m}$ and $\vecr_2 = - \vec{m}$.

\subsubsection{Unambiguous attack}

\begin{observation}\label{obs: qubits}
Given an unambiguous attack, the maximal randomness of a qubit state $\vecr$, with $r = \abs{\vecr \,}$, is given by the guessing probability
\begin{equation}
 \pud^{*} \mleft( \vecr \,\mright) = 1-r
 \end{equation}  
 and is achieved if and only if the measurement is unbiased to the diagonal basis of the state.
\end{observation}
The bound is an immediate consequence of Theorem \ref{thm: any state}, using $\lmin \mleft( \rho \mright)= \frac{1}{2} \left(1-r \right)$. That it is saturated by any measurement basis unbiased to the diagonal basis of the state (i.e. when $m=0$) follows, since the condition \eqref{eqn: suff cond} then holds. To show that Eve can achieve a strictly higher guessing probability when $m \neq 0$, consider the following decomposition (which is not necessarily optimal), 
\begin{equation}
     \vecr_0 = \sqrt{1-m^2} \,\vec{n} + m \vec{m}\,, \qquad \vecr_j = \left(-1 \right)^{j+1} \vec{m}\,,
\end{equation}
with
\begin{equation}
 p_0 = \frac{p}{\sqrt{1-m^2}}\,, \qquad p_j = \frac{1}{2} \left( 1- p_0 \right) \left(1 + \left( -1\right)^{j+1} m \right) \,,  
\end{equation}
for $j \in \{1, \, 2\}$. We find the bound
\begin{equation}\label{eqn: mt upper qubit}
    \pud \mleft( \vecr, \, \{\pm \vec{m}\} \mright) \geq 1 - \frac{p}{\sqrt{1-m^2}}\,.
\end{equation}
Using $m^2 = r^2 - p^2$, for fixed $r$, the right-hand side of \eqref{eqn: mt upper qubit} decreases monotonically with $p$, so when $p < r$ (i.e. when $m \neq 0$), we have
\begin{equation}
    \pud \mleft( \vecr, \, \{\pm \vec{m}\} \mright) > 1 - r\,.
\end{equation}
We now find the eavesdropper's unambiguous guessing probability for any state $\vecr$ and measurement $\{\pm \vec{m}\}$.
\begin{theorem}\label{thm: qubits all UD}
    Given a state $\vecr$, of the form \eqref{eqn: qub form}, measured in the basis $\{\pm \vec{m}\}$, the unambiguous guessing probability of an eavesdropper is 
\begin{equation}\label{eqn: mt qubit UD mine}
    \pud \mleft( \vecr, \, \{\pm \vec{m}\}\mright) = \begin{cases}
        1- p\,, \quad & m + p \leq 1\,, \vspace{0.1 cm}
        \\
        \frac{1-r^2}{2 \left( 1 -m \right)}\,, \quad & m + p > 1 \,.
    \end{cases}
    \end{equation}
\end{theorem}
We sketch the proof here, with full details in Appendix \ref{app: UD qub}. Let's split the optimal strategy of an eavesdropper into two cases. We first consider the case where $\vecr$ lives in the convex hull of $\vec{n}$ and $\{\pm \vec{m}\}$, which we write as $\vecr \in \conv \mleft(  \{\pm \vec{m}\} , \, \vec{n}\mright)$. This condition is equivalent to $m + p \leq 1$. By construction, the state $\vecr$ must admit some decomposition such that $\vecr_0=\vec{n}$ and $\vecr_{j}= \left( -1\right)^{j+1}\vec{m}$.
Since the vectors $\{\pm \vec{m}\}$ are antiparallel, the only valid decomposition is
\begin{equation}
    p_0 = p\,, \qquad p_j = \frac{1}{2} \left( 1 - p  + \left( -1\right)^{j+1} m \right) \qquad \textnormal{for all} \;\; j \in \{1, \, 2\}\,,
\end{equation}
which gives us the expected guessing probability.
This decomposition is shown in Figure \ref{fig: unamb in conv}.

In our second case, $\vecr \notin \conv \mleft(  \{\pm \vec{m}\} , \, \vec{n} \mright)$, no such decomposition is possible. Eve's optimal strategy is to ignore the outcome given by $- \vec{m}$ (which is the least likely outcome, since we assume $m \geq 0$) and choose $\vecr_0=\vec{u}$,
with 
\begin{equation}\label{eqn: mt UD decomp 2}
    p_0 \vec{u} = p \vec{n} + \left(m -p_1 \right) \vec{m}\,, \qquad  p_0 = 1 - p_1\,, \qquad p_1 = \frac{1-r^2}{2 \left( 1-m\right)}\,, \qquad p_2=0\,.
\end{equation}
This decomposition returns the expected guessing probability, and is shown in Figure \ref{fig: unamb out conv}.

We could map these results to a state discrimination scenario with any ensemble of two linearly independent qubit states $\{\eta_j, \, \ket{\phi_j}\}$ using
\begin{equation}\label{eqn: mt m and p}
m = \eta_1 - \eta_2\,, \qquad p = 2 \sqrt{\eta_1 \eta_2} \, \abs{\langle \phi_1 | \phi_2 \rangle}\,.
\end{equation}
Substituting \eqref{eqn: mt m and p} into \eqref{eqn: mt qubit UD mine} would return the results of \cite{Jaeger_1995}. 

\begin{figure}
  \begin{subfigure}{0.43\textwidth}
    \centering
\resizebox{\linewidth}{!}{\input{Tikz/unambin3}}
    \vspace*{5mm}
    \caption{When $\vecr \in \conv \mleft(\pm \{\vec{m}\}, \, \vec{n} \mright)$.}
    \label{fig: unamb in conv}
\end{subfigure}
\hfill
\begin{subfigure}{0.43\textwidth}
\centering
\resizebox{\linewidth}{!}{\input{Tikz/unambout3}}
    \vspace*{5mm}
    \caption{When $\vecr \notin \conv \mleft(\{\pm \vec{m}\}, \, \vec{n} \mright)$.}
    \label{fig: unamb out conv}
\end{subfigure}
\caption{\emph{Unambiguous attacks.} An eavesdropper's optimal decomposition of a qubit state $\vecr$, given a measurement in the basis $\{\pm \vec{m}\}$.}
    \label{fig: unamb qubits}
\end{figure}

\subsubsection{FRIO attack}
\begin{observation}\label{obs: qubits FRIO}
Given a FRIO attack, the maximal randomness of a qubit state $\vecr$, with $r = \abs{\vecr \,}$, is given by the guessing probability
\begin{equation}\label{eqn: obs frio}
 \pq^{*} \mleft( \vecr \,\mright) = \frac{1}{4} \left( \sqrt{1+r-2Q} + \sqrt{1-r} \right)^2\,, 
 \end{equation}  
 and is achieved if and only if the measurement is unbiased to the diagonal basis of the state.
\end{observation}
The parameter $t=1$ satisfies the constraint in \eqref{eqn: r cond},
so \eqref{eqn: obs frio} follows immediately from Theorem \ref{thm: any state FRIO}, using $\lambda_{1}= \frac{1}{2} \left( 1+r\right)$ and $\lambda_2= \frac{1}{2} \left(1-r \right)$. 
We now find the eavesdropper's FRIO guessing probability for any state $\vecr$ and measurement $\{\pm \vec{m}\}$.

The maximum probability $\qmax$ of an inconclusive outcome given a state $\vecr$ and a measurement $\{\pm \vec{m}\}$ is
\begin{equation}
\qmax \mleft( \vecr, \, \{\pm \vec{m}\} \mright) = 1 - \pud \mleft( \vecr, \, \{\pm \vec{m}\}\mright)\,,    
\end{equation}
so from Theorem \ref{thm: app qubits all UD}, we find
\begin{equation}
    \qmax \mleft( \vecr, \, \{\pm \vec{m} \}\mright) = \begin{cases}
        p\,, \quad & m+ p \leq 1\,,
        \\[2pt]
        \frac{p^2 + \left( 1-m\right)^2}{2 \left( 1-m\right)}\,, \quad & m + p >1\,.
    \end{cases} 
\end{equation}
(From hereon we assume the state and measurement are fixed, and drop the explicit dependence of $\qmax$ on $\vecr$ and $\{\pm \vec{m}\}$.) We also introduce a critical value for the FRIO parameter,
 \begin{equation}
     \qcrit = \frac{1}{2} \frac{1-r^2}{1-p}\,.
 \end{equation}
We then solve the guessing probability given a FRIO attack with parameter $Q$, where $0<Q < \qmax$. 
\begin{theorem}\label{thm: qubits FRIO}
Given a qubit state $\vecr$, of the form \eqref{eqn: qub form}, measured in the basis $\{\pm \vec{m}\}$, the guessing probability of an eavesdropper given a fixed rate $Q$ of inconclusive outcomes is
\begin{equation}\label{eqn: mt qubit Q}
    \pq \mleft( \vecr, \, \{\pm \vec{m}\}\mright) = \frac{1}{2}\begin{cases}
         1 - Q + \sqrt{ \left( 1- Q\right)^2 - \left(p -Q \right)^2}\,, \quad &  Q \leq \qcrit\,, \vspace{0.2 cm}
        \\
         1 - Q + m - \frac{Q }{r^2} \left( Am - p \sqrt{r^2 - A^2}\right)\,,  \quad &  Q > \qcrit\,,
    \end{cases}
\end{equation}
where 
\begin{equation}
  A = 1-\frac{\qcrit \left( 1-p\right)}{Q}\,.
\end{equation}
 \end{theorem}
 \begin{figure}
    \begin{subfigure}{0.43\textwidth}
       \centering
       \resizebox{\linewidth}{!}{\input{Tikz/Qin3}}
      \vspace{0.5 cm}
       \caption{When $\vecr \in \conv \mleft( \{\vec{q}_j\}, \, \vec{n} \mright)$.}
      \label{fig: FRIO in}
    \end{subfigure}
    \hfill
    \begin{subfigure}{0.43\textwidth}
        \centering
        \resizebox{\linewidth}{!}{\input{Tikz/Qout3}}
       \vspace{0.5 cm}
        \caption{When $\vecr \notin \conv \mleft( \{\vec{q}_j\}, \,  \vec{n} \mright)$.}
        \label{fig: FRIO out}
    \end{subfigure}
    \caption{\emph{FRIO attacks.} An eavesdropper's optimal decomposition of a qubit state $\vecr$, given a measurement in the basis $\{\pm \vec{m}\}$ and a fixed rate of inconclusive outcomes.}
    \label{fig: FRIO both}
\end{figure}
We sketch the proof here, with full details in Appendix \ref{app: FRIO qub}. As before, Eve's optimal strategy is split into two cases. We first define the states
\begin{equation}
    \vec{q}_j = \left( -1\right)^{j+1} \sqrt{1-q^2} \, \vec{m} + q \vec{n}\,, \qquad q = \frac{p-Q}{1-Q} \qquad \textnormal{for all} \;\; j \in \{1, \, 2\}\,.
\end{equation}
Our first case, where $Q \leq \qcrit$, is equivalent to $\vecr \in \{\vec{n}, \, \{\vec{q}_j\}\}$. By construction, then, $\vecr$ admits a decomposition with $\vecr_0 =\vec{n}$ and $\vecr_j= \vec{q}_j$. It is given by the probability distribution
\begin{equation}
p_0 = Q\,, \qquad p_j = \frac{1}{2} \left( 1 - Q +  \frac{ \left( -1 \right)^{j+1} m}{\sqrt{1-q^2}} \right)\,, \qquad j \in \{1, \, 2\}\,.     
\end{equation}
The guessing probability then follows. This decomposition is shown in Figure \ref{fig: FRIO in}. 

In the second case, when $\vecr \notin \conv \mleft( \{\vec{q}_j\} , \, \vec{n} \mright)$, no such decomposition is possible. Instead, Eve ignores the outcome represented by $-\vec{m}$, and uses the decomposition 
\begin{equation}
   \vecr_0 = \vec{u}\,, \qquad \vecr_1 = \vec{v}\,, \qquad p_0=Q\,, \qquad p_1 = 1-Q\,, \qquad p_2=0\,, 
\end{equation}
where 
\begin{equation}\label{eqn: mt v decomp}
    \vec{v} = \frac{1}{1-Q} \left( \vecr - Q \vec{u} \right)
\end{equation}
and
\begin{equation}\label{eqn: mt u decomp}
    \vec{u} = u_1 \vec{m} + u_2 \vec{n}\,, \qquad u_1=  \frac{Am - p \sqrt{r^2 - A^2}}{r^2 }\,, \qquad u_2 = \frac{Ap + m \sqrt{r^2 - A^2}}{r^2 }\,.  
\end{equation}
This decomposition returns the expected guessing probability, and is shown in Figure \ref{fig: FRIO out}. Swapping to a typical state discrimination scenario by substituting \eqref{eqn: mt m and p} into \eqref{eqn: mt qubit Q}, we can recover the expressions from \cite{Bagan_2012, Jiménez_2021}.

\subsection{Application: random bases in quantum key distribution}\label{ssec: random bases}
The unambiguous guessing probability considered here is, by construction, smaller than the standard `minimum-error' guessing probability. Why, then, would an eavesdropper ever choose such an attack, prioritizing the accuracy of her information over its quantity? 
On one hand, when quantifying the computational unpredictability of the measurement outcomes, it is helpful to consider a predictor who either guesses correctly or refrains from making a prediction \cite{Abbott_2015}. We are primarily motivated, though, by scenarios in which an eavesdropper must limit the disturbance she causes while extracting information from quantum systems. In the following, we show that an eavesdropper may favour an unambiguous attack in some prepare-and-measure quantum key distribution scenarios.

Quantum key distribution, a prominent application of quantum information science, is a task in which randomness does some heavy lifting \cite{Pirandola_2020, Portmann_2022}. We will use as a prototype the original QKD protocol (termed BB84 after its authors \cite{BB84}), starting with a brief outline. We have two spatially separated parties, Alice and Bob. Using a private and uniform source of randomness, whose outputs we label by $a \in \{0, \, 1\}$, Alice generates a key which she wishes to communicate to Bob. She encodes the outcome $a$ in a state $\ket{\psi_{a, \, x}}$, according to the output $x \in \{0, \, 1\}$ of a \emph{second} source of private random numbers. (In the original formulation, the elements $\{x\}$ follow a uniform distribution, but it was later shown that a non-uniform, but still private, source can be used instead to increase the rate of key generation \cite{Lo_2005}.) The trick is for Alice to prepare her states in one of two unbiased bases, e.g., $\{\ket{z_{\pm}}\}$ or $\{\ket{x_{\pm}}\}$, where $\ket{x_{\pm}}= \frac{1}{\sqrt{2}}\left( \ket{z_+} + \ket{z_-}\right)$ (see Table \ref{tab: BB84}).
\begin{table}
    \extrarowheight2.5pt
    \centering
    \begin{tabular}{@{}c|l|p{4em}p{6em}} 
    \multicolumn{4}{c}{}\\[-20pt]
    \multicolumn{2}{c}{} & \multicolumn{2}{c}{$x$}\\\cline{3-4}
    \multicolumn{2}{c}{}            & \centering 0  & \multicolumn{1}{c}{1} \\\cline{3-4}
    \multirow{2}{*}{$a$} & 0  & \centering $\ket{z_{+}}$    & \multicolumn{1}{c|}{$\ket{x_{+}}$ } \\ 
                             & 1 & \centering $\ket{z_-}$   & \multicolumn{1}{c|}{ $\ket{x_-}$} \\\cline{3-4}                        
    \end{tabular}
    \caption{Alice's preparation of the states $\{\ket{\psi_{a, \, x}}\}$ in the BB84 protocol.}
    \label{tab: BB84}
    \end{table}

Alice then sends the states $\{\ket{\psi_{a, \, x}}\}$ to Bob along an insecure quantum channel. Here, `insecure' means the eavesdropper can intercept the states and perform any operation on them that's allowed by quantum theory (she cannot perfectly clone them, for instance \cite{Wootters_1982}). Upon receiving the states, Bob uses a random number generator of his own, with outputs $\{y\}$, to choose a measurement basis, $\{\ket{z_{\pm}}\}$ or $\{\ket{x_{\pm}}\}$\footnote{Typically, Bob's outputs $\{y\}$ follow the same probability distribution as $\{x\}$.}. Bob records his measurement outcomes $\{b\}$ in a string, assigning $b=0$ for outcomes $\ket{z_+}$ or $\ket{x_+}$ and $b=1$ otherwise.  
Alice and Bob then carry out several steps of classical post-processing on their strings. First is the sifting step, where they publicly compare the bases that they used, discarding all entries of the key for which their choice did not coincide. Next comes parameter estimation, where one party, say Alice, uses yet another source of private randomness to choose a subset of key entries to share publicly. This allows them to compute the quantum bit error rate (QBER),
\begin{equation}
    \textnormal{QBER} = p \mleft( a \neq b | x =y \mright)\,,
\end{equation}  
based on which they either abort or continue with the protocol\footnote{We will not consider losses here, i.e. cases where the states do not reach Bob.}. If the error rate is sufficiently low, Alice and Bob proceed to perform error correction, or information reconciliation, to ensure that their final keys are identical, and privacy amplification \cite{Bennett1988, Bennett1995, Renner_composable}, to reduce the eavesdropper's correlations with the final key.

Under the assumption of perfect randomness (among others), this protocol has been proven secure under the most general attacks on the quantum channel \cite{Mayers_1996, Biham_2006, Shor_2000, Kraus_2005}. However, if imperfect randomness is used to encode the states $\{\ket{\psi_{a, \, x}}\}$ \cite{Li_2015, Sun_2020, Chen_2025} or for parameter estimation \cite{Bouda_2012}, the security of the key may be compromised. Let's now imagine that Bob has the qubit state $\vecr$, from \eqref{eqn: qub form}, which he measures in the basis $\{\pm \vec{m}\}$ to produce the outcomes $\{y\}$ (while we have singled out Bob here for convenience, one could instead imagine Alice using this noisy qubit QRNG to produce the outcomes $\{x\}$). An eavesdropper with access to a purification of the state $\vecr$ could perform unambiguous state discrimination on it\footnote{Note that Eve could perform these `state discrimination' measurements long before the protocol begins. Since she knows which states have been steered to Bob's QRNG, she can just wait until Alice sends the corresponding qubits.}. A fraction $\pud \mleft( \vecr, \, \{\pm \vec{m} \}\mright)$ of the time, she knows with certainty in which basis he will measure, while the rest of the time she knows nothing.
When the state $\ket{\psi_{a, \, x}}$ corresponds to an inconclusive round, Eve lets it pass untouched, choosing the most likely outcome as her guess $e$.  On a conclusive round, Eve measures the state in the correct basis. Since this is the basis in which Bob will eventually measure the state, Eve causes no noticeable disturbance, even if this basis is not the one used by Alice. Overall, Eve's guess $e$ is equal to Bob's result $b$ with probability 
\begin{equation}\label{eqn: Eve key}
\begin{aligned}
    p \mleft( e = b \mright) &= \pud \mleft( \vecr, \,  \{\pm \vec{m}\}\mright) + \frac{1}{2}\left( 1 + \vecr \cdot \vec{m} \right)\left( 1 - \pud \mleft( \vecr, \,  \{ \pm\vec{m}\}\mright) \right) 
    \\
    &= \frac{1}{2} \left( 1 + m + \left( 1-m\right) \pud \mleft( \vecr, \,  \{\pm \vec{m}\}\mright)  \right)  \,, 
    \end{aligned}
\end{equation}
where $\frac{1}{2} \left( 1 + \vecr \cdot \vec{m} \right)= \frac{1}{2} \left( 1 + m \right)$ is Bob's average probability of returning the (likelier) outcome $+\vec{m}$. The expression \eqref{eqn: Eve key} quantifies Eve's knowledge of Bob's unprocessed key in the zero-error regime, based on individual attacks, where she causes no disturbance ($\textnormal{QBER}=0$). Notably, this analysis would not be possible if we knew only the min-entropy of Bob's QRNG. 

In a realistic implementation, Alice and Bob must allow for noise in the quantum channel, so they would probably choose some (nonzero) maximum allowed value $Z$ for the QBER. This gives the eavesdropper more freedom in her attack. Let's take for simplicity the case where Alice and Bob's random outcomes $\{x\}$ and $\{y\}$ follow a uniform distribution (in Bob's case, this implies that $m=0$). Eve might now perform FRIO state discrimination, such that with probability $Q$, she receives an inconclusive result and knows nothing about Bob's outcome. With probability $1-Q$, however, she makes a guess. As in the previous strategy, Eve lets the corresponding states pass untouched on an inconclusive round, outputting $e$ at random. Otherwise, she measures them in the basis which she guesses to be correct. With probability $ \pq \mleft(\vecr, \, \{\pm \vec{m}\} \mright)$ she measures the state in the right basis, guessing Bob's outcome correctly without causing a disturbance, while with probability $1 - Q- \pq \mleft(\vecr, \, \{\pm \vec{m}\} \mright)$ she measures in the wrong basis and disturbs the state. When this happens, since the state Bob receives is then unbiased to the basis in which he measures, his output will differ half the time from that of Eve. In total, using \eqref{eqn: mt qubit Q}, we can express Eve's knowledge of Bob's unprocessed key as
\begin{equation}
    p \mleft( e = b \mright) = \pq \mleft( \vecr, \, \{\pm \vec{m}\}\mright) + \frac{1}{2} Q = \frac{1}{2} \left( 1 + \sqrt{ \left( 1- Q\right)^2 - \left(p -Q \right)^2}\right) \,.
\end{equation}
The $\qber$ is
\begin{equation}
    \qber = \frac{1}{2} \left( 1 - Q - \pq\right) = \frac{1}{4} \left( 1 - Q - \sqrt{ \left( 1- Q\right)^2 - \left(p -Q \right)^2} \right) \,.
\end{equation}
To make the most of the allowed error margin, Eve could choose the parameter
\begin{equation}
   Q = \begin{cases}
       p - 4 Z - 2 \sqrt{2 Z\left( 1-p \right)}\,, & \qquad Z \leq \frac{1}{4} \left( 1 - \sqrt{1-p^2}\right)\,,
        \vspace{0.1 cm}
       \\[3pt]
       0\,, & \qquad Z > \frac{1}{4} \left( 1 - \sqrt{1-p^2}\right)\,,
   \end{cases} 
\end{equation}
where $Z$ is the maximum allowed $\qber$. Note that this is not necessarily the optimal approach for Eve, even if she is constrained to measure each of the states individually. Since she is allowed to cause a limited disturbance, her measurements need not even be projective (she might, for example, choose a weak measurement, such that she could tune the resulting disturbance \cite{Silva_2015}). Finding her optimal attack given a noisy QRNG and a bounded $\qber$ remains an interesting open problem. 

Note, finally, that this discussion could be adapted to other prepare-and-measure protocols for quantum key distribution in which the receiver uses a source of private randomness to choose his measurement bases, such as the two-state \cite{Bennett1992} and six-state \cite{Bru__1998} protocols.

\subsection{Noise}\label{ssec: noisy}
Moving beyond qubits, we now consider noisy states in any finite dimension $d$. We define a noisy state $\rho_\ve$ as a pure state $\ket{\phi}$ that has been subjected to an amount $\ve$ of isotropic noise,
\begin{equation}\label{eqn: mt noisy state}
    \rho_\ve = \Delta_{\ve} \mleft( \ketbra{\phi}{\phi} \mright)\,, \qquad \Delta_{\ve} \mleft( \cdot \mright) = \left( 1- \ve \right) \left( \cdot \right) + \frac{\ve}{d} \id\,, \qquad 0 < \ve < 1\,,
\end{equation}
where $\Delta_{\ve} \mleft( \cdot \mright)$ is a depolarizing channel.

\subsubsection{Unambiguous attack}
\begin{observation}\label{obs: noisy}
    The maximal randomness of the noisy state $\rho_\ve$ under an unambiguous attack is given by the guessing probability
    \begin{equation}\label{eqn: UD noisy}
    \pud^{*} \mleft( \rho_\ve \mright) = \ve\,.
\end{equation}
\end{observation}
This follows immediately from Theorem \ref{thm: any state}, noting that the state $\rho_\ve$ has $d-1$ degenerate `smallest' eigenvalues with value $\frac{\ve}{d}$.
This bound is saturated, for example, when the state $\ket{\phi}$ is unbiased to the measurement basis $\{\ket{m_j}\}$, as then we could represent $\rho_\ve$ in a basis $\{\ket{u_j}\}$ that is unbiased with respect to $\{\ket{m_j}\}$, ensuring that $\abs{\langle \umin | m_j \rangle}^2= \frac{1}{d}$ for all $j$. In this work, we consider only these `unbiased' measurements in relation to noisy states.

\subsubsection{FRIO attack}
\begin{observation}\label{obs: noisy FRIO}
    The maximal randomness of the noisy state $\rho_\ve$ under a FRIO attack is given by the guessing probability
    \begin{equation}\label{eqn: mt noisy FRIO}
    \pq^{*} \mleft( \rho_\ve \mright) = \frac{1}{d^2} \left( \sqrt{d \left(1-Q \right) - \ve \left( d-1\right)} + \left( d-1\right) \sqrt{\ve} \right)^2 \,.
\end{equation}
\end{observation}
Note that 
\begin{equation}
    \lambda_1 - \lambda_2 = 1 - \ve = 1 - d \lmin \mleft( \rho\mright)\,,
\end{equation}
so one should choose $t=1$ to satisfy \eqref{eqn: r cond}.
The result then follows from Theorem \ref{thm: any state FRIO} by plugging in
\begin{equation}
    \lambda_1 = \frac{1}{d}\left(d - \left( d-1\right)\ve \right)\,, \qquad \lambda_{k} = \frac{\ve}{d} \qquad \textnormal{for all} \;\; k \in \{2, \, \hdots, \, d\}\,.
\end{equation}
The condition in \eqref{eqn: opt cond FRIO} holds if and only if the state $\ket{\phi}$ is unbiased to the measurement basis $\{m_j\}$.

\subsubsection{Randomness with shared noise}\label{ssec: joint noise}
Although the unambiguous guessing probability is typically smaller than the minimum-error one, we find a case where they coincide: above a certain noise level, given \emph{two} noisy, correlated devices, an eavesdropper achieves a perfect guessing probability, with no need for an inconclusive outcome. Concretely, we consider a scenario where Alice has a noisy state $\rho_\ve$ and a noisy measurement $\m_{\ve}= \{M_j\}_j$, with 
\begin{equation}\label{eqn: noisy meas elements}
    M_j = \Delta_{\ve} \mleft( \ketbra{m_j}{m_j}\mright)\,,
\end{equation}
and, as before, we assume that the basis $\{\ket{m_j}\}$ is unbiased to the state $\ket{\phi}$. In this picture, Alice trusts the description of her devices (i.e. she knows the form of $\rho_\ve$ and $\m_\ve$), but she has no idea \emph{how} the measurement is implemented. We follow a model introduced in \cite{frauchiger2013} and developed in \cite{Dai_2023,Senno_2023,curran2025}, where an eavesdropper has quantum correlations with both the state \emph{and} the measurement device. Eve's optimal strategy is difficult to solve in this case, but we can lower bound her guessing probability by finding that of a `classical' eavesdropper, who can \emph{simultaneously} decompose the state $\rho_\ve$ and the measurement $\m_\ve$ \cite{Senno_2023}. The eavesdropper sends the states $\{\ket{\psi_\mu}\}$ and the measurements $\{\mathcal{N}_\nu\}$ according to a joint probability distribution $\{p_{\mu, \, \nu}\}$, choosing the most probable outcome in each case, where $\mathcal{N}_\nu=\{N_{x, \, \nu}\}_x$ are POVMs ($N_{x, \,\nu} \sgeq 0$ and $\sum_x N_{x, \,\nu}=\id$ for all $\nu$).

\begin{figure}
    \centering
\begin{subfigure}{0.45\textwidth}
    \begin{tikzpicture}
\begin{scope}[scale=0.75]
    \begin{axis}[
        axis lines=middle,
        xlabel={$\delta$},
        ylabel={$P$},
        xlabel style={at={(ticklabel* cs:1.02)}, anchor=west},
        ylabel style={at={(ticklabel* cs:1.02)}, anchor=south},
        ymin=0, ymax=1,
        xmin=0, xmax=1,
        xtick={0,0.5,1},
        ytick={0,0.5,1},
        domain=0:1,
        samples=200,
        legend style={
    at={(0.95,0.05)},
    anchor=south east,
    font=\small},
        legend style={font=\small}]
        \addplot[
            thick,
            black
        ] {x < 0.5 ? 0.5*( 1+ 2*sqrt(x*(1-x))  )  : 1};
        \addlegendentry{shared noise};
        \addplot[
            dashed,
            black
        ] { pow( sqrt(0.5*(1 - sqrt(x*(2 - x)))) + sqrt(x) , 2 ) };
        \addlegendentry{single noise};
    \end{axis}
\end{scope}
\end{tikzpicture}
\vspace{0.5 cm}
    \caption{Minimum-error guessing.}
    \label{fig: tmax mt}
\end{subfigure}
\hfill
\begin{subfigure}{0.45\textwidth}
    \begin{tikzpicture}
\begin{scope}[scale=0.75]
    \begin{axis}[
        axis lines=middle,
        xlabel={$\delta$},
        ylabel={$P$},
        xlabel style={at={(ticklabel* cs:1.02)}, anchor=west},
        ylabel style={at={(ticklabel* cs:1.02)}, anchor=south},
        ymin=0, ymax=1,
        xmin=0, xmax=1,
        xtick={0,0.5,1},
        ytick={0,0.5,1},
        domain=0:1,
        samples=200,
        legend style={
    at={(0.95,0.05)},
    anchor=south east,
    font=\small},
        legend style={font=\small}]
        \addplot[black, thick, domain=0:0.491] { 2*(1 - sqrt(1-x))  };
        \addlegendentry{shared noise};
        \addplot[
            dashed,
            black
        ] {x};
        \addlegendentry{single noise};
        \addplot[thick, black, domain=0.5:1] {1};
    \addplot[only marks, mark=o, mark size=2pt, black] coordinates {(0.5, 0.58)};
    \addplot[only marks, mark=*, mark size=2pt, black] coordinates {(0.5,1)};
    \end{axis}
\end{scope}
\end{tikzpicture}
\vspace{0.5 cm}
    \caption{Unambiguous guessing.}
    \label{fig: T=0 mt}
\end{subfigure}
    \caption{\emph{Shared versus single noise}. Eve's optimal guessing probability $P$ for the single-noise case (dashed) and a lower bound on her guessing probability in the shared-noise case (undashed), both as a function of the total noise $\delta$.}
    \label{fig: joint noise}
\end{figure}
By choosing a joint strategy for the `classical' Eve, in Appendix \ref{app: noisy both}, we find a critical noise value $\ecrit$ such that Eve has a perfect guessing probability when $\ve \geq \ecrit$. For any finite dimension $d$,
\begin{equation}
    \ecrit = \frac{d}{2 \left( d + \sqrt{d}\right)}\,.
\end{equation}
The value of $\ecrit$ increases with $d$ and is always strictly less than a half. Generalizing a qubit example in \cite{curran2025}, we can also compare the case where both the state and the measurement have noise $\ve$ to the case where Alice has a perfect noiseless measurement $\m$ and a \emph{noisier} state $\rho_\delta$, where
\begin{equation}\label{eqn: tot noise}
    \delta = \ve \left( 2 - \ve \right)\,, \qquad \ve = 1 - \sqrt{1 - \delta}\,.
\end{equation}
The total noise parameter $\delta$ is chosen such that the observed measurement statistics are preserved across the two scenarios, i.e.
\begin{equation}
    \tr \mleft( \Delta_{\delta} \mleft( \ketbra{\phi}{ \phi} \mright) \, M_j \mright) = \tr \mleft( \Delta_{\ve} \mleft( \ketbra{\phi}{\phi} \mright) \, \Delta_{\ve} \mleft( M_j \mright) \mright) \qquad \textnormal{for all} \;\; j \in \{1, \, \hdots, \, d\}\,.
\end{equation}
We prove in Appendix \ref{app: noisy both} that Eve achieves a strictly higher guessing probability when she has joint correlations. This continues to hold if we impose an upper bound on her probability of error, or force her to guess unambiguously. The proof works by constructing a piecewise strategy for Eve in the joint-noise case (which is not necessarily optimal) and comparing it to her optimal guessing probability in the single-noise case, which we can derive from \eqref{eqn: UD noisy} and \eqref{eqn: mt noisy FRIO}. A comparison of the strategies is shown for the qubit case in Figure \ref{fig: joint noise}.

\section{Conclusions}\label{ssec: conc}
We have studied the randomness generated by a quantum state and measurement pair, as quantified by the guessing probability of a correlated eavesdropper who makes no error, but who sometimes returns an inconclusive outcome. Relaxing the `unambiguous' constraint, we have also formulated randomness given a fixed rate $Q$ of inconclusive outcomes. By optimizing over all projective measurements, we solved the maximal randomness that can be generated from any quantum state in each case. Both concepts draw from ideas in quantum state discrimination, and exploit the symmetries between this task and the scenario in which an experimentalist generates randomness using characterized, but imperfect, devices. Our results, which centre mostly on qubit states and noisy states in any finite dimension, bound the randomness generated in these typical device-dependent scenarios, given an eavesdropper with arbitrarily high guessing accuracy. In the qubit case, when the measurement outcomes are used to select bases for the prototypical BB84 protocol for quantum key distribution, we showed that an unambiguous attack allows an eavesdropper to gain knowledge about the secret key without causing any disturbance. Further, our example of a noisy state paired with an unbiased noisy measurement shows that correct modelling of noise in one's devices is crucial to avoid overestimating the amount of private randomness generated.  Our work invites future generalizations to states that are not full rank and to more general measurements, where the eavesdropper could use maximal confidence measurements \cite{Croke_2006} in the absence of unambiguous ones.


\section*{Acknowledgments}
Thanks to Antonio Acín for providing feedback on this manuscript. I acknowledge funding
from the Government of Spain (Severo Ochoa CEX2019-000910-S and FUNQIP), Fundació Cellex, Fundació Mir-Puig, Generalitat de Catalunya (CERCA program), ERC
AdG CERQUTE and Ayuda PRE2022-101448 financiada por MCIN/AEI/ 10.13039/501100011033 y por el
FSE+.

\bibliographystyle{quantum}
\bibliography{refs}

@article{Mannalatha_2023,
   title={A comprehensive review of quantum random number generators: concepts, classification and the origin of randomness},
   volume={\, 22},
   ISSN={1573-1332},
   DOI={10.1007/s11128-023-04175-y},
   number={12},
   journal={Quantum Information Processing},
   publisher={Springer Science and Business Media LLC},
   author={Mannalatha, Vaisakh and Mishra, Sandeep and Pathak, Anirban},
   year={2023},
   month=dec,
eprint={2203.00261},
      archivePrefix={arXiv},
      primaryClass={quant-ph},
      url={https://arxiv.org/abs/1206.4145}}

@article{Herrero_Collantes_2017,
   title={Quantum random number generators},
   volume={\, 89},
   ISSN={1539-0756},
   DOI={10.1103/revmodphys.89.015004},
   number={1},
   journal={Reviews of Modern Physics},
   publisher={American Physical Society (APS)},
   author={Herrero-Collantes, Miguel and Garcia-Escartin, Juan Carlos},
   year={2017},
   month=feb,
eprint={1604.03304},
      archivePrefix={arXiv},
      primaryClass={quant-ph},
      url={https://arxiv.org/abs/1206.4145}}

@article{Meng_2024,
  title = {Maximal intrinsic randomness of a quantum state},
  author = {Meng, Shuyang and Curran, Fionnuala and Senno, Gabriel and Wright, Victoria J. and Farkas, M\'at\'e and Scarani, Valerio and Ac\'{\i}n, Antonio},
  journal = {Physical Review A},
  volume = {110},
  issue = {1},
  pages = {L010403},
  numpages = {6},
  year = {2024},
  month = {Jul},
  publisher = {American Physical Society},
  doi = {10.1103/PhysRevA.110.L010403},
eprint={2307.15708},
      archivePrefix={arXiv},
      primaryClass={quant-ph},
      url={https://arxiv.org/abs/1206.4145}
}

@article{Chefles_1998,
   title={Unambiguous discrimination between linearly independent quantum states},
   volume={239},
   ISSN={0375-9601},
   DOI={10.1016/s0375-9601(98)00064-4},
   number={6},
   journal={Physics Letters A},
   publisher={Elsevier BV},
   author={Chefles, Anthony},
   year={1998},
   month=mar, 
pages={339–347},
eprint={quant-ph/9807022},
      archivePrefix={arXiv},
      primaryClass={quant-ph}}

@article{Bagan_2012,
   title={Optimal discrimination of quantum states with a fixed rate of inconclusive outcomes},
   volume={\, 86},
   ISSN={1094-1622},
   DOI={10.1103/physreva.86.040303},
   number={4},
   journal={Physical Review A},
   publisher={American Physical Society (APS)},
   author={Bagan, E. and Muñoz-Tapia, R. and Olivares-Rentería, G. A. and Bergou, J. A.},
   year={2012},
   month=oct,
eprint={1206.4145
},
      archivePrefix={arXiv},
      primaryClass={quant-ph},
      url={https://arxiv.org/abs/1206.4145}}

@article{Senno_2023,
   title={Quantifying the Intrinsic Randomness of Quantum Measurements},
   volume={131},
   ISSN={1079-7114},
   DOI={10.1103/physrevlett.131.130202},
   number={13},
   journal={Physical Review Letters \,},
   publisher={American Physical Society (APS)},
   author={Senno, Gabriel and Strohm, Thomas and Acín, Antonio},
   year={2023},
   month=sep,
    eprint={2211.03581v1},
    archivePrefix={arXiv},
    primaryClass={quant-ph},
    url={https://arxiv.org/abs/2211.03581v1}}

@book{Skrzypczyk_2023,
author = {Skrzypczyk, Paul and Cavalcanti, Daniel},
title = {Semidefinite Programming in Quantum Information Science},
publisher = {IOP Publishing},
year = {2023},
series = {2053-2563},
isbn = {978-0-7503-3343-6},
url = {https://dx.doi.org/10.1088/978-0-7503-3343-6},
doi = {10.1088/978-0-7503-3343-6}
}

@book{waltrous2018info,
  author={Watrous, John},
  title={The Theory of Quantum Information},
  publisher={Cambridge University Press},
year={2018},
doi = {
10.1017/9781316848142},
address = {Cambridge},
URL = {https://www.cambridge.org/core/books/theory-of-quantum-information/AE4AA5638F808D2CFEB070C55431D897}
}

@article{Eldar_UD_2003,
   title={A semidefinite programming approach to optimal unambiguous discrimination of quantum states},
    author={Eldar, Yonina C.},
   volume={49},
   ISSN={0018-9448},
   DOI={10.1109/tit.2002.807291},
   number={2},
   journal={IEEE Transactions on Information Theory},
   publisher={Institute of Electrical and Electronics Engineers (IEEE)},
   year={2003},
   month=feb, 
    pages={446–456},
    eprint={quant-ph/0206093v2},
    archivePrefix={arXiv},
    primaryClass={quant-ph}}

@article{Chefles_inter_1998,
author = {Anthony Chefles and Stephen M. Barnett},
title = {Strategies for discriminating between non-orthogonal quantum states},
journal = {Journal of Modern Optics},
volume = {45},
number = {6},
pages = {1295--1302},
year = {1998},
publisher = {Taylor \& Francis},
doi = {10.1080/09500349808230919}
}

@book{horn2012matrix,
  author={Horn, Roger A. and Johnson, Charles R.},
  title={Matrix Analysis},
  publisher={Cambridge University Press},
year={2012},
doi = {
10.1017/CBO9780511810817},
address = {Cambridge},
edition   = {2nd},
URL = {https://www.cambridge.org/core/books/matrix-analysis/9CF2CB491C9E97948B15FAD835EF9A8B}
}

@article{Jaeger_1995,
title = {Optimal distinction between two non-orthogonal quantum states},
journal = {Physics Letters A},
volume = {197},
number = {2},
pages = {83-87},
year = {1995},
issn = {0375-9601},
doi = {https://doi.org/10.1016/0375-9601(94)00919-G},
author = {Gregg Jaeger and Abner Shimony}
}

@article{Schrödinger_1935, title={Discussion of Probability Relations between Separated Systems}, 
volume={31}, 
DOI={10.1017/S0305004100013554}, 
number={4}, 
journal={Mathematical Proceedings of the Cambridge Philosophical Society}, 
author={Schrödinger, Erwin}, 
year={1935}, 
pages={555–563}}

@article{Hughston_1993,
title = {A complete classification of quantum ensembles having a given density matrix},
journal = {Physics Letters A},
volume = {183},
number = {1},
pages = {14-18},
year = {1993},
issn = {0375-9601},
doi = {https://doi.org/10.1016/0375-9601(93)90880-9},
url = {https://www.sciencedirect.com/science/article/pii/0375960193908809},
author = {Lane P. Hughston and Richard Jozsa and William K. Wootters}
}

@article{gisin_1989, 
title={Stochastic quantum dynamics and relativity}, 
nolink={},
volume={62}, 
ISSN={0018-0238}, 
number={4}, 
journal={Helvetica Physica Acta}, 
author={Gisin, Nicolas}, 
year={1989}, 
month={May}, 
pages={363–371} }

@misc{frauchiger2013,
      title={True randomness from realistic quantum devices}, 
      author={Daniela Frauchiger and Renato Renner and Matthias Troyer},
      year={2013},
      eprint={1311.4547},
      archivePrefix={arXiv},
      primaryClass={quant-ph},
      url={https://arxiv.org/abs/1311.4547}, 
}

@article{Dai_2023,
  title = {Intrinsic randomness under general quantum measurements},
  author = {Dai, Hao and Chen, Boyang and Zhang, Xingjian and Ma, Xiongfeng},
  journal = {Physical Review Research},
  volume = {5},
  issue = {3},
  pages = {033081},
  numpages = {16},
  year = {2023},
  month = {Aug},
  publisher = {American Physical Society},
  doi = {10.1103/PhysRevResearch.5.033081},
eprint={2203.08624},
    archivePrefix={arXiv},
    primaryClass={quant-ph}
}

@misc{curran2025,
      title={Maximal intrinsic randomness of noisy quantum measurements}, 
      author={Fionnuala Curran and Morteza Moradi and Gabriel Senno and Magdalena Stobinska and Antonio Acín},
      year={2025},
      eprint={2506.22294},
      archivePrefix={arXiv},
      primaryClass={quant-ph},
      url={https://arxiv.org/abs/2506.22294}, 
}

@article{Ivanovic_1987,
title = {How to differentiate between non-orthogonal states},
journal = {Physics Letters A},
volume = {123},
number = {6},
pages = {257-259},
year = {1987},
issn = {0375-9601},
doi = {https://doi.org/10.1016/0375-9601(87)90222-2},
author = {I. D. Ivanovic}
}

@article{Dieks_1988,
title = {Overlap and distinguishability of quantum states},
journal = {Physics Letters A},
volume = {126},
number = {5},
pages = {303-306},
year = {1988},
issn = {0375-9601},
doi = {https://doi.org/10.1016/0375-9601(88)90840-7},
author = {D. Dieks}
}

@article{Peres_1988,
title = {How to differentiate between non-orthogonal states},
journal = {Physics Letters A},
volume = {128},
number = {1},
pages = {19},
year = {1988},
issn = {0375-9601},
doi = {https://doi.org/10.1016/0375-9601(88)91034-1},
author = {Peres, Asher}
}

@article{Chefles_and_Barnett_1998,
   title={Optimum unambiguous discrimination between linearly independent symmetric states},
    DOI={10.1016/s0375-9601(98)00827-5},
   volume={250},
   ISSN={0375-9601},
   number={4–6},
   journal={Physics Letters A},
   publisher={Elsevier BV},
   author={Chefles, Anthony and Barnett, Stephen M.},
   year={1998},
   month=dec, 
pages={223–229},
eprint={9807023
},
archivePrefix={arXiv},
primaryClass={quant-ph}
}

@article{Croke_2006,
   title={Maximum Confidence Quantum Measurements},
   volume={96},
   ISSN={1079-7114},
   DOI={10.1103/physrevlett.96.070401},
   number={7},
   journal={Physical Review Letters \,},
   publisher={American Physical Society (APS)},
   author={Croke, Sarah and Andersson, Erika and Barnett, Stephen M. and Gilson, Claire R. and Jeffers, John},
   year={2006},
   month=feb,
eprint={quant-ph/0604026v1},
      archivePrefix={arXiv},
      primaryClass={quant-ph}}

@article{Zhang_inter_1999,
   title={General strategies for discrimination of quantum states},
   volume={261},
   ISSN={0375-9601},
   DOI={10.1016/s0375-9601(99)00566-6},
   number={1–2},
   journal={Physics Letters A},
   publisher={Elsevier BV},
   author={Zhang, Chuan-Wei and Li, Chuan-Feng and Guo, Guang-Can},
   year={1999},
   month=oct, pages={25–29},
eprint={quant-ph/9908001},
    archivePrefix={arXiv},
    primaryClass={quant-ph}}

@article{Eldar_inter_2003,
   title={Mixed-quantum-state detection with inconclusive results},
   volume={\,67},
   ISSN={1094-1622},
   DOI={10.1103/physreva.67.042309},
   number={4},
   journal={Physical Review A},
   publisher={American Physical Society (APS)},
   author={Eldar, Yonina C.},
   year={2003},
   month=apr,
eprint={quant-ph/0211121},
      archivePrefix={arXiv},
      primaryClass={quant-ph}}

@article{Hayashi_2008,
   title={State discrimination with error margin and its locality},
   volume={\, 78},
   ISSN={1094-1622},
   url={http://dx.doi.org/10.1103/PhysRevA.78.012333},
   DOI={10.1103/physreva.78.012333},
   number={1},
   journal={Physical Review A},
   publisher={American Physical Society (APS)},
   author={Hayashi, A. and Hashimoto, T. and Horibe, M.},
   year={2008},
   month=jul,
eprint={0804.4349},
      archivePrefix={arXiv},
      primaryClass={quant-ph}}

@book{Nesterov1994,
author = {Nesterov, Yurii and Nemirovskii, Arkadii},
title = {Interior-Point Polynomial Algorithms in Convex Programming},
publisher = {Society for Industrial and Applied Mathematics},
year = {1994},
doi = {10.1137/1.9781611970791},
address = {},
edition   = {},
URL = {https://epubs.siam.org/doi/abs/10.1137/1.9781611970791}
}

@article{Law_2014,
   title={Quantum randomness extraction for various levels of characterization of the devices},
   volume={47},
   ISSN={1751-8121},
   DOI={10.1088/1751-8113/47/42/424028},
   number={42},
   journal={Journal of Physics A: Mathematical and Theoretical},
   publisher={IOP Publishing},
   author={Law, Yun Zhi and Thinh, Le Phuc and Bancal, Jean-Daniel and Scarani, Valerio},
   year={2014},
   month=oct, 
   pages={424028},
   eprint={1401.4243},
    archivePrefix={arXiv},
    primaryClass={quant-ph}}

@article{Konig_2009,
	doi = {10.1109/tit.2009.2025545},
  year = 2009,
month = {September},
publisher = {Institute of Electrical and Electronics Engineers ({IEEE})},
volume = {\, 55},
number = {9},
author = {Robert Konig and Renato Renner and Christian Schaffner},
title = {The Operational Meaning of Min- and Max-Entropy},
journal = {{IEEE} Transactions on Information Theory\,},
eprint={0807.1338},
      archivePrefix={arXiv},
      primaryClass={quant-ph}
}

@misc{renner2006thesis,
      title={Security of Quantum Key Distribution}, 
      author={Renato Renner},
      year={2006},
      eprint={quant-ph/0512258},
      archivePrefix={arXiv},
      primaryClass={quant-ph},
      url={https://arxiv.org/abs/quant-ph/0512258}, 
}

@article{Konig_2011,
   title={Sampling of Min-Entropy Relative to Quantum Knowledge},
   volume={57},
   ISSN={1557-9654},
   DOI={10.1109/tit.2011.2146730},
   number={7},
   journal={IEEE Transactions on Information Theory},
   publisher={Institute of Electrical and Electronics Engineers (IEEE)},
   author={Konig, Robert and Renner, Renato},
   year={2011},
   month=jul, pages={4760–4787},
eprint={0712.4291},
      archivePrefix={arXiv},
      primaryClass={quant-ph}}

@misc{anco_2024secure,
      title={How much secure randomness is in a quantum state?}, 
      author={Kriss Gutierrez Anco and Tristan Nemoz and Peter Brown},
      year={2024},
      eprint={2410.16447},
      archivePrefix={arXiv},
      primaryClass={quant-ph}
}

@article{Touzel_2007,
   title={Optimal bounded-error strategies for projective measurements in nonorthogonal-state discrimination},
   volume={\, 76},
   ISSN={1094-1622},
   DOI={10.1103/physreva.76.062314},
   number={6},
   journal={Physical Review A},
   publisher={American Physical Society (APS)},
   author={Touzel, M. A. P. and Adamson, R. B. A. and Steinberg, A. M.},
   year={2007},
   month=dec,
eprint={0708.1540},
      archivePrefix={arXiv},
      primaryClass={quant-ph}}

@article{Gräfe_2014,
   title={On-chip generation of high-order single-photon $\textnormal{W}$-states },
   volume={\, 8},
   DOI={10.1038/nphoton.2014.204},
   issue={10},
   journal={Nature Photonics},
   author={Gräfe, Markus and Heilmann, René  and Perez-Leija, Armando and Keil, Robert and Dreisow, Felix  and Heinrich, Matthias and Moya-Cessa, Hector and Nolte, Stefan and Christodoulides, Demetrios N. and Szameit, Alexander},
   year={2014},
   month=aug}

@article{Gao_2023,
   title={Scalable Generation and Detection of on-Demand W States in Nanophotonic Circuits},
   volume={23},
   ISSN={1530-6992},
   url={http://dx.doi.org/10.1021/acs.nanolett.3c01551},
   DOI={10.1021/acs.nanolett.3c01551},
   number={11},
   journal={Nano Letters},
   publisher={American Chemical Society (ACS)},
   author={Gao, Jun and Santos, Leonardo and Krishna, Govind and Xu, Ze-Sheng and Iovan, Adrian and Steinhauer, Stephan and Gühne, Otfried and Poole, Philip J. and Dalacu, Dan and Zwiller, Val and Elshaari, Ali W.},
   year={2023},
   month=may, 
pages={5350–5357}, 
eprint={2307.06116},
      archivePrefix={arXiv},
      primaryClass={quant-ph}}

@article{Sun_2002,
  title = {Mathematical nature of and a family of lower bounds for the success probability of unambiguous discrimination},
  author = {Sun, Xiaoming and Zhang, Shengyu and Feng, Yuan and Ying, Mingsheng},
  journal = {Physical Review A},
  volume = {65},
  issue = {4},
  pages = {044306},
  numpages = {3},
  year = {2002},
  month = {Apr},
  publisher = {American Physical Society},
  doi = {10.1103/PhysRevA.65.044306}}

@article{Herzog_2012,
   title={Optimal state discrimination with a fixed rate of inconclusive results: Analytical solutions and relation to state discrimination with a fixed error rate},
   volume={\, 86},
   ISSN={1094-1622},
   DOI={10.1103/physreva.86.032314},
   number={3},
   journal={Physical Review A},
   publisher={American Physical Society (APS)},
   author={Herzog, Ulrike},
   year={2012},
   month=sep,
eprint={1206.4412},
      archivePrefix={arXiv},
      primaryClass={quant-ph}}

@article{Fiur_ek_2003,
   title={Optimal discrimination of mixed quantum states involving inconclusive results},
   volume={\, 67},
   ISSN={1094-1622},
   url={http://dx.doi.org/10.1103/PhysRevA.67.012321},
   DOI={10.1103/physreva.67.012321},
   number={1},
   journal={Physical Review A},
   publisher={American Physical Society (APS)},
   author={Fiurášek, Jaromír and Ježek, Miroslav},
   year={2003},
   month=jan,
eprint={quant-ph/0208126v1},
      archivePrefix={arXiv},
      primaryClass={quant-ph}
}

@article{Sugimoto_2009,
   title={Discrimination with error margin between two states: Case of general occurrence probabilities},
   volume={\, 80},
   ISSN={1094-1622},
   DOI={10.1103/physreva.80.052322},
   number={5},
   journal={Physical Review A},
   publisher={American Physical Society (APS)},
   author={Sugimoto, H. and Hashimoto, T. and Horibe, M. and Hayashi, A.},
   year={2009},
   month=nov,
eprint={0906.4884v2},
      archivePrefix={arXiv},
      primaryClass={quant-ph}}

@article{Ekert_1994,
  title = {Eavesdropping on quantum-cryptographical systems},
  author = {Ekert, Artur K. and Huttner, Bruno and Palma, G. Massimo and Peres, Asher},
  journal = {Physical Review A},
  volume = {50},
  issue = {2},
  pages = {1047--1056},
  numpages = {0},
  year = {1994},
  month = {Aug},
  publisher = {American Physical Society},
  doi = {10.1103/PhysRevA.50.1047},
  url = {https://link.aps.org/doi/10.1103/PhysRevA.50.1047}
}

@article{Dusek_2000,
   title={Unambiguous state discrimination in quantum cryptography with weak coherent states},
   volume={\, 62},
   ISSN={1094-1622},
   DOI={10.1103/physreva.62.022306},
   number={2},
   journal={Physical Review A},
   publisher={American Physical Society (APS)},
   author={Dušek, Miloslav and Jahma, Mika and Lütkenhaus, Norbert},
   year={2000},
   month=jul,
eprint={quant-ph/9910106},
      archivePrefix={arXiv},
      primaryClass={quant-ph}}

@book{helstrom_1976,
   title={Quantum Detection and Estimation Theory},
   publisher={Academic Press, New York},
   author={Helstrom, Carl W.},
   nolink={},
   year={1976}}

@article{Holevo_1973,
title = {Statistical decision theory for quantum systems},
journal = {Journal of Multivariate Analysis},
volume = {3},
number = {4},
pages = {337-394},
year = {1973},
issn = {0047-259X},
doi = {https://doi.org/10.1016/0047-259X(73)90028-6},
author = {Holevo, A. S.}
}

@Article{Jiménez_2021,
author = {Jiménez, Omar and Solís–Prosser, Miguel Angel and Neves, Leonardo and Delgado, Aldo},
title = {Mutual Information and Quantum Discord in Quantum State Discrimination with a Fixed Rate of Inconclusive Outcomes},
journal = {Entropy},
volume = {\,23},
year = {2021},
number = {1},
article-number = {73},
PubMedID = {33418984},
ISSN = {1099-4300},
DOI = {10.3390/e23010073}
}

@article{Bennett1992,
  title = {Quantum cryptography using any two nonorthogonal states},
  author = {Bennett, Charles H.},
  journal = {Physical Review Letters},
  volume = {68},
  issue = {21},
  pages = {3121--3124},
  numpages = {0},
  year = {1992},
  month = {May},
  publisher = {American Physical Society},
  doi = {10.1103/PhysRevLett.68.3121},
  url = {https://link.aps.org/doi/10.1103/PhysRevLett.68.3121}
}

@article{Yuen_1996,
doi = {10.1088/1355-5111/8/4/015},
url = {https://doi.org/10.1088/1355-5111/8/4/015},
year = {1996},
month = {aug},
publisher = {},
volume = {8},
number = {4},
pages = {939},
author = {Horace P. Yuen},
title = {Quantum amplifiers, quantum duplicators and quantum cryptography},
journal = {Quantum and Semiclassical Optics: Journal of the European Optical Society Part B}
}

@article{Curty2007,
  author       = {Curty, Marcos and
                  Zhang, Lucy Liuxuan and
                 Lo, Hoi{-}Kwong and
                  L{\"{u}}tkenhaus, Norbert},
  title        = {Sequential attacks against differential-phase-shift quantum key distribution
                  with weak coherent states},
  journal      = {Quantum Information and Computation },
  volume       = {7},
  number       = {7},
  pages        = {665--688},
  year         = {2007},
  doi          = {10.26421/QIC7.7-7},
  eprint={quant-ph/0609094},
      archivePrefix={arXiv},
      primaryClass={quant-ph}
}

@article{Branciard_2007,
   title={Zero-error attacks and detection statistics in the coherent one-way protocol for quantum cryptography},
   volume={7},
   ISSN={1533-7146 },
   DOI={10.26421/QIC7.7-6},
   number={7},
   journal={Quantum Information and Computation},
   publisher={Rinton Press},
   author={Branciard, Cyril and Gisin, Nicolas and L{\"{u}}tkenhaus, Norbert and Scarani, Valerio},
   year={2007},
   month=sep, 
   pages={639-664},
   eprint={quant-ph/0609090},
      archivePrefix={arXiv},
      primaryClass={quant-ph}
   }

@article{Brask_2017,
   title={Megahertz-Rate Semi-Device-Independent Quantum Random Number Generators Based on Unambiguous State Discrimination},
   volume={\, 7},
   ISSN={2331-7019},
   url={http://dx.doi.org/10.1103/PhysRevApplied.7.054018},
   DOI={10.1103/physrevapplied.7.054018},
   number={5},
   journal={Physical Review Applied},
   publisher={American Physical Society (APS)},
   author={Brask, Jonatan Bohr and Martin, Anthony and Esposito, William and Houlmann, Raphael and Bowles, Joseph and Zbinden, Hugo and Brunner, Nicolas},
   year={2017},
   month=may,
   eprint={1612.06566v2},
      archivePrefix={arXiv},
      primaryClass={quant-ph}}

@inbook{Abbott_2015,
   title={On the Unpredictability of Individual Quantum Measurement Outcomes},
   ISBN={9783319235349},
   ISSN={1611-3349},
   url={http://dx.doi.org/10.1007/978-3-319-23534-9_4},
   DOI={10.1007/978-3-319-23534-9_4},
   booktitle={Fields of Logic and Computation II},
   publisher={Springer International Publishing},
   author={Abbott, Alastair A. and Calude, Cristian S. and Svozil, Karl},
   year={2015},
   pages={69–86},
   eprint={1403.2738v4},
      archivePrefix={arXiv},
      primaryClass={quant-ph}}

@inproceedings{BB84,
author = {Bennett, Charles H. and Gilles Brassard},
title = {Quantum cryptography: Public key distribution and coin tossing},
booktitle = {Proceedings of IEEE International Conference on Computers, Systems, and Signal Processing},
year = {1984},
pages = {175–179},
nolink={}
}

@article{Pirandola_2020,
   title={Advances in quantum cryptography},
   volume={12},
   ISSN={1943-8206},
   DOI={10.1364/aop.361502},
   number={4},
   journal={Advances in Optics and Photonics},
   publisher={Optica Publishing Group},
   author={Pirandola, S. and Andersen, U. L. and Banchi, L. and Berta, M. and Bunandar, D. and Colbeck, R. and Englund, D. and Gehring, T. and Lupo, C. and Ottaviani, C. and Pereira, J. L. and Razavi, M. and Shamsul Shaari, J. and Tomamichel, M. and Usenko, V. C. and Vallone, G. and Villoresi, P. and Wallden, P.},
   year={2020},
   month=dec, pages={1012},
   eprint={1906.01645v1},
      archivePrefix={arXiv},
      primaryClass={quant-ph}}

@article{Portmann_2022,
   title={Security in quantum cryptography},
   volume={\, 94},
   ISSN={1539-0756},
   url={http://dx.doi.org/10.1103/RevModPhys.94.025008},
   DOI={10.1103/revmodphys.94.025008},
   number={2},
   journal={Reviews of Modern Physics},
   publisher={American Physical Society (APS)},
   author={Portmann, Christopher and Renner, Renato},
   year={2022},
   month=jun,
   eprint={2102.00021},
      archivePrefix={arXiv},
      primaryClass={quant-ph}}

@article{Lo_2005,
   title={Efficient Quantum Key Distribution Scheme and a Proof of Its Unconditional Security},
   volume={\, 18},
   DOI={10.1007/s00145-004-0142-y},
   number={4},
   journal={Journal of Cryptology},
   author={Lo, H.K. and Chau, H. and Ardehali, M.},
   year={2005},
   pager={133–165},
   eprint={quant-ph/0011056v3},
      archivePrefix={arXiv},
      primaryClass={quant-ph}
   }

@article{Wootters_1982,
   title={A single quantum cannot be cloned},
   volume={299},
   DOI={10.1038/299802a0},
   journal={Nature},
   author={Wootters, W. and Zurek, W.},
   year={1982},
   pages={802–803}}

@article{Bennett1988,
author = {Bennett, Charles H. and Brassard, Gilles and Robert, Jean-Marc},
title = {Privacy Amplification by Public Discussion},
journal = {SIAM Journal on Computing},
volume = {17},
number = {2},
pages = {210-229},
year = {1988},
doi = {10.1137/0217014}
}

@article{Bennett1995,
  author={Bennett, C. H. and Brassard, G. and Crepeau, C. and Maurer, U. M.},
  journal={IEEE Transactions on Information Theory}, 
  title={Generalized privacy amplification}, 
  year={1995},
  volume={41},
  number={6},
  pages={1915-1923},
  doi={10.1109/18.476316}}

@InProceedings{Renner_composable,
author="Renner, Renato
and K{\"o}nig, Robert",
title="Universally Composable Privacy Amplification Against Quantum Adversaries",
booktitle="Theory of Cryptography",
year="2005",
publisher="Springer Berlin Heidelberg",
pages="407--425",
eprint={quant-ph/0403133},
      archivePrefix={arXiv},
      primaryClass={quant-ph},
      doi={10.1007/978-3-540-30576-7_22}
}

@InProceedings{Mayers_1996,
author="Mayers, Dominic",
title="Quantum Key Distribution and String Oblivious Transfer in Noisy Channels",
booktitle="Advances in Cryptology --- CRYPTO '96",
year="1996",
publisher="Springer Berlin Heidelberg",
address="Berlin, Heidelberg",
pages="343--357",
eprint={quant-ph/9606003
},
      archivePrefix={arXiv},
      primaryClass={quant-ph},
      doi={10.1007/3-540-68697-5_26}
}

@article{Shor_2000,
   title={Simple Proof of Security of the {BB84} Quantum Key Distribution Protocol},
   volume={85},
   ISSN={1079-7114},
   url={http://dx.doi.org/10.1103/PhysRevLett.85.441},
   DOI={10.1103/physrevlett.85.441},
   number={2},
   journal={Physical Review Letters},
   publisher={American Physical Society (APS)},
   author={Shor, Peter W. and Preskill, John},
   year={2000},
   month=jul, pages={441–444},
   eprint={quant-ph/0003004
},
      archivePrefix={arXiv},
      primaryClass={quant-ph}}

@article{Kraus_2005,
   title={Lower and Upper Bounds on the Secret-Key Rate for Quantum Key Distribution Protocols Using One-Way Classical Communication},
   volume={\, 95},
   ISSN={1079-7114},
   DOI={10.1103/physrevlett.95.080501},
   number={8},
   journal={Physical Review Letters},
   publisher={American Physical Society (APS)},
   author={Kraus, B. and Gisin, N. and Renner, R.},
   year={2005},
   month=aug,
   eprint={quant-ph/0410215},
      archivePrefix={arXiv},
      primaryClass={quant-ph}}

@article{Biham_2006,
   title={A Proof of the Security of Quantum Key Distribution},
   volume={19},
   DOI={10.1007/s00145-005-0011-3},
   journal={Journal of Cryptography},
   publisher={Springer Nature},
   author={Biham, Eli  and Boyer, Michel and Boykin, P. Oscar  and Mor, Tal  and Roychowdhury, Vwani},
   year={2006},
   pages={381–439},
   eprint={quant-ph/0511175},
      archivePrefix={arXiv},
      primaryClass={quant-ph}}

@article{Chen_2025,
   title={Improving the performance of quantum key distribution with weak-randomness basis selection},
   volume={\, 24},
   DOI={10.1007/s11128-025-04780-z},
   number={162},
   journal={ Quantum Information Processing},
   publisher={Springer Nature},
   author={Chen, Zhi-Jiang and Hao, Chen-Peng  and Gong, Li and  Guo, Jian-Sheng and Wang, Yang and  Zhang, Chun-Mei and  Li, Hong-Wei},
   year={2025}}

@article{Sun_2020,
   title={Security evaluation of quantum key distribution with weak basis-choice flaws},
   volume={\, 10},
   ISSN={2045-2322},
   url={http://dx.doi.org/10.1038/s41598-020-75159-6},
   DOI={10.1038/s41598-020-75159-6},
   number={1},
   journal={Scientific Reports},
   publisher={Springer Science and Business Media LLC},
   author={Sun, Shi-Hai and Tian, Zhi-Yu and Zhao, Mei-Sheng and Ma, Yan},
   year={2020},
   month=oct,
   eprint={2008.06290},
      archivePrefix={arXiv},
      primaryClass={quant-ph}}

@article{Li_2015,
   title={Randomness determines practical security of {BB84} quantum key distribution},
   volume={\, },
   DOI={10.1038/srep16200},
   number={16200},
   journal={Scientific Reports},
   author={Li, Hong-Wei and Yin, Zhen-Qiang  and Wang, Shuang and Qian, Yong-Jun and Chen, Wei and Guo, Guang-Can  and Han, Zheng-Fu },
   year={2015}, 
   eprint={1508.06396},
      archivePrefix={arXiv},
      primaryClass={quant-ph}}

@article{Bouda_2012,
   title={Weak randomness seriously limits the security of quantum key distribution},
   volume={\,86},
   ISSN={1094-1622},
   DOI={10.1103/physreva.86.062308},
   number={6},
   journal={Physical Review A},
   publisher={American Physical Society (APS)},
   author={Bouda, Jan and Pivoluska, Matej and Plesch, Martin and Wilmott, Colin},
   year={2012},
   month=dec,
   eprint={1206.1287},
      archivePrefix={arXiv},
      primaryClass={quant-ph}}

@article{Silva_2015,
   title={Multiple Observers Can Share the Nonlocality of Half of an Entangled Pair by Using Optimal Weak Measurements},
   volume={\, 114},
   ISSN={1079-7114},
   url={http://dx.doi.org/10.1103/PhysRevLett.114.250401},
   DOI={10.1103/physrevlett.114.250401},
   number={25},
   journal={Physical Review Letters},
   publisher={American Physical Society (APS)},
   author={Silva, Ralph and Gisin, Nicolas and Guryanova, Yelena and Popescu, Sandu},
   year={2015},
   month=jun,
   eprint={1408.2272},
      archivePrefix={arXiv},
      primaryClass={quant-ph}}

@article{Nakahiri_2012,
  title = {Discrimination between geometrically uniform quantum states with inconclusive results},
  author = {Nakahira, Kenji and Usuda, Tsuyoshi Sasaki and Kato, Kentaro},
  journal = {Physical Review A},
  volume = {86},
  issue = {3},
  pages = {032316},
  numpages = {9},
  year = {2012},
  month = {Sep},
  publisher = {American Physical Society},
  doi = {10.1103/PhysRevA.86.032316},
  url = {https://link.aps.org/doi/10.1103/PhysRevA.86.032316}
}

@article{Bru__1998,
   title={Optimal Eavesdropping in Quantum Cryptography with Six States},
   volume={81},
   ISSN={1079-7114},
   url={http://dx.doi.org/10.1103/PhysRevLett.81.3018},
   DOI={10.1103/physrevlett.81.3018},
   number={14},
   journal={Physical Review Letters},
   publisher={American Physical Society (APS)},
   author={Bruß, Dagmar},
   year={1998},
   month=Oct, pages={3018–3021},
   eprint={quant-ph/9805019},
    archivePrefix={arXiv},
      primaryClass={quant-ph}}

\appendix

\setcounter{theorem}{0}

\section{SDPs}\label{sec: SDPs}
\subsection{General form}
Following the conventions of \cite{Skrzypczyk_2023, waltrous2018info},
we will consider Hermitian operators, say $F$, which satisfy $F=F^{\dagger}$, and hermiticity-preserving maps $\Lambda \mleft(\cdot \mright)$, such that $\Lambda \mleft( F \mright)$ is Hermitian whenever $F$ is Hermitian. The adjoint map of $\Lambda \mleft( \cdot \mright)$, defined as $\Lambda^{\dagger} \mleft( \cdot \mright)$, is the unique map that satisfies
\begin{equation}
    \tr \mleft( \Lambda \mleft( X \mright) \, Y  \mright) = \tr \mleft( X \, \Lambda^{\dagger} \mleft( Y \mright)  \mright)\qquad \textnormal{for all} \;\; X, \, Y\,.   
\end{equation}
Given Hermitian operators $A$, $\{B_i\}$ and $\{C_j\}$ and hermiticity-preserving maps $\{\Phi_{i} \mleft( \cdot \mright) \}$ and $\{\Gamma_j \mleft( \cdot \mright)\}$, our primal and dual semidefinite programming problems are\footnote{The primal and dual SDPs in \eqref{eqn: standard SDP} are equivalent to those in \cite[Equations 2.1 and 2.22]{Skrzypczyk_2023} if we replace $C_j \rightarrow - C_j$ and $\Gamma_j \mleft( \cdot \mright) \rightarrow - \Gamma_j \mleft( \cdot \mright)$ for all $j$.}
\begin{align}\label{eqn: standard SDP}
  \begin{array}{rlcrl}
    \multicolumn{2}{c}{} & \qquad \quad  & 
      \multicolumn{2}{c}{} \vspace{0.1cm} \\
  \underset{  X   }{\text{maximize}}  &  \tr \mleft( A X \mright) & & \underset{ \{ Y_i \}\,, \,\{  Z_j     \}   }{\text{minimize}}    &  \underset{i}{\sum} \tr \mleft( Y_i B_i  \mright) - \underset{j}{\sum} \tr \mleft( Z_j C_j  \mright)  \vspace{0.1cm}
    \\
    \textrm{subject to} &  \Phi_i \mleft( X \mright) = B_i   \quad \text{for all} \;\; i   \vspace{0.1cm}  & & \textrm{subject to} &  \underset{j }{\sum  } \, \Gamma_j^{\dagger} \mleft( Z_j \mright) =   \underset{i }{\sum  } \Phi_i^{\dagger} \mleft( Y_i \mright) - A 
    \\
    & \Gamma_j \mleft( X \mright) \sgeq C_j \; \;\;\, \textnormal{for all} \;\; j \,    &&  &  Z_j \sgeq 0 \qquad \; \;\,  \textnormal{for all} \;\; j  \,.
  \end{array} 
\end{align}
We say that the primal problem is strictly feasible if there exists some variable $X$ such that $\Phi_i \mleft( X \mright) = B_i$ for all $i$ and such that $\Gamma_j \mleft( X\mright)- C_j$ is \emph{strictly positive} for all $j$; similarly, the dual problem is strictly feasible if there exist variables $\{Y_i\}$ and $\{Z_j\}$ that satisfy the equality condition and such that $Z_j$ is strictly positive for all $j$. From \cite[Conic Duality Theorem]{Nesterov1994}, if both dual and primal problems are strictly feasible, the duality gap is zero, so we are free to use either the primal or dual formulations to solve our problem. 

The complementary slackness conditions, which are necessary for a solution to be optimal, are 
\begin{equation}
    Z_j \left( \Gamma_j \mleft( X\mright) - C_j \right) = 0 \qquad \textnormal{for all} \;\; j \,.
\end{equation}
These conditions are often useful for deriving the optimal variables for an SDP.

 \subsection{Unambiguous attacks}\label{ssec: unamb rand}
 \subsubsection{Figure of merit}\label{ssec: unamb fom}
Let's imagine an eavesdropper who sends an ensemble $\{p_\mu, \, \rho_\mu\}$ to Alice, with $\sum_\mu p_\mu \rho_\mu = \rho$. Alice then measures her state $\rho$ in an orthonormal basis $\{\ket{m_j}\}$ of dimension $d$, where we write her POVM as $\m=\{M_j\}_j$, $M_j = \ketbra{m_j}{m_j}$. Since the guessing probability must be unambiguous, Eve is forced to send the states $\rho_j = \ketbra{m_j}{m_j}$, and to reserve one state $\rho_0$ for her inconclusive outcome, such that
\begin{equation}
    p_0 \rho_0 + \sum_{j=1}^{d} p_j \rho_j = \rho\,. 
\end{equation}
Since all the states must be positive semidefinite, we have the condition
\begin{equation}
   p_0 \rho_0 = \rho - \sum_{j=1}^{d} p_j \rho_j = \rho - \sum_{j=1}^{d} p_j \ketbra{m_j}{m_j} \sgeq 0\,.
\end{equation}
Our optimization problem is then 
\begin{equation}\label{eqn: UD rand}
\begin{aligned}
  \pud \mleft( \rho, \, \m \mright) = \;\;     \underset{\{ p_j \} }{\text{maximize}}  \, \quad & \sum_{j=1}^{d} p_j  && {}
        \\
         \text{subject to} \quad & p_j \sgeq 0 \quad \textnormal{for all} \;\; j \in \{1, \, \hdots, \, d\} &&   
         \\
          \quad & \rho \sgeq \sum_{j=1}^{d} p_j \ketbra{m_j}{m_j} \,. \quad &&  {}
    \end{aligned}
\end{equation}
This can be converted to an SDP of the standard form by introducing the matrix
\begin{equation}
    L = \sum_{j=1}^{d} p_j \ketbra{m_j}{m_j}
\end{equation}
and enforcing the conditions 
\begin{equation}
    L \sgeq 0\,, \qquad \langle m_i | L | m_j \rangle = 0 \qquad \textnormal{for all} \;\; i \neq j\,, \quad i, \, j \in \{1, \, \hdots, \, d\}\,.
\end{equation}
The SDP is 
\begin{equation}\label{eqn: UD SDP rand}
\begin{aligned}
  \pud \mleft( \rho, \, \m \mright) = \;\;     \underset{L }{\textnormal{maximize}}  \, \quad & \tr L  && {}
        \\
         \text{subject to} \quad & L \sgeq 0 \!\!\!  &&  
         \\
          \quad & \rho \sgeq L &&  {}
          \\
          \quad & \langle m_i | L | m_j \rangle = 0 \!\!\!  &&  \textnormal{for all} \;\; i \neq j\,, \;\; i, \, j \in \{1, \,  \hdots, \, d\}\,.
    \end{aligned}
\end{equation}
Comparing this SDP to the standard form in \eqref{eqn: standard SDP}, we identify the map $\Phi_{ij} \mleft( \cdot \mright) = \langle m_i | L | m_j \rangle$ for $i \neq j$, and its adjoint map $\Phi^{\dagger}_{ij} \mleft( \cdot \mright) = \left( \cdot\right) \ketbra{m_j}{m_i}$. The dual problem is then
\begin{equation}\label{eqn: dual UD SDP rand}
\begin{aligned}
  \pud \mleft( \rho, \, \m \mright) = \;\;     \underset{Z, \; \{y_{ij}\} }{\textnormal{minimize}}  \, \quad & \tr \mleft( Z \rho\mright)  && {}
        \\
         \text{subject to} \quad & Z \sgeq 0 \!\!\!  &&  
         \\
          \quad & Z \sgeq \id - \underset{i \neq j}{\sum_{i, \, j=1}^{d}} y_{ij} \ketbra{m_j}{m_i}\,, &&  {}
    \end{aligned}
\end{equation}
where $\{y_{ij}\}$ are real numbers. The primal problem is strictly feasible, as we could use the variables
\begin{equation}
L = \left( \lmin \mleft( \rho \mright)- \epsilon \right)\id\,, \qquad  0 <\epsilon < \lmin \mleft( \rho \mright)\,,    
\end{equation}
and similarly for the dual, we have the strictly feasible variables
\begin{equation}
 Z = \alpha \id\,, \qquad   \alpha > 1\,, \qquad y_{ij}=0 \qquad \textnormal{for all} \;\; i,\, j \in \{1, \, \hdots, \, d\}\,.   
\end{equation}
Given an optimal variable $L$, we can recover Eve's optimal mixing probability distribution as 
\begin{equation}
 p_{0} \rho = \rho - L\,, \qquad   p_j = \langle m_j | L | m_j \rangle \qquad \textnormal{for all} \;\; j \in \{1, \, \hdots, \, d\}\,.
\end{equation}

\subsubsection{Maximal randomness}\label{ssec: UD max rand}
We assume throughout that Alice makes a measurement $\m$ in an othnonormal basis $\{\ket{m_j}\}$. We can show that Eve's guessing probability would not decrease if Alice coarse-grained her measurement, i.e. if she performed a measurement $\hat{\m}=\{\hat{M}_a\}_a$ with $n <d$ elements
\begin{equation}\label{eqn: cg meas}
    \hat{M}_a = \sum_{j \in S_a} M_j\,,
\end{equation}
where $M_j=\ketbra{m_j}{m_j}$ and $\{S_a\}$ are mutually exclusive subsets of $\{1, \, \hdots, \, d\}$ such that 
\begin{equation}
   \sum_{a=1}^{n} \, \sum_{j \in S_a} =\sum_{j=1}^{d} \,.
\end{equation}
Note that the inconclusive outcome 0 is not an element of any set $S_a$.
If $\{p_j, \, \rho_j\}$ is Eve's optimal decomposition for the original measurement $\m$, we can define a coarse-grained decomposition $\{\hat{p}_a, \, \hat{\rho}_a \}$ with
\begin{equation}\label{eqn: cg decomp}
    \hat{p}_a \hat{\rho}_a = \sum_{j \in S_a} p_j \rho_j\,.
\end{equation}
This decomposition still satisfies the unambiguous condition, as
\begin{equation}
    \tr \mleft( \hat{p}_a \, \hat{\rho}_a  \hat{M}_b \mright) = \sum_{j \in S_a} \, \sum_{k \in S_b} p_j \tr \mleft( \rho_j M_k \mright) = \hat{p}_a \delta_{a, \, b} \qquad \textnormal{for all} \;\; a, \, b \in \{1, \, \hdots, \, n\}    \,.
\end{equation}
Eve's guessing probability is then lower bounded by
\begin{equation}
\pud \mleft( \rho, \, \hat{\m} \mright) \geq  \sum_{a=1}^{n} \hat{p}_a \tr \mleft( \hat{\rho}_a \hat{M}_a \mright) = \sum_{a=1}^{n} \hat{p}_a = \sum_{j=1}^{d} p_j = \pud \mleft( \rho, \, \m \mright)\,.
\end{equation}
When deriving the \emph{maximal} randomness of a state in the unambiguous case, then, which is optimized over all projective measurements, we are free to restrict to rank-one projective measurements, since coarse-graining a projective measurement won't decrease the unambiguous guessing probability.

\subsection{Randomness with a fixed rate of inconclusive outcomes}\label{ssec: rand Q}
\subsubsection{Figure of merit}\label{sssec: frio fom}
Here, we will again imagine our eavesdropper Eve sending the ensemble $\{ p_\mu, \, \rho_{\mu} \}$
to Alice, who measures in the orthonormal basis $\{\ket{m_j}\}$. This time, Eve will return an inconclusive outcome with probability $p_0=Q$, with $0 < Q < \qmax$, where $\qmax$ is the optimal $Q$ in the unambiguous case. The optimization problem is the following, 
\begin{equation}\label{eqn: opt Q rand}
\begin{aligned}
  \pq \mleft( \rho, \, \m \mright) = \;\;     \underset{ \{p_\mu, \, \rho_\mu \} }{\text{maximize}}  \, \quad & \sum_{j=1}^{d} p_j  \langle m_j | \rho_j | m_j \rangle  && {}
        \\
         \text{subject to} \quad & \rho_\mu \sgeq 0 \quad \textnormal{for all} \;\; \mu \in \{0, \, 1, \, \hdots, \, d\}  &&  
         \\
          \quad &  \sum_{\mu=0}^{d} p_\mu \rho_\mu = \rho \,\quad &&  {}
          \\
          \quad &  p_0 = Q\,. \quad &&  {}
    \end{aligned}
\end{equation}
This can be converted to a standard SDP by defining 
\begin{equation}\label{eqn: rho tilde}
     \tilde{\rho} = \sum_{\mu=0}^{d} p_\mu \rho_\mu \otimes \ketbra{\mu}{\mu} + \underset{\mu \neq \nu}{\sum_{\mu, \, \nu =0}^{d}} G_{\mu \nu} \otimes \ketbra{\mu}{\nu}\,, \qquad \tilde{M} = \sum_{j=1}^d \ketbra{m_j}{m_j} \otimes \ketbra{j}{j}\,,
\end{equation}
where $\{G_{\mu \nu}\}$ are arbitrary Hermitian matrices that don't contribute to the guessing probability. The matrix $\tilde{\rho}$ is Hermitian, and can only be positive semidefinite if all of its principal submatrices are positive semidefinite (see e.g. \cite[Observation 7.12]{horn2012matrix}), so to make sure that $\rho_\mu \sgeq 0$ for all $\mu$, it is sufficient to enforce $\tilde{\rho} \sgeq 0$. To ensure that the new variable $\tilde{\rho}$ has the form \eqref{eqn: rho tilde}, we also have to impose the constraints
\begin{equation}
    \begin{aligned}
       \sum_{\mu=0}^{d}  \left( \id \otimes \bra{\mu} \right) \; \tilde{\rho} \; \left( \id \otimes \ket{\mu} \right)   &= \rho\,,
       \\
       \tr \mleft( \tilde{\rho} \; \id \otimes \ketbra{0}{0} \mright) &= Q\,. 
    \end{aligned}
\end{equation}
The SDP in standard form is then
\begin{equation}\label{eqn: SDP Q rand}
\begin{aligned}
  \pq \mleft( \rho, \, \m \mright) = \;\;     \underset{ \tilde{\rho} }{\text{maximize}}  \, \quad & \tr \mleft( \tilde{\rho} \, \tilde{M} \mright)  && {}
        \\
         \text{subject to} \quad & \tilde{\rho} \sgeq 0 \!\!\!  &&  
         \\
          \quad &   \sum_\mu   \left( \id \otimes \bra{\mu} \right) \; \tilde{\rho} \; \left( \id \otimes \ket{\mu} \right)   = \rho &&  {}
          \\
          \quad &  \tr \mleft( \tilde{\rho} \; \id \otimes \ketbra{0}{0} \mright) = Q\,.  \quad &&  {}
    \end{aligned}
\end{equation}
We identify the maps
\begin{equation}
    \begin{aligned}
        \Lambda \mleft( \cdot  \mright) &= \sum_\mu   \left( \id \otimes \bra{\mu} \right) \; \cdot \; \left( \id \otimes \ket{\mu} \right)\,,
        \\
       \Omega \mleft( \cdot \mright) &= \tr \mleft(  \left( \cdot \right) \; \id \otimes \ketbra{0}{0} \mright)
    \end{aligned}
\end{equation}
and their adjoint maps
\begin{equation}
    \begin{aligned}
        \Lambda^{\dagger} \mleft( \cdot  \mright) &=  \left( \cdot \right) \otimes \id\,,
        \\
       \Omega^{\dagger} \mleft( \cdot \mright) &= \left(\cdot\right) \id \otimes \ketbra{0}{0}\,. 
    \end{aligned}
\end{equation}
We can then write the dual as 
\begin{equation}\label{eqn: dual Q SDP rand 1}
\begin{aligned}
  \pq \mleft( \rho, \, \m\mright) = \;\;     \underset{ G, \, w}{\textnormal{minimize}}  \, \quad &  \tr \mleft( G \rho\mright) - w Q  && {}
        \\
         \text{subject to} \quad & G \otimes \id - w \id \otimes \ketbra{0}{0} - \tilde{M} \sgeq 0\,, &&  
    \end{aligned}
\end{equation}
where $w$ is a real number.
Noting that all of the terms in the constraint are diagonal in the second Hilbert space $\{\ket{\mu}\}$, we can instead write
\begin{equation}\label{eqn: dual Q SDP rand}
\begin{aligned}
  \pq \mleft( 
  \rho, \, \m \mright) = \;\;     \underset{ G, \; w}{\textnormal{minimize}}  \, \quad &  \tr  \mleft( G \rho \mright) - w Q  && {}
        \\
         \text{subject to} \quad & G \sgeq w \id  && 
         \\
          \quad & G \sgeq \ketbra{m_j}{m_j}  && \textnormal{for all} \;\; j \in \{1, \, \hdots, \, d\}\,.
    \end{aligned}
\end{equation}
Both primal and dual problems are strictly feasible, as we could use the variables
\begin{equation}
 \rho_{\mu}=\rho\,, \qquad   p_0= Q\,, \qquad p_j = \frac{1-Q}{d} \qquad \textnormal{for all}\;\; \mu \in \{0, \, 1, \, \hdots, \, d\}\,, \quad j \in \{ 1, \, \hdots , \, d\}\,,   
\end{equation}
for the primal and 
\begin{equation}
 G= \left( 1 + \epsilon\right) \id\,, \qquad \epsilon>0\,, \qquad  w=0   
\end{equation}
for the dual. The complementary slackness condition is
\begin{equation}
 \tilde{\rho} \left( G \otimes \id - w \id \otimes \ketbra{0}{0} - \tilde{M} \right) =0\,,  
\end{equation}
or, equivalently,
\begin{equation}
  p_0  \rho_0 \left( G - w \id \right) =0\,, \qquad p_j \rho_j \left( G - \ketbra{m_j}{m_j} \right) =0 \qquad \textnormal{for all} \;\;  j \in \{1, \, \hdots, \, d \}\,.
\end{equation}
Summing the condition $G \sgeq \ketbra{m_j}{m_j}$ over all $j \in \{1, \, \hdots, \, d \}$, we find $ G \sgeq \frac{\id}{d}$, so $G$ must have full rank. Since $G$ is positive definite, $G-\ketbra{m_j}{m_j}$ can have at most one non-positive eigenvalue (see e.g. \cite[Corollary 4.3.9]{horn2012matrix}), so we conclude that all nontrivial states $\rho_j$ must be rank one for all $j \in \{1, \, \hdots, \, d\}$ (otherwise we have $p_j=0$).

\subsubsection{Maximal randomness}\label{ssec: max rand FRIO}
We will show that Eve's FRIO guessing probability would not decrease if Alice coarse-grained her measurement, as in \eqref{eqn: cg meas}.
As before, the inconclusive outcome 0 is not an element of any set $S_a$.
Given the optimal decomposition $\{p_\mu, \, \rho_\mu\}$ for the original measurement $\m$, the coarse-grained decomposition $\{\hat{p}_a, \, \hat{\rho}_a\}$ in \eqref{eqn: cg decomp} is valid for $\hat{\m}$ because $p_0$ is unchanged.
Eve's guessing probability is then lower bounded by
\begin{equation}
\pq \mleft( \rho, \, \hat{\m} \mright) \geq  \sum_{a=1}^{n} \hat{p}_a \tr \mleft( \hat{\rho}_a \hat{M}_a \mright) = \pq \mleft( \rho, \, \m\mright) + \sum_{a=1}^{n} \, \underset{j \neq k}{\sum_{j, \, k \in \mathcal{S}_a}} p_j \tr \mleft( \rho_j \, M_k \mright)  \geq \pq \mleft( \rho, \, \m \mright)\,.
\end{equation}
When deriving the \emph{maximal} randomness of a state under a FRIO attack, since coarse-graining a projective measurement won't decrease the eavesdropper's guessing probability, we are free to restrict to rank-one projective measurements.

\section{Randomness and state discrimination}\label{app: rand and SD}
From the Schr\"{o}dinger-Gisin-Hughston-Jozsa-Wootters theorem \cite{Schrödinger_1935, gisin_1989, Hughston_1993}, we know that any ensemble of states can be steered by local measurements performed by a distant party on the purification of a state. In the following, we will use this fact to convert our randomness scenario to a state discrimination scenario \cite{Holevo_1973, helstrom_1976}.
\paragraph{Randomness.}
In the randomness scenario, Alice holds a state $\rho$, which has rank $d$, on which she performs a measurement in a $d$-dimensional orthonormal basis $\{\ket{m_j}\}$. Her state is decomposed as an ensemble $\{p_\mu, \, \rho_\mu\}$, with $\sum_{\mu=0}^{d}p_\mu \rho_\mu=\rho$. Consider the operators
\begin{equation}\label{eqn: F POVM}
    F_{\mu} = p_\mu \rho^{- \frac{1}{2}} \rho_\mu \rho^{- \frac{1}{2}}\,, \qquad \mu = \{0, \, 1, \, \hdots, \, d\}\,.
\end{equation}
The terms $\{F_\mu\}$ are all positive semidefinite and they sum to the identity, so $\{F_\mu\}_\mu$ is a POVM. Define the purification of the state $\rho$ as $\ket{\Psi}$, with
\begin{equation}\label{eqn: bipartite pure}
    \ket{\Psi} = \sum_{j=1}^{d} \left( \sqrt{\rho} \otimes \id \right) \ket{u_j , \, u_j}= \sum_{j=1}^{d} \left(  \id \otimes \sqrt{\rho} \right) \ket{u_j , \, u_j}\,,
\end{equation}
where $\{\ket{u_j}\}$ is the eigenbasis of $\rho$. We can see that Alice would receive the ensemble $\{p_\mu, \, \rho_\mu \}$ if Eve measured her part of $\ket{\Psi}$ with the POVM $\{E_\mu\}_\mu = \{F_{\mu}^{*}\}_\mu$, where $\left( *\right)$ denotes complex conjugation in the basis $\{\ket{u_j}\}$. The POVM elements $E_\mu$ and the states $\rho_\mu$ are related by
\begin{equation}
    p_\mu \rho_\mu = \sqrt{\rho}\,  E_\mu^{*} \sqrt{\rho}\,.
\end{equation}
Since we know, in Eve's optimal decomposition for a fixed rate of inconclusive outcomes, that the states $\rho_j$ are rank one for $j \in \{1, \, \hdots, \, d\}$, we conclude that the POVM elements $E_j$ are also rank one, $E_j = \ketbra{e_j}{e_j}$ (note that the vectors $\{\ket{e_j}\}$ are not necessarily normalized). We can then write the guessing probability \eqref{eqn: opt Q rand} in terms of Eve's POVM $\{E_\mu\}_\mu$ as
\begin{equation}\label{eqn: opt Q rand 2}
\begin{aligned}
  \pq \mleft( \rho, \, \m \mright) = \;\;     \underset{  \{E_\mu\}_\mu }{\text{maximize}}  \, \quad & \sum_{j=1}^{d}   \abs{\langle m_j | \sqrt{\rho} | e_j^{*} \rangle}^2  && {}
        \\
         \text{subject to} \quad & E_\mu \sgeq 0 \quad \textnormal{for all} \;\; \mu \in \{0, \, 1, \, \hdots, \, d\}  &&  
         \\
          \quad &  \sum_{\mu=0}^{d} E_\mu =\id \,\quad &&  {}
          \\
          \quad &  \tr \mleft( E_0 \,\rho\mright) = Q\,. \quad &&  {}
    \end{aligned}
\end{equation}
Meanwhile, when Alice measures her local state in $\{\ket{m_j}\}$, Eve receives the steered states 
\begin{equation}\label{eqn: steered}
    \sqrt{\eta_j} \ket{\phi_j} = \sum_{i=1}^{d} \langle m_j | u_i\rangle \sqrt{\rho} \ket{u_i} = \sqrt{\rho} \ket{m_j^{*}}\,.
\end{equation}
 Since the state $\rho$ is full rank in the space spanned by the measurement basis $\{\ket{m_j}\}$, every probability $\eta_j$ is nonzero, and the states $\{\ket{\phi_j}\}$ are linearly independent.

\paragraph{State discrimination.}
Let's say we have an ensemble of $d$ linearly independent states $\{\eta_j, \,\ket{\phi_j}\}$ satisfying
\begin{equation}
\sum_{j=1}^{d} \eta_j \ketbra{\phi_j}{\phi_j} = \rho\,.
\end{equation}
The state $\rho$ has rank $d$ because the states $\{\ket{\phi_j}\}$ are linearly independent, so we can invert it to define the set
\begin{equation}
    K_j = \eta_j \rho^{- \frac{1}{2}}  \ketbra{\phi_j}{\phi_j} \rho^{- \frac{1}{2}}\,.
\end{equation}
The elements $K_j$ are rank one and positive and they sum to the identity, so, since there are $d$ of them spanning a $d$-dimensional basis, $\{K_j\}_j$ must be a measurement in some orthonormal basis $\{\ket{\xi_j}\}$, where we write $ K_j = \ketbra{\xi_j}{\xi_j}$. We can then write the states as
\begin{equation}\label{eqn: k states}
    \sqrt{\eta_j} \ket{\phi_j} = \sqrt{\rho} \ket{\xi_j}\,.
\end{equation}
Comparing this form to \eqref{eqn: steered}, we see that the ensemble $\{\eta_j, \, \ket{\phi_j}\}$ could have arisen from a measurement in the basis $\{\ket{m_j}\}=\{\ket{\xi_j^{*}}\}$ applied to Alice's part of the purification \eqref{eqn: bipartite pure}. From \eqref{eqn: k states}, we can also conclude that
\begin{equation}\label{eqn: ortho}
    \sqrt{\eta_j \eta_l} \, \langle \phi_j | \rho^{-1} | \phi_j \rangle = \delta_{j, \, l} \qquad \textnormal{for all} \;\; j, \, l \in \{1, \, \hdots , \, d\}\,.
\end{equation}
To unambiguously discriminate her states, Eve must make a measurement $\{E_\mu\}_\mu$ of the form \cite{Chefles_1998}
\begin{equation}
    E_{j} = P_j  \ketbra{\phi^{\perp}_j}{\phi^{\perp}_j} \qquad \textnormal{for all} \;\; j \in \{1, \, \hdots, \, d\}\,, \qquad E_{0} = \id - \sum_{j=1}^{d} E_j\,,
\end{equation}
where $\{\ket{\phi^{\perp}_j}\}$ are the reciprocal states of $\{\ket{\phi_j}\}$ \cite{Chefles_1998, Eldar_UD_2003}, i.e. the unique (unnormalized) states that satisfy
\begin{equation}
    \langle \phi^{\perp}_i | \phi_k \rangle = \delta_{i\, k} \, \qquad \textnormal{for all} \;\; i, \, k  \in \{1, \, \hdots, \, d\}\,.
\end{equation}
From \cite[Equation 2.10]{Chefles_1998}, the SDP for her optimal guessing probability is then
\begin{equation}\label{eqn: chef UD}
\begin{aligned}
  \pud \mleft( \{ \eta_j , \, \ket{\phi_j} \} \mright) = \;\;     \underset{\{ P_j \} }{\text{maximize}}  \, \quad & \sum_{j=1}^{d} \eta_j P_j  && {}
        \\
         \text{subject to} \quad & P_j \geq 0 \quad \textnormal{for all} \;\; j \in \{1, \, \hdots, \, d\}  &&  
         \\
          \quad & \id \sgeq \sum_{j=1}^{d} P_{j} \ketbra{\phi^{\perp}_j}{\phi^{\perp}_j} \,.\quad &&  {}
    \end{aligned}
\end{equation}
Using \eqref{eqn: k states} and \eqref{eqn: ortho}, we have $\ket{\phi_j^{\perp}} = \eta_j \rho^{-1} \ket{\phi_j} =\sqrt{\eta_j}\rho^{- \frac{1}{2}} \ket{m_j^{*}}$,
such that Eve's POVM operators can be rewritten as 
\begin{equation}\label{eqn: new SD meas}
    E_j = \eta_j P_j\, \rho^{- \frac{1}{2}} \ketbra{m_j^{*}}{m_j^{*}} \rho^{-\frac{1}{2}} \,, \qquad E_{0} = \id - \sum_{j=1}^{d} E_j
\end{equation}
and the SDP as
\begin{equation}\label{eqn: chef 2 UD}
\begin{aligned}
  \pud \mleft( \{ \eta_j , \, \ket{\phi_j} \} \mright) = \;\;     \underset{\{ P_j \} }{\text{maximize}}  \, \quad & \sum_{j=1}^{d} \eta_j P_j  && {}
        \\
         \text{subject to} \quad & P_j \geq 0 \quad \textnormal{for all} \;\; j \in \{1, \, \hdots, \, d\}  &&  
         \\
          \quad & \rho \sgeq \sum_{j=1}^{d} \eta_j P_{j} \ketbra{m_j^{*}}{m_j^{*}} \,. \quad &&  {}
    \end{aligned}
\end{equation}
If we replace $p_j = \eta_j P_j$,
we recover exactly the form of the SDP \eqref{eqn: UD rand} for randomness. This means if we find a mixing probability distribution $\{p_j\}$ that's optimal for Eve when Alice has the state $\rho$ and makes a measurement in the basis $\{\ket{m_j}\}$, we also find optimal variables $\{ \eta_j P_j \}$ to define Eve's measurement for state discrimination, of the form \eqref{eqn: new SD meas}, when she receives the states $\sqrt{\eta_j} \ket{ \phi_j}= \sqrt{\rho} \ket{m_j^{*}}$. 

Similarly, given an ensemble of pure states $\{\eta_j, \, \ket{\phi_j}\}$, the SDP for state discrimination with a fixed rate $Q$ of inconclusive outcomes is \cite{Bagan_2012, Chefles_inter_1998} 
\begin{equation}\label{eqn: opt Q discrim}
\begin{aligned}
  \pq \mleft( \{ \eta_j , \, \ket{\phi_j} \} \mright) = \;\;     \underset{\{ E_\mu \}_\mu }{\text{maximize}}  \, \quad & \sum_{j=1}^{d} \eta_j \langle \phi_j | E_j | \phi_j \rangle  && {}
        \\
         \text{subject to} \quad & E_\mu \sgeq 0 \quad \textnormal{for all} \;\; \mu \in \{1, \, \hdots, \, d\}  &&  
         \\
          \quad &  \sum_{\mu=0}^{d} E_\mu = \id \,\quad &&  {}
          \\
          \quad &   \tr \mleft( E_0 \,\rho\mright)  = Q\,. \quad &&  {}
    \end{aligned}
\end{equation}
Subbing in $\sqrt{\eta_j} \ket{\phi_j}= \sqrt{\rho} \ket{m_j^{*}}$, we have 
\begin{equation}\label{eqn: rho opt Q discrim}
\begin{aligned}
  \pq \mleft( \{\eta_j , \, \ket{\phi_j}\} \mright) = \;\;     \underset{\{ E_\mu \}_\mu }{\text{maximize}}  \, \quad & \sum_{j=1}^{d}  \langle m_j^{*} | \sqrt{\rho} \,  E_j \sqrt{\rho} | m_j^{*} \rangle  && {}
        \\
         \text{subject to} \quad & E_\mu \sgeq 0 \quad \textnormal{for all} \;\; \mu \in \{1, \, \hdots, \, d\}  &&  
         \\
          \quad &  \sum_{\mu=0}^{d} E_\mu = \id \,\quad &&  {}
          \\
          \quad &   \tr \mleft( E_0 \,\rho\mright)  = Q\,. \quad &&  {}
    \end{aligned}
\end{equation}
Following similar reasoning to that in Appendix \ref{sssec: frio fom}, we can derive the corresponding dual problem \cite{Fiur_ek_2003, Eldar_inter_2003}
\begin{equation}\label{eqn: rho dual Q SDP discrim}
\begin{aligned}
  \pq \mleft( \{\eta_j , \, \ket{\phi_j}\} \mright) = \;\;     \underset{ F, \, y}{\textnormal{minimize}}  \, \quad &  \tr F - y Q  && {}
        \\
         \text{subject to} \quad & F \sgeq y \rho  && 
         \\
          \quad & F \sgeq \sqrt{\rho} \ketbra{m_j^{*}}{m_j^{*}} \sqrt{\rho}  && \textnormal{for all} \;\; j \in \{1, \, \hdots, \, d\}\,,
    \end{aligned}
\end{equation}
where $y$ is a real variable.
The complementary slackness conditions are
\begin{equation}
    E_0 \left( F - y \rho \right) =0\,, \qquad E_j \left( F - \sqrt{\rho} \ketbra{m_j^{*}}{m_j^{*}} \sqrt{\rho}  \right) =0 \qquad \textnormal{for all} \;\; j \in \{1, \, \hdots, \, d\}\,.
\end{equation}
Summing up all the inequality constraints on $F$ over all $j \in \{1, \, \hdots, \, d\}$, we find $F \sgeq \frac{1}{d}\rho$, so $F$ must be full rank. Since $F$ is positive definite, $F- \sqrt{\rho}\ketbra{m_j^{*}}{m_j^{*}}\sqrt{\rho}$ can have at most one non-positive eigenvalue (see e.g. \cite[Corollary 4.3.9]{horn2012matrix}), so we conclude that all nontrivial POVM elements $E_j$ must be rank one for all $j$, $E_j = \ketbra{e_j}{e_j}$. We can then rewrite the primal problem as
\begin{equation}\label{eqn: rho opt Q discrim 2}
\begin{aligned}
  \pq \mleft( \{\eta_j , \, \ket{\phi_j}\} \mright) = \;\;     \underset{\{ E_\mu \}_\mu }{\text{maximize}}  \, \quad & \sum_{i=1}^{d}  \abs{\langle m_j^{*} | \sqrt{\rho} |  e_j \rangle}^2  && {}
        \\
         \text{subject to} \quad & E_\mu \sgeq 0 \quad \textnormal{for all} \;\; \mu \in \{1, \, \hdots, \, d\}  &&  
         \\
          \quad &  \sum_{\mu=0}^{d} E_\mu = \id \,\quad &&  {}
          \\
          \quad &   \tr \mleft( E_0 \,\rho\mright)  = Q\,, \quad &&  {}
    \end{aligned}
\end{equation}
which is exactly equivalent to \eqref{eqn: opt Q rand 2}. Returning to \eqref{eqn: F POVM} and \eqref{eqn: steered}, this means that if a decomposition $\{p_\mu, \, \rho_\mu \}$ is optimal for the randomness problem \eqref{eqn: opt Q rand}, when Alice holds $\rho$ and measures it in $\{\ket{m_j}\}$, then the POVM $\{E_\mu\}_\mu$ will be optimal for Eve in the state discrimination problem \eqref{eqn: opt Q discrim}, when she receives a linearly independent set of $d$ states $\{\eta_j, \, \ket{\phi_j}\}$, where
\begin{equation}
    E_\mu = p_\mu \rho^{- \frac{1}{2}} \rho_\mu \rho^{- \frac{1}{2}}\,, \qquad \sqrt{\eta_j} \ket{\phi_j} = \sqrt{\rho} \ket{m_j}\,.
\end{equation}

\section{Maximal randomness for any state}\label{sec: any state}
\subsection{Unambiguous attacks}\label{ssec: any state UD}
We want to find the maximal randomness of any quantum state $\rho$ under an unambiguous attack, given by the minimization of the unambiguous guessing probability \eqref{eqn: UD rand} over all rank-one projective measurements $\m$, 
\begin{equation}
    \pud^{*} \mleft( \rho\mright) = \min_{\m} \pud \mleft( \rho, \, \m \mright)\,.
\end{equation}
Here we prove Theorem \ref{thm: any state} from the main text in two steps. 
\begin{theorem}\label{thm: app any state}
The maximal randomness of any state $\rho$ under an unambiguous attack is given by the guessing probability 
\begin{equation}\label{eqn: app opt UD any rho}
 \pud^{*} \mleft( \rho\mright) = d \lmin \mleft( \rho \mright)\,. \end{equation}     \end{theorem}
\begin{proof}
\mbox{}\\
\\
\textbf{Lower bound.} For any measurement basis $\{\ket{m_j}\}$, Eve could use the decomposition
\begin{equation}
    p_0 \rho_0 = \rho - \lmin \mleft( \rho \mright) \id\,, \qquad p_j \rho_j = \lmin \mleft( \rho \mright) \ketbra{m_j}{m_j} \qquad \textnormal{for all} \;\; j \in \{1, \, \hdots, \, d\}\,.
\end{equation}
This gives us a lower bound 
\begin{equation}
    \pud \mleft( \rho, \, \m\mright) \geq d \lmin \mleft( \rho\mright)\,.
\end{equation}
\mbox{}\\
\textbf{Upper bound.} Denote the eigenvector of $\rho$ corresponding to its minimum eigenvalue as $\ket{u_\textnormal{min}}$, and choose a measurement $\m_{\textnormal{min}}$ such that the basis $\{\ket{m_j}\}$ is unbiased to $\ket{\umin}$, i.e. such that
\begin{equation}
    \abs{\langle \umin | m_j \rangle}^2 = \frac{1}{d} \qquad \textnormal{for all} \;\; j \in \{1, \, \hdots, \, d\}\,.
\end{equation}
The projector $\ketbra{\umin}{\umin}$ can then be expanded as
\begin{equation}
    \ketbra{\umin}{\umin} = \frac{\id}{d} + \underset{i \neq j}{\sum_{i, \, j=1}^{d}} \langle m_i | \umin \rangle \langle \umin | m_j \rangle \ketbra{m_i}{m_j}\,.
\end{equation}
The following are valid variables for the dual problem \eqref{eqn: dual UD SDP rand}, 
\begin{equation}
    Z = d \ketbra{\umin}{\umin}\,, \qquad y_{ij} = - d \langle m_j | \umin \rangle \langle \umin | m_i \rangle\,.
\end{equation}
These variables give us the lower bound 
\begin{equation}
    \pud \mleft( \rho, \, \m_{\textnormal{min}}\mright) \leq d \lmin \mleft( \rho\mright)\,.
\end{equation}
Since the upper and lower bounds match, we have proven Theorem \ref{thm: any state}. 
\end{proof}
\paragraph{Sufficient condition.} We have shown that having $\ket{\umin}$ unbiased to the measurement basis $\{\ket{m_j}\}$ is a sufficient condition to achieve maximal randomness under an unambiguous attack, but in the following example, we show that it is not necessary. Consider the qutrit state $\rho = \sum_{i=1}^{3} \lambda_i \ketbra{u_i}{u_i}$, with eigenvalues arranged in decreasing order, and with degeneracy such that $\lambda_2 = \lambda_3= \lmin \mleft( \rho \mright)$.
Take the measurement basis $\{\ket{m_j}\}$ with
\begin{equation}
    \begin{aligned}
        \ket{m_1} &= \frac{1}{\sqrt{3}} \ket{u_1} + \frac{1}{\sqrt{2}} \ket{u_2} + \frac{1}{\sqrt{6}} \ket{u_3}\,,
        \\
        \ket{m_2} &= -\frac{1}{\sqrt{3}} \ket{u_1} + \frac{1}{\sqrt{2}} \ket{u_2} - \frac{1}{\sqrt{6}} \ket{u_3}\,,
        \\
        \ket{m_3} &= \frac{1}{\sqrt{3}} \ket{u_1} - \sqrt{\frac{2}{3}} \ket{u_3} \,.
    \end{aligned}
\end{equation}
We notice that the \emph{maximum} eigenvector $\ket{u_1}$ is unbiased with respect to the basis $\{\ket{m_j}\}$, but we cannot construct a minimum eigenvector
\begin{equation}
    \ket{\umin} = \alpha \ket{u_2} + \beta \ket{u_3}\,, \qquad \abs{\alpha}^2 + \abs{\beta}^2 =1
\end{equation}
such that $\ket{\umin}$ is unbiased to $\{\ket{m_j}\}$. Nonetheless, we can prove that we saturate the bound in Theorem \ref{thm: any state} with this state-measurement pair by choosing the following variables for the dual problem \eqref{eqn: dual UD SDP rand},
\begin{equation}
    Z = \frac{3}{2} \left( \id - \ketbra{u_1}{u_1}\right)\,, \qquad y_{ij} = \frac{3}{2} \langle m_j | u_1\rangle \langle u_1 | m_i \rangle\,.
\end{equation}

\paragraph{State discrimination.} Let's take any linearly indendent ensemble of $d$ pure states $\{\eta_j, \, \ket{\phi_j}\}$ which sum to $\rho$. Our previous result implies that 
 \begin{equation}
     \pud \mleft( \{\eta_j, \, \ket{\phi_j}\}\mright) \geq d \lmin \mleft( \rho \mright)\,, 
 \end{equation}
 which proves Theorem \ref{thm: SD} in the main text. The decomposition $\{p_\mu, \, \rho_\mu\}$ in the previous section is equivalent to the measurement $\{E_\mu\}_\mu$, with
 \begin{equation}
     E_0 = \id - \lmin \mleft( \rho \mright)  \rho^{-1}\,, \qquad E_j = \lmin \mleft( \rho \mright) \eta_j \rho^{-1} \ketbra{\phi_j}{\phi_j} \rho^{-1} \quad \textnormal{for all} \;\; j \in \{1, \, \hdots, \, d\}\,.
 \end{equation}
 The sufficient condition for optimality becomes
 \begin{equation} \eta_j \abs{ \langle \umin  | \phi_j\rangle }^2 = \frac{\lmin \mleft( \rho \mright)}{d} \qquad \textnormal{for all} \;\; j \in \{1, \, \hdots, \, d\}\,.
 \end{equation}
 Theorem \ref{thm: SD} could also have been proven by subbing $q_i=\eta_i$ for all $i \in \{1, \, \hdots, \, d\}$ into \cite[Theorem 1]{Sun_2002}, noting from \eqref{eqn: k states} that the state overlaps satisfy 
 \begin{equation}
  \langle \phi_i | \phi_j\rangle = \frac{1}{\sqrt{\eta_i \eta_j}} \langle \xi_i | \rho | \xi_j \rangle \qquad \textnormal{for all} \;\; i, \, j \in \{1, \, \hdots, \, d\} \,,
 \end{equation}
where $\{\ket{\xi_i}\}$ is an orthonormal basis.

\subsection{FRIO attacks}\label{ssec: any state FRIO}
We now want to find the maximal randomness of any quantum state $\rho$ under FRIO attacks, given by the minimization of the FRIO guessing probability \eqref{eqn: mt opt Q rand} over all rank-one projective measurements $\m$, 
\begin{equation}
    \pq^{*} \mleft( \rho\mright) = \min_{\m} \pq \mleft( \rho, \, \m \mright)\,.
\end{equation}
Here we prove Theorem \ref{thm: any state FRIO} from the main text in two steps. 
\setcounter{theorem}{2}
\begin{theorem}
Let $\{\lambda_k\}$ be the eigenvalues of the state $\rho$, arranged in non-increasing order.
The maximal randomness of $\rho$ under a FRIO attack is given by the guessing probability 
\begin{equation}\label{eqn: app opt FRIO any rho}
 \pq^{*} \mleft( \rho\mright) = \frac{1}{d}\left( \sqrt{t \left( \, \sum_{k=1}^{t}\lambda_k - Q\right)} + \sum_{l=t+1}^{d} \sqrt{\lambda_l} \right)^2 \,,
 \end{equation} 
 where $t \in \{1, \, \hdots, \, d-1\}$ is chosen such that 
 \begin{equation}\label{eqn: r cond app}
    \sum_{j=1}^{t-1} \left( \lambda_{j} - \lambda_{t} \right) < Q \leq \sum_{j=1}^{t} \left( \lambda_{j} - \lambda_{t+1} \right)\,.
\end{equation}
 \end{theorem}
 \begin{proof}
 \mbox{}\\
\\
\textbf{Lower bound.} Given any measurement basis $\{\ket{m_j}\}$, consider the following decomposition, which we draw from \cite[Theorem 2]{Nakahiri_2012},
\begin{equation}\label{eqn: FRIO decomp}
     p_0 \rho_0 = \chi \, \rho \, \chi\,, \qquad p_j \rho_j = \Lambda \sqrt{\rho} \ketbra{m_j}{m_j} \sqrt{\rho} \, \Lambda \qquad \textnormal{for all} \;\; j \in \{1, \, \hdots, \, d\}\,,
 \end{equation}
where
\begin{equation}
    \chi = \sum_{j=1}^{d} \sqrt{\chi_j} \ketbra{u_j}{u_j}\,, \qquad \Lambda = \sum_{j=1}^{d} \sqrt{1 - \chi_j} \ketbra{u_j}{u_j}\,,
\end{equation}
where $\{\ket{u_j}\}$ is the eigenbasis of $\rho$ and
\begin{equation}
    \chi_{j} = \begin{cases}
        1 - \frac{1}{t \, \lambda_{j}} \left( \sum_{k=1}^{t}\lambda_{k} - Q \right)\,, \qquad & j \in \{1, \, \hdots, \, t \}\,,
        \\[4 pt]
        0\,, \qquad & j \in \{t+1, \, \hdots, \, d\}\,.
    \end{cases}
\end{equation}
Note that, from \eqref{eqn: r cond app}, we have
\begin{equation}\label{eqn: FRIO ineqs}
  0 <  t  \, \lambda_{t + 1} \leq \sum_{k=1}^{t} \lambda_{k} - Q < t \, \lambda_{t} \leq t \lambda_{j}
\end{equation}
for all $j \in \{1, \, \hdots, \, t\}$, where we use the fact that $\rho$ has full rank.
This implies that $0 \leq \chi_{j} \leq 1$ for all $j \in \{1, \, \hdots, \, d\}$, such that $\chi \sgeq 0$ and $\Lambda \sgeq 0$, so the decomposition in \eqref{eqn: FRIO decomp} is valid because
\begin{equation}
        p_0 = \tr \mleft( \chi \rho \chi \mright) = \sum_{j=1}^{d} \chi_j \lambda_{j} = \sum_{j=1}^{t} \left( \lambda_{j} - \frac{1}{t} \left( \, \sum_{k=1}^{t} \lambda_{k} - Q \right) \right) =Q\,
\end{equation}
and
\begin{equation}
    p_0 \rho_0 + \sum_{j=1}^{d} p_j \rho_{j} = \chi \rho \chi + \Lambda \rho \Lambda = \sum_{j=1}^{d} \chi_j \ketbra{u_j}{u_j} + \sum_{j=1}^{d} \left( 1 - \chi_j\right) \ketbra{u_j}{u_j} = \rho\,.
\end{equation}
It gives us the lower bound 
\begin{equation}\label{eqn: gen FRIO bound}
    \pq \mleft( \rho, \, \m \mright) \geq \sum_{j=1}^{d} \langle m_j | \Lambda \sqrt{\rho} | m_j \rangle^2
\end{equation}
for any basis $\{\ket{m_j}\}$. 
Notice that 
\begin{equation}
    \Lambda \rho = \sum_{k=1}^{d} \sqrt{1- \chi_{k}} \sqrt{\lambda_k} \ketbra{u_k}{u_k} \sgeq 0
\end{equation}
and 
\begin{equation}
    \sum_{j=1}^{d} \langle m_j | \Lambda \sqrt{\rho} | m_j \rangle = \tr \mleft( \Lambda \sqrt{\rho} \mright)\,.
\end{equation}
Let's define the difference 
\begin{equation}
    \ve_{j} = \langle m_j | \Lambda \sqrt{ \rho} | m_j \rangle - \frac{\tr \mleft( \Lambda \sqrt{\rho} \mright)}{d} \qquad \textnormal{for all} \;\; j \in \{1, \, \hdots, \, d\}\,,
\end{equation}
such that $\sum_{j=1}^{d} \ve_j =0$. We can rewrite the right-hand side of \eqref{eqn: gen FRIO bound} as
\begin{equation}
 \sum_{j=1}^{d} \langle m_j | \Lambda \sqrt{\rho} | m_j \rangle^2 =    \sum_{j=1}^{d} \left( \left( \frac{\tr \mleft( \Lambda \sqrt{\rho}\mright)}{d} \right)^2 + \frac{2}{d} \tr \mleft(  \Lambda \sqrt{\rho}\mright) \ve_{j} + \ve_{j}^2 \right) = \frac{1}{d} \left( \tr \mleft( \Lambda \sqrt{ \rho} \mright) \right)^2 + \sum_{j=1}^{d} \ve_j^2\,.
\end{equation}
Since $\sum_{j=1}^{d} \ve_j^2$ is strictly positive unless all $\ve_j=0$, we find
\begin{equation}
 \pq \mleft( \rho, \, \m \mright) \geq \frac{1}{d} \left( \tr \mleft( \Lambda \sqrt{\rho} \mright) \right)^2  = \frac{1}{d}\left( \sqrt{t \left( \, \sum_{k=1}^{t}\lambda_k - Q\right)} + \sum_{l=t+1}^{d} \sqrt{\lambda_l} \right)^2\,.  
\end{equation}
This inequality is strict unless the following condition is met, \begin{equation}\label{eqn: frio cond even}
  \langle m_j | \Lambda \sqrt{ \rho} | m_j \rangle = \frac{\tr \mleft( \Lambda \sqrt{\rho} \mright)}{d} \qquad \textnormal{for all} \;\; j \in \{1, \, \hdots, \, d\}\,.  
\end{equation}    
\mbox{}\\
\textbf{Upper bound.} Here, we consider only the case where condition \eqref{eqn: frio cond even} is satisfied. Take the following variables for the dual problem \eqref{eqn: dual Q SDP rand}, 
\begin{equation}
    G= w \sum_{j=1}^{t} \ketbra{u_j}{u_j} + \beta \sum_{j=t+1}^{d} \frac{1}{\sqrt{\lambda_j}} \ketbra{u_j}{u_j}\,, \qquad  w = \beta \sqrt{\frac{t}{\sum_{k=1}^{t}\lambda_k - Q }}\,,
\end{equation}
where
\begin{equation}
     \beta = \frac{1}{d}\tr \mleft( \Lambda \sqrt{\rho} \mright) =\frac{1}{d} \left( \sqrt{t \left( \,\sum_{k=1}^{t}\lambda_k - Q \right)} + \sum_{l=t+1}^{d} \sqrt{\lambda_l}   \right)\,.
\end{equation}
Using \eqref{eqn: FRIO ineqs}, we have
\begin{equation}
    \sum_{k=1}^{t} \lambda_{k} - Q \geq t \lambda_{t+1} > 0 \,,
\end{equation}
so $w$ is well defined. Since both $\beta$ and $w$ are strictly positive, $G$ is positive definite. To prove that $G - w \id \sgeq 0$, then, we need only show that 
\begin{equation}
    \frac{\beta}{\sqrt{\lambda_j}} - w \geq 0 \qquad \textnormal{for all} \;\; j \in \{t+1, \, \hdots, \, d\}
\,.
\end{equation}
Expanding and multiplying both sides by $\sqrt{\lambda_j}$, we find
\begin{equation}
\beta - w \sqrt{\lambda_j} =   \beta \left( 1 - \sqrt{\frac{t \lambda_j}{\sum_{k=1}^{t}\lambda_{k} - Q } }  \right) \geq \beta \left( 1 - \sqrt{\frac{t \lambda_{t+1}}{\sum_{k=1}^{t}\lambda_{k} - Q } }  \right) \geq 0\,,
\end{equation}
where in the first inequality we use $\lambda_{j} \leq \lambda_{t+1}$ for all $j \in \{t+1, \, \hdots, \, d\}$ and in the second we use \eqref{eqn: FRIO ineqs}. 

We now only have to prove 
\begin{equation}
    G - \ketbra{m_j}{m_j} \sgeq 0 \qquad \textnormal{for all} \;\; j \in \{1, \, \hdots, \, d\}\,.
\end{equation}
Since $G$ is positive definite, $G- \ketbra{m_j}{m_j}$ can have at most one non-positive eigenvalue for any $j$ (see e.g. \cite[Corollary 4.3.9]{horn2012matrix}). Note that
\begin{equation}
    G \Lambda \sqrt{\rho} = \left( \sum_{k=1}^{t} \sqrt{1- \chi_k} \sqrt{\lambda_k} + \beta \sum_{k=t+1}^{d} \right)\ketbra{u_k}{u_k} = \beta \id\,.
\end{equation}
The vector $\Lambda \sqrt{\rho} \ket{m_j}$ is then a zero eigenvector for $G- \ketbra{m_j}{m_j}$ for any $j \in \{1, \, \hdots, \, d\}$, as
\begin{equation}
    \left( G - \ketbra{m_j}{m_j} \right) \Lambda \sqrt{\rho} \ket{m_j} = G \Lambda \sqrt{\rho} \ket{m_j} - \langle m_j | \Lambda \sqrt{\rho} | m_j \rangle = \left( \beta - \frac{\tr \mleft( \Lambda \sqrt{\rho}  \mright)}{d} \right) \ket{m_j} =0
\end{equation}
for all $j \in \{1, \, \hdots, \, d\}$, where we use condition \eqref{eqn: frio cond even}.

Since $G$ and $w$ are valid variables, we can use them to bound the FRIO guessing probability,
\begin{equation}
\begin{aligned}
    \pq \mleft( \rho, \, \m \mright) &\leq \tr \mleft( G \rho \mright) - w Q =w \left( \sum_{j=1}^{t} \lambda_j -Q \right) + \beta \sum_{j=t+1}^{d} \sqrt{\lambda_j}
    \\
    &= \beta \left( \sqrt{t \left(\sum_{k=1}^{t}\lambda_k - Q \right)} + \sum_{l=t+1}^{d} \sqrt{\lambda_l}   \right)
    \\
    &= \frac{1}{d} \left( \sqrt{t \left(\sum_{k=1}^{t}\lambda_k - Q \right)} + \sum_{l=t+1}^{d} \sqrt{\lambda_l}   \right)^2\,.
    \end{aligned}
\end{equation}
Since the upper and lower bounds match, we have proven Theorem \ref{thm: any state FRIO}.
\end{proof}

\section{Qubits}
\subsection{General properties}
We will represent our qubit state $\rho$ as a Bloch vector $\vecr$,
\begin{equation}
    \rho = \frac{1}{2} \left( \id + \vecr \cdot \vec{\sigma}\right)\,, \qquad \abs{\vecr \,}^2 < 1\,, 
\end{equation}
where $\vec{\sigma} = \left( \sigma_x, \, \sigma_y, \, \sigma_z\right)$ is a vector of Pauli matrices. Let's fix the measurement basis as $\pm \vec{m}$ and introduce a `normal' state $\vec{n}$ to which $\vec{m}$ is perpendicular, i.e.
\begin{equation}\label{eqn: qubit meas}
    \ketbra{m_j}{m_j} = \frac{1}{2} \left( \id + \vec{m}_j   \cdot \vec{\sigma}\right)\,,  \qquad \vec{m}_j = \left( -1\right)^{j+1} \vec{m}\,, \qquad \underset{i \neq j}{\sum_{i, \,j=1}^{2}}\ketbra{m_i}{m_j} = \vec{n} \cdot \vec{\sigma}\,.
\end{equation}
Without loss of generality, we will write our state as
\begin{equation}\label{eqn: qubit}
    \vecr = m \vec{m} + p \vec{n}\,, \qquad  r^2= p^2 + m^2  < 1\,,
\end{equation}
with $0 \leq p, \, m < 1$, since the state must be full rank. The square roots $\rho$ and $\rho^{- \frac{1}{2}}$ are given by
\begin{equation}
    \sqrt{\rho} = \frac{1}{ 2\sqrt{\gamma }} \Big( \id + \gamma \vecr \cdot \vec{\sigma} \Big)\,, \qquad \rho^{- \frac{1}{2}} = \frac{\sqrt{\gamma}}{1-\gamma} \left( \id - \gamma \vecr \cdot \vec{\sigma}\right)\,, \qquad \gamma = \frac{1 - \sqrt{1-r^2}}{r^2}\,,
\end{equation}
where we have
\begin{equation}
    \gamma^2 r^2 = 2 \gamma -1\,, \qquad \left( \vecr \cdot \vec{\sigma} \right) \left( \vecr \cdot \vec{\sigma} \right) = r^2 \id\,.
\end{equation}
Note also the useful property
\begin{equation}
    \big( \vec{a} \cdot \vec{\sigma} \big) \big( \vec{b}  \cdot \vec{\sigma} \big) = \vec{a} \cdot \vec{b} \, \id + i \big( \vec{a} \times \vec{b} \big) \cdot \vec{\sigma}
\end{equation}
for any Bloch vectors $\vec{a}$ and $\vec{b}$, such that 
\begin{equation}
    \left( \id + \vec{a} \cdot \vec{\sigma} \right) \left( \id + \vec{b} \cdot \vec{\sigma} \right) \left( \id + \vec{a} \cdot \vec{\sigma} \right) = \left( 1 + 2 \vec{a} \cdot \vec{b} + a^2 \right) \id + \left( 2\left( 1 +  \vec{a} \cdot \vec{b} \, \right) \vec{a} + \left( 1 - a^2 \right) \vec{b}  \,  \right) \cdot \vec{\sigma}\,.
\end{equation}
Eve receives the ensemble of states $\{\eta_j , \, \ket{\phi_j}\}$ given by
\begin{equation}
    \eta_{j} \ketbra{\phi_j}{\phi_j} = \sqrt{\rho} \ketbra{m_j}{m_j} \sqrt{\rho} = \frac{1 + \left( -1 \right)^{j+1} m}{4} \left( \id + \vec{s}_j \cdot \vec{\sigma}\right)\,,
\end{equation}
where
\begin{equation}
    \vec{s}_j = \frac{ \gamma \left( 1 + \left( -1\right)^{j+1} \gamma m\right) \vecr + \left( -1 \right)^{j+1} \left( 1 - \gamma\right) \vec{m}   }{\gamma \left( 1 + \left( -1\right)^{j+1} m \right)}\,
\end{equation}
(note that complex conjugation was not needed here since we work in the real plane of the Bloch ball).
The prior probabilities for the states are then
\begin{equation}
    \eta_j = \frac{1 + \left( -1 \right)^{j+1} m}{2}\,,
\end{equation}
while the squared overlap between the states is
\begin{equation}
    \abs{\langle \phi_1 | \phi_2 \rangle}^2 = \frac{1}{2} \left( 1 + \vec{s}_1 \cdot \vec{s}_2\right) = \frac{1}{2}\left( 1 - \frac{1-r^2 - p^2}{1-m^2}\right) =  \frac{p^2}{1-m^2}\,,
\end{equation}
such that 
\begin{equation}\label{eqn: qub overlap}
    \abs{\langle \phi_1 | \phi_2 \rangle} = \frac{p}{\sqrt{1-m^2}}\,.
\end{equation}
We can then write the parameters $p$ and $m$ in terms of $\eta_1$, $\eta_2$ and $\abs{\langle \phi_1 | \phi_2 \rangle}$ as
\begin{equation}\label{eqn: m and p}
    m = \eta_1 - \eta_2\,, \qquad p = 2 \sqrt{\eta_1 \eta_2} \, \abs{\langle \phi_1 | \phi_2 \rangle}\,.
\end{equation}
In terms of the steered states $\vec{s}_1$ and $\vec{s}_2$, we have
\begin{equation}
      \vec{m} = \frac{a_2 \eta_1 \, \vec{s}_1 - a_1 \eta_2 \, \vec{s}_2 }{  1 - \gamma } \,,\qquad \vec{n} = \frac{b_1 \eta_2 \, \vec{s}_2 - b_2 \eta_1 \,\vec{s}_1}{p \left( 1 - \gamma \right) }\,,
\end{equation}
with 
\begin{equation}
    a_j =  \gamma \left( 1 + \left(-1 \right)^{j+1}\gamma m \right)\,, \qquad b_j =      m \gamma \left( 1 + \left( -1 \right)^{j+1} \gamma m \right) + \left( -1 \right)^{j+1} \left( 1 - \gamma\right)\,.
\end{equation}
Given the optimal decomposition $\{p_\mu, \, \vecr_\mu\}$ in the randomness picture, the optimal state discrimination measurement has elements
\begin{equation}
E_\mu = p_\mu \rho^{- \frac{1}{2}} \rho_{\mu} \rho^{- \frac{1}{2}} = \frac{p_\mu  \left( 1 - \vecr \cdot \vecr_\mu \right)}{1- r^2} \left(  \id - \vec{e}_\mu \cdot \vec{\sigma} \right)  \,,  
\end{equation}
with
\begin{equation}
    \vec{e}_\mu = \frac{1}{\gamma \left( 1 - \vecr \cdot \vecr_\mu\right)} \left( \gamma \left( 1 - \gamma \left(\vecr \cdot \vecr_\mu \right)\right) \vecr - \left( 1 - \gamma\right) \vecr_\mu  \right)\,.
\end{equation}

\subsection{Unambiguous guessing probability}\label{app: UD qub} 

\setcounter{theorem}{2}
From the conditions on the decomposition $\{p_\mu, \, \rho_\mu\}$ in \eqref{eqn: UD rand}, we find that $\vecr$ is decomposed in terms of Bloch vectors as 
\begin{equation}
    \vecr = p_0 \vecr_0 + p_1 \vec{m}_1 + p_2 \vec{m_2} \,.
\end{equation}
Observation \ref{obs: qubits}, proved in the main text, tells us that the maximal randomness of a qubit state $\vecr$ under an unambiguous attack is given by the guessing probability 
\begin{equation}
 \pud^{*} \mleft( \vecr \, \mright) = 1-r 
 \end{equation} 
and is achieved only when $m=0$. We will now prove Theorem \ref{thm: qubits all UD} from the main text, which gives an eavesdropper's unambiguous guessing probability for any qubit state $\vecr$ and measurement $\{\pm \vec{m}\}$.
\setcounter{theorem}{3}
\begin{theorem}\label{thm: app qubits all UD}
    Given a state $\vecr$, of the form \eqref{eqn: qub form}, measured in the basis $\{\pm \vec{m}\}$, the unambiguous guessing probability of an eavesdropper is 
\begin{equation}\label{eqn: qubit UD mine}
    \pud \mleft( \vecr, \, \{\pm \vec{m}\}\mright) = \begin{cases}
        1- p\,, \quad & m + p \leq 1\,, \vspace{0.1 cm}
        \\
        \frac{1-r^2}{2 \left( 1 -m \right)}\,, \quad & m + p > 1 \,.
    \end{cases}
    \end{equation}
\end{theorem}
\begin{proof}
 We split the proof of \eqref{eqn: qubit UD mine} into two cases.
\mbox{}\\
\\
    \textbf{When $m + p \leq 1$.} This condition implies that   $\vecr \in \conv \mleft( \{\pm \vec{m}\}, \, \vec{n} \,\mright)$, so we can find a decomposition of $\vecr$ into $\pm \, \vec{m}$ and $\vec{n}$, 
\begin{equation}
        \vecr = p_0 \vec{n} +p_1 \vec{m}_1 + p_2 \vec{m}_2\,,
    \end{equation}
    where we set $\vecr_0 = \vec{n}$. Since $\vec{m}_1$ and $\vec{m}_2$ are antiparallel, we must have $p_0 =p$. This gives us the lower bound
    \begin{equation}
        \pud \mleft( \vecr, \, \{\vec{m}_j\}\mright) \geq 1-p\,.
    \end{equation}
Choosing the following variables for the dual problem \eqref{eqn: dual UD SDP rand},
\begin{equation}
    Z = \id - \vec{n} \cdot \vec{\sigma}\,, \qquad y_{12}= y_{21} = 1\,,
\end{equation}
it is easily shown that $Z \sgeq 0$ and that 
\begin{equation}
    Z - \id + \underset{i \neq j}{\sum_{i, \, j =1 }^{2}} y_{ij}\ketbra{m_j}{m_i} = Z - \id + \vec{n} \cdot \vec{\sigma} =0\,,
\end{equation}
so these variables are valid. They give the upper bound 
\begin{equation}
 \pud \mleft( \vecr, \, \{\vec{m}_j\}\mright) \leq \frac{1}{2}\tr \mleft( Z \left( \id + \vecr \cdot \vec{\sigma} \right) \mright) = 1-p\,.   
\end{equation}
Since the upper and lower bounds match, we have found the optimal unambiguous guessing probability. 
\mbox{}\\
\\
\textbf{When $m + p >1$.} In this case, we neglect the second outcome entirely, setting $p_2=0$, such that the decomposition is
\begin{equation}\label{eqn: UD decomp 1}
    \vecr_0 = \vec{u}\,, \qquad \vecr_1 = \vec{m}\,,
\end{equation}
where 
\begin{equation}\label{eqn: UD decomp 2}
    p_0 \vec{u} = p \vec{n} + \left(m -p_1 \right) \vec{m}\,, \qquad p_1 = \frac{1-r^2}{2 \left( 1-m\right)}\,, \qquad p_0 = 1 - p_1\,.
\end{equation}
We can then lower bound the guessing probability by 
  \begin{equation}
        \pud \mleft( \vecr, \, \{\vec{m}_j\}\mright) \geq \frac{1-r^2}{2 \left( 1-m\right)}\,.
    \end{equation}  
Consider the following variables for the dual problem \eqref{eqn: dual UD SDP rand}, 
\begin{equation}
    Z = \frac{1-p_1}{1-m } \left( \id - \frac{ \vecr -p_1 \vec{m} }{1-p_1} \cdot \vec{\sigma} \right)\,, \qquad y_{12}=y_{21}= \frac{p}{1-m}\,.
\end{equation}
We have $Z \sgeq 0$, since 
\begin{equation}
    \abs{\frac{ \vecr -p_1 \vec{m} }{1-p_1}}^2 =1\,,
\end{equation}
and
\begin{equation}
    Z - \id + \underset{i \neq j}{\sum_{i, \, j=1}^{2}} y_{ij} \ketbra{m_j}{m_i} = \frac{m-p_1}{1-m} \left(  \id - \vec{m} \cdot \vec{\sigma}\right) \sgeq 0\,, 
\end{equation}
where positive semidefiniteness holds because 
\begin{equation}
    m - p_1 = \frac{p^2 - \left( 1-m\right)^2}{2 \left( 1-m\right)} = \frac{ \left( 1+p -m \right) \left( m+p-1\right)}{2 \left( 1-m\right)} > 0\,,
\end{equation}
where we use the inequality $m+ p > 1$. These variables give us the upper bound 
\begin{equation}
        \pud \mleft( \vecr, \, \{\vec{m}_j\}\mright) \leq \frac{1}{2}\tr \mleft( Z \left( \id + \vecr \cdot \vec{\sigma} \right) \mright) = \frac{1-r^2}{2 \left( 1 -m \right)}\,.
    \end{equation} 
    Since the upper and lower bounds match, we have found the optimal unambiguous guessing probability, and proven Theorem \ref{thm: qubits all UD}.
        \end{proof}
We can translate the result \eqref{eqn: qubit UD mine} to the state discrimination case by subbing in the parameters from \eqref{eqn: qub overlap} and \eqref{eqn: m and p}. This lets us recover the results of \cite{Jaeger_1995},
\begin{equation}\label{eqn: qubit UD jaeger}
    \pud \mleft( \{\eta_j, \, \ket{\phi_j} \}\mright) = \begin{cases}
        1- 2 \sqrt{\eta_1 \eta_2} \, \abs{\langle \phi_1 | \phi_2 \rangle}\,, \quad & \sqrt{\eta_1} \abs{\langle \phi_1 | \phi_2 \rangle} \leq \sqrt{\eta_2} \vspace{0.15 cm}
        \\[2pt]
        \eta_1 \left( 1 - \abs{  \langle \phi_1 | \phi_2 \rangle}^2 \right)\,, \quad & \sqrt{\eta_1} \abs{\langle \phi_1 | \phi_2 \rangle} > \sqrt{\eta_2} \,.
    \end{cases}
    \end{equation}

\subsection{Fixed rate of inconclusive outcomes}\label{app: FRIO qub}
Since we now know the eavesdropper's unambiguous guessing probability given a qubit state $\rho$, described by \eqref{eqn: qubit}, and a measurement in an orthonormal basis $\{\ket{m_j}\}$, described by \eqref{eqn: qubit meas}, we can define $\qmax$, the maximum rate of inconclusive outcomes given the optimal guessing strategy. From \eqref{eqn: qubit UD mine}, we have
\begin{equation}
    \qmax = \begin{cases}
        p\,, \quad & m+ p \leq 1\,,
        \\[2pt]
        \frac{p^2 + \left( 1-m\right)^2}{2 \left( 1-m\right)}\,, \quad & m + p >1\,.
    \end{cases} 
\end{equation}
We will also define a critical value for the FRIO parameter,
 \begin{equation}
     \qcrit = \frac{1}{2} \frac{1-r^2}{1-p}\,.
 \end{equation}
 Note that
\begin{equation}
     \qmax - \qcrit = \begin{cases}
         \frac{ \left( m + p -1\right) \left( 1 + m -p\right)  }{2 \left( 1-p \right) }  \;\, \leq 0 \,,  \quad & m + p \leq 1\,, \vspace{0.15 cm}
         \\
         \frac{\left( m + p -1\right) \left( m + p -r^2\right) }{ 2 \left( 1-m\right) \left( 1-p\right) } > 0\,, \quad & m + p > 1\,.
     \end{cases}
 \end{equation}
Since $Q < \qmax$, this means $Q < \qcrit$ whenever $m+p \leq 1$, while when $m+p>1$, there are two distinct regions of $Q$, $0 < Q \leq \qcrit$ and $\qcrit \leq Q < \qmax$. Notice, further, that $Q < p$ holds whenever $Q \leq \qcrit$. When $m+p\leq 1$, this is trivial, since $\qmax=p$ and $Q < \qmax$. When $m+p>1$, we have
\begin{equation}
     \qcrit - p = -\frac{\left( m + p -1 \right) \left( 1+m -p \right) }{2 \left( 1-p\right)}  < 0\,,
 \end{equation}
so $Q \leq \qcrit < p$.
We are now ready to prove Theorem \ref{thm: qubits FRIO} from the main text.
\begin{theorem}\label{thm: app qubits FRIO}
    Given a qubit state $\vecr$, of the form \eqref{eqn: qub form}, measured in the basis $\{\pm \vec{m}\}$, the guessing probability of an eavesdropper given a fixed rate $Q$ of inconclusive outcomes is
\begin{equation}\label{eqn: qubit Q}
    \pq \mleft( \vecr, \, \{\pm \vec{m}\}\mright) = \frac{1}{2}\begin{cases}
         1 - Q + \sqrt{ \left( 1- Q\right)^2 - \left(p -Q \right)^2}\,, \quad &  Q \leq \qcrit\,, \vspace{0.2 cm}
        \\
         1 - Q + m - \frac{Q }{r^2} \left( Am - p \sqrt{r^2 - A^2}\right)\,,  \quad &  Q > \qcrit\,,
    \end{cases}
\end{equation}
where 
\begin{equation}
  A = 1-\frac{\qcrit \left( 1-p\right)}{Q}\,.
\end{equation}
 \end{theorem}

\begin{proof}
\mbox{}\\
\\
\textbf{When $Q \leq \qcrit$.} As argued above, $Q < p$ in this region. Choosing $p_0=Q$ and $\vecr_0 = \vec{n}$, we decompose the state $\vecr$ as
\begin{equation}\label{eqn: in Qcrit decomp}
    \vecr= Q \vec{n} +p_1 \vecr_1 + p_2 \vecr_2 \,, 
\end{equation}
where 
\begin{equation}
    \vecr_j =  \sqrt{1-q^2} \,\vec{m}_j + q \vec{n}\,, \qquad q = \frac{p-Q}{1-Q}\,, \qquad p_j = \frac{1}{2} \left( 1 - Q +  \frac{ \left( -1 \right)^{j+1} m}{\sqrt{1-q^2}} \right)
\end{equation}
for $j \in \{1, \, 2\}$.
This decomposition is valid because $0 < q < 1$ and
\begin{equation}
    p_2 =  \frac{1}{2} \left( 1 - Q -  \frac{  m}{\sqrt{1-q^2}} \right) = \frac{\left(1-p\right) \left( \qcrit - Q\right)}{ \sqrt{1-q^2} \left( \left(1 - Q \right) \sqrt{1-q^2} +m  \right)}\geq 0\,.
\end{equation}
We can then bound the guessing probability from below by
\begin{equation}
        \pq \mleft( \vecr, \, \{\vec{m}_j\}\mright) \geq \frac{1}{2}\sum_{j=1}^{2} p_j \left( 1 + \vecr_j \cdot \vec{m}_j\right) =  \frac{1}{2} \left( 1 - Q + \sqrt{ \left( 1- Q\right)^2 - \left(p -Q \right)^2}\right)    \,.
    \end{equation}
Consider the following variables for the dual problem \eqref{eqn: dual Q SDP rand}, 
\begin{equation}
    G = \frac{1+ \sqrt{1-q^2}}{2 \sqrt{1-q^2}} \left( \id + \vec{g} \cdot \vec{\sigma } \right)\,, \qquad \vec{g} = - \frac{1 - \sqrt{1-q^2}}{q} \vec{n}\,, \qquad w = \frac{1}{2} \left( 1 + \sqrt{\frac{1-q}{1+q}} \right)\,.
\end{equation}
These variables satisfy the necessary constraints, since we have 
\begin{equation}
    G - \ketbra{m_j}{m_j} = \frac{1}{2 \sqrt{1-q^2}} \left( \id - \sqrt{1-q^2} \left( \frac{q}{\sqrt{1-q^2}} \vec{n} +\left( -1 \right)^{j+1} \vec{m} \right) \cdot \vec{\sigma} \right) \sgeq 0\,,
\end{equation}
which is positive semidefinite because 
\begin{equation}
    \abs{\frac{q}{\sqrt{1-q^2}} \vec{n} - \left( -1 \right)^{j+1} \vec{m}}^2 = \frac{1}{1-q^2}
\end{equation}
and 
\begin{equation}
    G - w \id = \frac{q}{2 \sqrt{1-q^2}} \left( \id - \vec{n} \cdot \vec{\sigma} \right) \sgeq 0\,.
\end{equation}
These variables give the upper bound
\begin{equation}
 \pq \mleft( \vecr, \, \{\vec{m}_j\}\mright) \leq \frac{1}{2}\tr \mleft( G \left( \id + \vecr \cdot \vec{\sigma} \right) \mright) - w Q =  \frac{1}{2} \left( 1 - Q + \sqrt{ \left( 1- Q\right)^2 - \left(p -Q \right)^2 }\right)\,.   \end{equation}
Since the upper and lower bounds coincide, we conclude that this is the optimal guessing probability. 
\mbox{}\\
\\
\textbf{When $Q > \qcrit$.} This condition implies $m+p >1$. For convenience, we define
\begin{equation}
  A = 1 - \frac{1-r^2}{2Q} =1-\frac{\qcrit \left( 1-p\right)}{Q}\,.
\end{equation}
The variable $A$ increases with $Q$, so in the range $\qcrit < Q < \qmax$, we have
\begin{equation}
   p < A < 1 - \frac{\left( 1-m\right)\left( 1-r^2\right) }{p^2 + \left(1-m \right)^2 } \,.
\end{equation}
We decompose $\vecr$ as
\begin{equation}
    \vecr= Q \vec{u} +\left( 1-Q\right) \vec{v} \,,
\end{equation}
where we choose $\vecr_0= \vec{u}$ and $\vecr_1=\vec{v}$, with
\begin{equation}\label{eqn: v decomp}
    \vec{v} = v_1 \vec{m} + v_2 \vec{n} =\frac{1}{1-Q} \left( \vecr - Q \vec{u} \right)
\end{equation}
and
\begin{equation}\label{eqn: u decomp}
    \vec{u} = u_1 \vec{m} + u_2 \vec{n}\,, \qquad u_1=  \frac{Am - p \sqrt{r^2 - A^2}}{r^2 }\,, \qquad u_2 = \frac{Ap + m \sqrt{r^2 - A^2}}{r^2 }\,.  
\end{equation}
We first prove that the square root $\sqrt{r^2 - A^2}$ is well defined. This is so because 
\begin{equation}
    r - A = \frac{\left( 1-r\right)\left( 1+ r -2Q \right)}{2Q} >  \frac{\left( 1-r\right)\left( 1+ r -2\qmax \right)}{2Q} =\frac{\left( 1-r\right)^2 \left( r+m\right)}{2Q \left( 1-m \right)} >0\,.
\end{equation}
We also have $\vecr \cdot \vec{u}=A$, such that $\vec{v}$ is pure. The variable $u_1$ increases continuously with $A$, so, writing it explicitly in terms of $Q$ as
\begin{equation}
    u_1 \mleft( Q\mright) = \frac{1}{2Qr^2} \left( \left( 2Q -1 + r^2\right)m - p \sqrt{ \left( 1-r^2\right) \left( 4Q \left(1-Q \right) -1 +r^2\right)  }\right)\,,
\end{equation}
we have 
\begin{equation}
    u_1 \mleft( Q \mright) > u_1 \mleft( \qcrit\mright) = \frac{p \left( m - \sqrt{r^2-p^2}\right)}{r^2} =0
\end{equation}
and 
\begin{equation}
    u_1 \mleft( Q \mright) < u_1 \mleft( \qmax \mright) = \frac{p^2 - \left( 1-m \right)^2}{2 \qmax \left(1-m \right)} = \frac{m -1 + \qmax}{\qmax}\,.
\end{equation}
Similarly, the variable $u_2$ decreases continuously with $A$, because
\begin{equation}
    \frac{\partial u_2}{\partial A} = \frac{  p^2 - A^2 }{ \sqrt{r^2 - A^2} \left( p \sqrt{r^2-A^2} +Am \right) } < 0\,,
\end{equation}
so, expressing $u_2$ as
\begin{equation}
    u_2 \mleft( Q \mright) = \frac{1}{2Qr^2} \left( \left( 2Q -1 + r^2\right)p + m \sqrt{ \left( 1-r^2\right) \left( 4Q \left(1-Q \right) -1 +r^2\right)  }\right)\,,
\end{equation}
we have
\begin{equation}
    u_2 \mleft( Q \mright) < u_2 \mleft( \qcrit \mright) = 1\,, \qquad u_2 \mleft( Q \mright) \geq u_2 \mleft( \qmax \mright) = \frac{p}{\qmax}\,.
\end{equation}
We can see then that the decomposition in \eqref{eqn: v decomp} and \eqref{eqn: u decomp} interpolates between the decomposition \eqref{eqn: in Qcrit decomp} at the critical point $\qcrit$ and the decomposition \eqref{eqn: UD decomp 1} and \eqref{eqn: UD decomp 2}, for the unambiguous case, at $\qmax$. Moreover, we have
\begin{equation}
\begin{aligned}
 m -   u_1 \mleft( Q\mright) &> m -   u_1 \mleft( \qmax \mright) = \frac{\left(1-m \right) \left( 1- \qmax\right)}{\qmax} >0\,, 
 \\
u_2 \mleft( Q \mright) -p   &>  u_2 \mleft( \qmax \mright) -p = \frac{p \left( 1- \qmax\right)}{\qmax} > 0\,,
 \end{aligned}
\end{equation}
so, from \eqref{eqn: v decomp}, we find $u_1 < v_1$ and $u_2 > v_2 $, such that
\begin{equation}
    u_2 v_1 - u_1 v_2 >0\,.
\end{equation}
The decomposition in \eqref{eqn: v decomp} and \eqref{eqn: u decomp} gives us the following lower bound on the guessing probability,
\begin{equation}
\begin{aligned}
        \pq \mleft( \vecr, \, \{\vec{m}_j\}\mright) &\geq \frac{1}{2} \left( 1- Q\right) \left( 1 + \vec{v} \cdot \vec{m}\right) =  \frac{1}{2} \left( 1 - Q + m - Q u_1 \right)  
        \\[2pt]
        &= \frac{1}{2} \left( 1 - Q + m - \frac{Q}{r^2} \left( Am - p \sqrt{r^2 - A^2} \right) \right) \,.
        \end{aligned}
    \end{equation}
For the upper bound, consider the following variables for the dual problem \eqref{eqn: dual Q SDP rand}
\begin{equation}
    G= \frac{g}{2}\left( \id + \vec{g} \cdot \vec{\sigma}     \right)\,, \qquad \vec{g} =\frac{\vec{m} - \left( g-1\right)\vec{v} }{g}\,,
    \end{equation}
where
\begin{equation}
  g = 1 + \frac{u_2}{u_2 v_1 -u_1 v_2 }\,, \qquad w = \frac{1}{2} \left( 1 + \frac{u_2 -v_2}{u_2 v_1 - u_1 v_2 } \right)\,. 
\end{equation}
Notice that the vector $\vec{g}$ is not normalized, and that it has overlaps
\begin{equation}
\begin{aligned}
    g \vec{g} \cdot \vec{m} &= 1 - \left( g-1\right) v_1 = - \frac{u_1 v_2}{ u_2 v_1 - u_1 v_2 }\,,
    \\
    g \vec{g} \cdot \vec{n} &= - \left( g-1\right) v_2 = - \frac{u_2 v_2}{u_2 v_1 - u_1 v_2}\,,
    \end{aligned}
\end{equation}
so we can write $g \vec{g}$ as
\begin{equation}\label{eqn: g vec}
    g \vec{g} = - \frac{v_2}{u_2 v_1 - u_1 v_2} \vec{u} = - \left( g - 2w \right) \vec{u}\,.
\end{equation}
These variables satisfy
\begin{equation}
    G - \ketbra{m_1}{m_1} = \frac{g-1}{2} \left( \id + \frac{g \vec{g} - \vec{m}}{g-1} \cdot \vec{\sigma} \right) = \frac{g-1}{2} \left( \id - \vec{v} \cdot \vec{\sigma} \right) \sgeq 0\,,
\end{equation}
since $g-1 > 0$, and
\begin{equation}
 G - \ketbra{m_2}{m_2} = \frac{g-1}{2} \left( \id + \frac{g \vec{g} + \vec{m}}{g-1} \cdot \vec{\sigma} \right) = \frac{g-1}{2} \left( \id + \frac{2\vec{m} -\left( g-1\right)\vec{v} }{g-1}  \cdot \vec{\sigma} \right) \sgeq 0\,,
\end{equation}
since 
\begin{equation}
    \abs{2 \vec{m} - \left(g-1 \right) \vec{v} \,}^2 = 4 \left( 1 - \left(g-1 \right)v_1   \right) + \left( g-1\right)^2  < \left(g-1 \right)^2\,.
\end{equation}
Finally, using the form \eqref{eqn: g vec}, we have
\begin{equation}
    G - w \id = \frac{g-2w}{2} \left( \id + \frac{g \vec{g} }{g - 2w} \cdot \vec{\sigma} \right) = \frac{g-2w}{2} \left( \id  - \vec{u} \cdot \vec{\sigma} \right) \sgeq 0\,,
\end{equation}
since $g - 2w >0$. The variables are valid, so we can use them to upper bound the guessing probability. First, we remark that
\begin{equation}\label{eqn: square root}
    u_2 v_1 - u_1 v_2 = \frac{m u_2 - p u_1}{1-Q} = \frac{\sqrt{r^2 - A^2}}{1-Q}\,.
\end{equation}
Finally, then, we have 
\begin{equation}
    \begin{aligned}
        \pq \mleft( \vecr, \, \{\vec{m}_j\}\mright) &\leq \tr \left( G \rho\right) - w Q = \frac{1}{2} \left( g - 2wQ+ g \vec{g} \cdot \vecr \right) 
       \\[2pt]
       &= \frac{1}{2} \left( 1 - Q + m + \frac{u_2 \left( 1-Q - \vec{v} \cdot \vecr \right) +Q v_2  }{u_2 v_1 - u_1 v_2} \right)
       \\[2pt]
       &= \frac{1}{2} \left( 1 - Q + m + \frac{Q \left(p - A  u_2\right)}{  \left( 1 - Q \right)\left(u_2  v_1 - u_1 v_2\right) } \right)
       \\[2pt]
       &= \frac{1}{2} \left( 1 - Q + m + \frac{Q \left( p \sqrt{r^2 - A^2} - Am\right)}{  r^2  } \right)\,,
    \end{aligned}
\end{equation}
where in the third line we use \eqref{eqn: v decomp} and $\vec{u} \cdot \vecr =A $, and in the final line we use \eqref{eqn: square root}. 
Since the upper and lower bounds match, we have found the optimal guessing probability. 
\end{proof}
Swapping to a state discrimination scenario by substituting \eqref{eqn: m and p} into \eqref{eqn: qubit Q}, we can recover the expressions from \cite{Bagan_2012, Jiménez_2021}. The probability of error is $\perr = 1 - Q - \pq \mleft( \vecr, \, \{\pm \vec{m}\}\mright)$. If we fix $\perr=T$, we can invert \eqref{eqn: qubit Q} to find $Q$ in terms of $T$ \cite{Herzog_2012}. Let's first find the extreme points of $T$. From $Q < \qmax$, we find $T > 0$, while from $Q > 0$, we find \cite{Meng_2024}
\begin{equation}
    T < \tmax = \frac{1}{2} \left( 1 - \sqrt{1-p^2}\right) = \frac{1}{4} \left( \sqrt{1+p} - \sqrt{1-p}\right)^2\,.
\end{equation}
We can also identify the critical point
\begin{equation}
    \tcrit = \frac{1}{2} \left( 1 - m - \qcrit\right) = \frac{\left( m+p-1\right)^2}{4 \left(1-p \right)}\,.
\end{equation}
When $Q \leq \qcrit$, we can invert \eqref{eqn: qubit Q} to find 
\begin{equation}\label{eqn: Q first}
  Q \mleft( T \mright) = p - 2 T - 2\sqrt{T \left( 1-p \right) }\,.  
\end{equation}
This function decreases continuously with $T$, so we have
\begin{equation}
    Q \mleft( T \mright) > Q \mleft( \tmax \mright) = - 1 + p + \sqrt{1-p^2} - \sqrt{1-p} \left( \sqrt{1+p} - \sqrt{1-p} \right) =0\,.
\end{equation}
When $m+p \leq 1$, we have
\begin{equation}
    Q \mleft( T \mright) < Q \mleft( 0 \mright) = p\,,
\end{equation}
while when $m+p >1$ (still with $Q \leq \qcrit$), we have
\begin{equation}
    Q \mleft( T \mright) \leq Q \mleft( \tcrit \mright) = \frac{1-r^2}{2 \left( 1-p\right)} = \qcrit \,,
\end{equation}
so the function $Q \mleft( T\mright)$ in \eqref{eqn: Q first} lives in the correct range. Using \eqref{eqn: Q first}, the guessing probability for a fixed $T \geq \tcrit$ is then
\begin{equation}
    \pt \mleft( \vecr, \, \{\pm \vec{m} \} \mright) = 1 - T - Q \mleft( T\mright) = \left( \sqrt{1-p} + \sqrt{T} \right)^2\,.
\end{equation}
When $Q > \qcrit$, we can invert \eqref{eqn: qubit Q} to find 
\begin{equation}\label{eqn: Q second}
    Q \mleft( T \mright) = 1 - T - \frac{1}{2 \left(1-m \right)^2} \left( p \sqrt{2 T} + \sqrt{\left(1-r^2 \right) \left( 1-m-2T\right) } \right)^2\,.
\end{equation}
For convenience, let's define the function
\begin{equation}
    f \mleft( T \mright) = p \sqrt{2 T} + \sqrt{ \left(1 - r^2 \right) \left( 1- m - 2T\right) }\,.
\end{equation}
Its derivative is
\begin{equation}
\begin{aligned}
    \frac{\partial f \mleft( T \mright)}{\partial T} &= \frac{\left( 1-m\right) \left( p^2 - 2 T \left(1+m \right)\right) }{ \sqrt{ 2T \left( 1 - m - 2 T \right)} \left( p \sqrt{1-m-2T} + \sqrt{ 2 T \left(1 - r^2 \right) }\right)   }
    \\
    &> \frac{ \left( 1-m\right)  \left( 1-r^2\right) \left( \left( m + p -1\right) +p \right) }{ 2 \left( 1-p\right)\sqrt{ 2T \left( 1 - m - 2 T \right)} \left( p \sqrt{1-m-2T} + \sqrt{ 2 T \left(1 - r^2 \right) }\right)   } >0\,,
    \end{aligned}
\end{equation}
where we use $T < \tcrit$ in the second line. We find then that the $Q \mleft( T\mright)$ decreases strictly with $T$, as expected. We can then bound $Q \mleft( T\mright)$ by
\begin{equation}
    Q \mleft( T \mright) < Q \mleft( 0 \mright) = 1 - \frac{1-r^2}{2 \left( 1 - m\right)} = \qmax\,.
\end{equation}
Using the property
\begin{equation}
\begin{aligned}
 \frac{\left( 1-m\right) \left( 1-p+m\right) }{\sqrt{2 \left( 1-p\right)}} &=  \frac{p \left( m+p-1 \right) + 1 - r^2 }{\sqrt{2 \left( 1-p\right)}}= p \sqrt{2 \tcrit} + \sqrt{ \left( 1-r^2\right) \left( 1- m - 2 \tcrit\right)^2 }\,,
    \end{aligned}
\end{equation}
we also have 
\begin{equation}
  Q \mleft( T\mright) >  Q \mleft( \tcrit \mright) = 1 - \tcrit - \frac{\left( 1-p+m\right)^2}{4 \left( 1-p \right)} = \qcrit\,.
\end{equation}
The function $Q \mleft( T\mright)$ from \eqref{eqn: Q second} then lives in the correct range. The guessing probability when $T < \tcrit$ is then
\begin{equation}
    \pt \mleft( \vecr, \, \{\pm \vec{m}\}\mright) = 1 -  T - Q \mleft( T\mright) =\frac{1}{2 \left(1-m \right)^2} \left( p \sqrt{2 T} + \sqrt{\left(1-r^2 \right) \left( 1-m-2T\right) } \right)^2\,.
\end{equation}
Summarizing the above, for a fixed error probability $T$, the eavesdropper's optimal guessing probability given a qubit state $\vecr$ and a measurement $\{\pm \vec{m}\}$ is
\begin{equation}\label{eqn: qubit T}
    \pt \mleft( \vecr, \, \{\pm \vec{m}\}\mright) = \begin{cases}
    \left( \sqrt{1-p} + \sqrt{T} \right)^2\,, \quad & T \geq \tcrit\,, \vspace{0.2 cm}
     \\
     \frac{1}{2 \left(1-m \right)^2} \left( p \sqrt{2 T} + \sqrt{\left(1-r^2 \right) \left( 1-m-2T\right) } \right)^2\,, \quad & T < \tcrit\,.
    \end{cases}
\end{equation}
 One could convert to a state discrimination scenario by subbing in
\begin{equation}
    m = \eta_1 - \eta_2\,, \qquad p = 2 \sqrt{\eta_1 \eta_2} \abs{\langle \phi_1 | \phi_2 \rangle}\,, \qquad r^2 = 1 - 4 \eta_1 \eta_2 \left(1 -  \abs{\langle \phi_1 | \phi_2 \rangle}^2  \right)\,,
\end{equation}
from \eqref{eqn: m and p}, into \eqref{eqn: qubit T}, recovering the results of \cite{Sugimoto_2009, Hayashi_2008}.

\section{Noisy states}

In any finite dimension $d$, we consider states of the form
\begin{equation}\label{eqn: noisy state}
    \rho_\ve = \left( 1 - \ve \right) \ketbra{\phi}{\phi} + \frac{\ve}{d} \id\,.
\end{equation}
For future convenience, we define the terms
\begin{equation}
    \aep = d- \ve \left( d-1 \right)\,, \qquad \id_{\neq \phi} = \id - \ketbra{\phi}{\phi}\,,
\end{equation}
such that 
\begin{equation}\label{eqn: sqrt noisy}
    \rho_\ve = \frac{1}{d} \left( \aep \ketbra{\phi}{\phi} + \ve \id_{\neq \phi}\right)\,, \qquad \sqrt{\rho_{\ve}} = \frac{1}{\sqrt{d}} \left( \sqrt{\aep}
    \ketbra{\phi}{\phi} + \sqrt{\ve} \, \id_{\neq \phi}\right)\,.
\end{equation}
For any basis $\{\ket{m_j}\}$ that's unbiased to the state $\ket{\phi}$, we have the useful properties
\begin{equation}
    \langle m_j | \rho_{\ve} | m_j \rangle= \frac{1}{d} \,, \qquad \langle m_j | \sqrt{\rho_\ve} | m_j \rangle = \frac{\tr \! \sqrt{\rho_\ve}}{d} \qquad \textnormal{for all} \;\; j \in \{1, \, \hdots, \, d\}\,.
\end{equation}

\subsection{Unambiguous guessing probability}
As a special case of Theorem \ref{thm: any state}, we know that 
\begin{equation}
    \pud^{*} \mleft( \rho_\ve \mright) = \ve\,.
\end{equation}
This bound is saturated, for example, when the state $\ket{\phi}$ is unbiased to the measurement basis $\{\ket{m_j}\}$, as then we could represent $\rho_\ve$ in a basis $\{\ket{u_j}\}$ that is unbiased with respect to $\{\ket{m_j}\}$, which ensures that $\abs{\langle \umin | m_j \rangle}^2= \frac{1}{d}$ for all $j$. This proves Observation \ref{obs: noisy}.

We know, as a special case of \cite[Theorem 1]{Meng_2024}, that the maximal randomness from a  noisy state in the minimum-error ($Q=0$) scenario is
\begin{equation}
    \pme^{*} \mleft( \rho_\ve\mright) = \frac{1}{d} \left( \tr \sqrt{\rho_\ve}\right)^2\,,
\end{equation}
so Eve's probability of error, given the optimal strategy, is in the range
\begin{equation}
    0 < \perr < 1 - \frac{1}{d} \left( \tr \sqrt{\rho_\ve}\right)^2\,.
\end{equation}

\subsection{Fixed rate of inconclusive outcomes}\label{app: noisy FRIO}
The eigenvalues of $\rho_\ve$ are as follows, 
\begin{equation}
    \lambda_1 = \frac{A_{\ve}}{d}\,, \qquad \lambda_{j} = \frac{\ve}{d} \qquad \textnormal{for all} \;\; j \in \{2, \, \hdots, \, d\}\,.
\end{equation}
Choosing $t=1$, the constraint \eqref{eqn: r cond} then becomes
\begin{equation}
    0 < Q \leq \lambda_1 - \lambda_2 = 1 - \ve\,.
\end{equation}
Since this covers the whole allowed range of $Q$, we are free to choose $t=1$ for any noisy state $\rho_\ve$.
As a special case of Theorem \ref{thm: any state FRIO}, then, we find 
\begin{equation}\label{eqn: app Pq noisy}
    \pq^{*} \mleft( \rho_\ve \mright) = \frac{1}{d} \left( \sqrt{\lambda_1 -Q} + \sum_{j=2}^{d} \sqrt{\lambda_l} \right) = \frac{1}{d^2} \left( \sqrt{d \left(1-Q \right) - \left(d-1 \right)\ve} + \left( d-1\right) \sqrt{\ve} \right)^2\,.
\end{equation}
We have 
\begin{equation}
    \chi_{j} = \begin{cases}
        1 - \frac{A_{\ve} - Q d}{A_{\ve}} \qquad & j=1\,,
        \\[2pt]
        0 \qquad & j \in \{2, \, \hdots, \, d\}\,,
    \end{cases}
\end{equation}
such that 
\begin{equation}
    \Lambda = \sqrt{\frac{A_{\ve} - Qd}{A_{\ve}}} \ketbra{u_1}{u_1} + \sum_{j=2}^{d} \ketbra{u_j}{u_j}
\end{equation}
and 
\begin{equation}
    \Lambda \sqrt{\rho} = \left( \sqrt{\frac{A_{\ve} - Qd}{d}} - \sqrt{\frac{\ve}{d}} \right) \ketbra{u_1}{u_1} + \sqrt{\frac{\ve}{d}} \id\,.
\end{equation}
The condition in \eqref{eqn: opt cond FRIO} holds if and only if the state $\ket{\phi}$ is unbiased to the measurement basis $\{m_j\}$. This proves Observation \ref{obs: noisy FRIO}.

Let's now translate this result to the case where the eavesdropper has a  fixed probability $T$ of making an error. Fixing $\m= \{M_j\}_j$, where $M_{j}=\ketbra{m_j}{m_j}$ and $\{\ket{m_j}\}$ is unbiased to the state $\ket{\phi}$, we can use \eqref{eqn: app Pq noisy} to write
\begin{equation}\label{eqn: easier noisy FRIO}
 \pq \mleft( \rho_\ve, \, \m \mright) = \frac{1-Q}{d} \left( \tr \sqrt{\rho_\gamma}\right)^2\,, \qquad \gamma = \frac{\ve}{1-Q}\,,  \end{equation}
 where $0 < Q < \qmax$ and $\qmax = 1 - \ve$.
The probability of error is then 
\begin{equation}
    \perr = 1-Q - \frac{1-Q}{d} \left( \tr \sqrt{\rho_\gamma}\right)^2\,.
\end{equation}
The extreme points of $T$ are $T=0$, at $\qmax$, and $\tmax$, corresponding to $Q=0$, where
\begin{equation}
    \tmax = 1 - \frac{1}{d} \left( \tr \sqrt{\rho_\ve}\right)^2 = \frac{d-1}{d^2} \left( \sqrt{d - \ve \left( d-1\right)} - \sqrt{\ve} \right)^2\,.
\end{equation}
If we fix $\perr=T$, we can invert \eqref{eqn: easier noisy FRIO} to find $Q$ in terms of $T$ as
\begin{equation}
 Q \mleft( T \mright) =   1 -T - \frac{1}{d-1} \left( \sqrt{T} + \sqrt{\ve \left(d-1 \right)} \right)^2\,.
\end{equation}
This function decreases continuously with $T$, as expected, so we have 
\begin{equation}
    Q \mleft( T \mright) < Q \mleft( 0\mright) = 1 - \ve = \qmax
\end{equation}
and 
\begin{equation}
    Q \mleft( T \mright) > Q \mleft( \tmax \mright) = 1 - \tmax - \frac{1}{d^2} \left( \sqrt{d - \ve \left( d-1\right)} + \left( d-1\right) \sqrt{\ve} \right)^2 =0\,,
\end{equation}
so $Q \mleft( T \mright)$ has the correct range. 
We can then solve the eavesdropper's optimal guessing probability with a fixed probability $T$ of error,  
\begin{equation}\label{eqn: noisy Q T}
  \pt \mleft( \rho_\ve, \, \m \mright) =  \frac{1}{d-1} \left( \sqrt{T} + \sqrt{\ve \left(d-1 \right)} \right)^2  \,, \qquad 0 < T < \tmax\,.
\end{equation}

\section{Noisy states and measurements}\label{app: noisy both}
We will now consider a scenario where Alice holds the noisy state $\rho_\ve$ \emph{and} a noisy measurement, given by $\m_\ve = \{M_{x, \, \ve}\}_x$ with
\begin{equation}
  M_{x, \, \ve} =  \left( 1 - \ve \right) \ketbra{m_x}{m_x} + \frac{\ve}{d} \id \qquad \textnormal{for all} \;\; x \in \{1, \, \hdots, \, d\}\,,
\end{equation}
where $\{\ket{m_x}\}$ is an orthonormal basis and, as before, 
\begin{equation}
    \ket{\phi} = \frac{1}{\sqrt{d}} \sum_{x=1}^{d} \ket{m_x}\,.
\end{equation}
Note that 
\begin{equation}
    \langle \phi | M_{x, \, \ve} | \phi \rangle = \frac{1}{d}\,, \qquad \langle \phi | \sqrt{M_{x, \, \ve}} | \phi \rangle = \frac{\tr \sqrt{\rho_{\ve}}}{d} \qquad \textnormal{for all} \;\; x \in \{1, \, \hdots, \, d\}\,,
\end{equation}
and that
\begin{equation}
    \sqrt{\rho_\ve} \ket{m_x} = \frac{1}{d} \left( \left( \sqrt{A_\ve} - \sqrt{\ve} \right) \ket{\phi} + \sqrt{d \ve} \ket{m_x} \right) \qquad \textnormal{for all} \;\; x \in \{1, \, \hdots, \, d\}\,.
\end{equation}
We assume that Eve can simultaneously decompose $\rho_\ve$ into states $\{\ket{\psi_\mu}\}$ and $\m_\ve$ into POVMs $\{\mathcal{N}_{\nu}\}$ according to a joint probability distribution $\{p_{\mu, \, \nu}\}$. The measurements $\mathcal{N}_\nu=\{N_{x, \, \nu}\}_{x}$ must be valid, i.e. 
\begin{equation}
\begin{aligned}
  N_{x, \, \nu} \sgeq 0\,, \qquad \sum_{x=1}^{d} N_{x, \, \nu} = \id \qquad \textnormal{for all} \;\; x \in \{1, \, \hdots , \, d\}\,, \quad \nu \in \{0, \, \hdots, \, d\}\,, 
\end{aligned}
\end{equation}
while the joint conditions are (see \cite[Equation 5]{Senno_2023})
\begin{equation}\label{eqn: joint constraints}
\begin{aligned}
    &\sum_{\mu, \, \nu=0}^{d} p_{\mu, \, \nu} \ketbra{\psi_\mu}{\psi_\mu} = \rho_{\ve}\,,
    \\
    &\sum_{\mu, \, \nu =0}^{d} p_{\mu, \, \nu} N_{x, \, \nu} = M_{x, \, \ve}\,,
    \\
    & \sum_{\mu, \, \nu =0}^{d} p_{\mu, \, \nu} \langle \psi_\mu | N_{x, \, \nu} | \psi_\mu \rangle = \tr \mleft(\rho_\ve \, M_{x, \, \ve} \mright)\,.
\end{aligned}    
\end{equation}
For convenience, we denote Eve's decomposition by the shorthand $\mathcal{D} = \{\{p_{\mu, \, \nu}\}, \, \{\ket{\psi_\mu} \}, \, \{ \mathcal{N_\nu} \}   \}$. Given some decomposition $\mathcal{D}$ that satisfies the constraints \eqref{eqn: joint constraints}, Eve chooses her outcome deterministically based on the inputs $\mu$ and $\nu$ using a guessing function 
\begin{equation}
    f \mleft( \mu, \, \nu \mright) \in \{0, \, 1, \, \hdots , \, d\}\,. 
\end{equation}
The outcome $f \mleft( \mu, \, \nu \mright) =0$ is inconclusive, as before, so Eve's guessing probability is given by 
\begin{equation}
    \pj \mleft( \rho, \, \m | \mathcal{D}, \, \, f \mright) =  \sum_{\mu, \, \nu =0}^{d} p_{\mu, \, \nu} \, p \mleft( f \mleft( \mu, \, \nu \mright) | \mu, \, \nu \mright)\,,
\end{equation}
where $p \mleft( y \, | \mu, \, \nu \mright)$ is the conditional probability of achieving outcome $y$ given inputs $\mu$ and $\nu$. When $y=0$, the conditional probability is trivially zero for any $\mu$ or $\nu$, since there is no outcome 0, while for all $x \in \{1, \, \hdots, \, d\}$, we have 
\begin{equation}
    p \left( x | \mu, \, \nu\right) = \langle \psi_\mu | N_{x, \, \nu} | \psi_{\mu} \rangle\,.
\end{equation}
Eve's guessing probability probability when she makes a `classical decomposition', as in this framework, is never greater than if she had joint quantum correlations with both the state and the measurement \cite{Senno_2023}, according to a model introduced in \cite{frauchiger2013} and developed in \cite{Senno_2023,Dai_2023,curran2025}. We will not attempt to find Eve's optimal (quantum or classical) guessing probability in the joint-noise case; instead, we will choose a decomposition $\mathcal{D}$ and a guessing function $f$ to \emph{lower bound} Eve's joint guessing probability, given any upper bound $T$ on her error probability.

We begin by introducing a critical value for the noise $\ve$,
\begin{equation}
    \ecrit = \frac{d}{2 \left( d+\sqrt{d}\right)}\,.
\end{equation}
Generalizing a joint decomposition from \cite{curran2025}, we will use the following states $\ket{\psi_\mu}$ and measurements $\mathcal{N}_\nu=\{N_{x, \, \nu}\}_{x}$ with different probability distributions $\{p_{\mu, \, \nu}\}$, 
\begin{equation}\label{eqn: decomp}
    \begin{aligned}
     \ket{\psi_\mu} &= \begin{cases}
         \ket{\phi} \,, \qquad \hspace{7.35 cm} & \mu =0\,,
         \vspace{0.1 cm}
         \\
         \sqrt{d \rho_\gamma} \ket{m_\mu}  \,, \qquad & \mu \in \{1, \, \hdots, \, d\}\,,
     \end{cases}  
     \\
     \\
     N_{x, \, \nu} &= \begin{cases}
         \ketbra{m_x}{m_x}\,, \quad & \nu =0\,,
         \vspace{0.1 cm}
         \\
         d \sqrt{M_{x, \, \gamma}} \ketbra{\phi}{\phi} \sqrt{M_{x, \, \gamma}} + \frac{d-1}{d} \gamma \left( \id_{\neq x} - \ketbra{\psi_{\neq x}}{\psi_{\neq x}} \right)\,, \quad & \nu = x\,,
         \vspace{0.1 cm}
         \\
         \sqrt{M_{x, \,\gamma}} \left( \ket{m_x} - \ket{m_\nu} \right) \left( \bra{m_x} - \bra{m_\nu} \right) \sqrt{M_{x, \,\gamma}}\,, \quad & \nu \neq x\,, \;\;\; \nu \in \{1, \, \hdots, \, d\}\,,
     \end{cases}
    \end{aligned}
\end{equation}
where 
\begin{equation}
    \gamma = \frac{\ve}{1-Q}\,, \qquad \id_{\neq x} = \id - \ketbra{x}{x}\,, \qquad \ket{\psi_{\neq x}} = \frac{1}{\sqrt{d-1}}  \underset{y \neq x}{\sum_{y=1}^{d}} \ket{y}\,.
\end{equation}
Here, $Q$ is just a parameter in the range $0 \leq Q \leq 1-\ve$ that's chosen by Eve. Notice that, when $\nu=0$, we have
\begin{equation}
   \sum_{x=1}^{d} N_{x, \, 0} = \id\,,
\end{equation}
while for all $\nu \in \{1, \, \hdots, \, d\}$, we have
\begin{equation}
\underset{x \neq \nu}{\sum_{x=1}^{d}} N_{x, \, \nu} =   \frac{1}{d} \left( A_\gamma \id_{\neq \nu} - \sqrt{A_\gamma \gamma \left( d-1\right)} \left( \ketbra{\psi_{\neq \nu}}{m_\nu} + \ketbra{m_\nu}{\psi_{\neq \nu}}\right) + \gamma \left( d-1 \right) \ketbra{m_\nu}{m_\nu} \right)\,,
\end{equation}
and for all $x \in \{1, \, \hdots, \, d\}$, we have
\begin{equation}
  \sqrt{M_{x, \, \gamma}} \ket{\phi} = \frac{1}{d} \left( \sqrt{A_\gamma} \ket{m_x} + \sqrt{\gamma \left( d-1\right)} \ket{\psi_{\neq x}} \right)\,.
\end{equation}
The POVMs $\mathcal{N}_\nu= \{N_{x, \, \nu}\}_x$ in \eqref{eqn: decomp} are valid, then, because all of the elements $N_{x, \, \nu}$ are positive semidefinite and
\begin{equation}
    \sum_{x=1}^{d} N_{x, \, \nu} = \left( 1 - \delta_{\nu, \, 0}\right) N_{\nu, \, \nu} + \underset{x \neq \nu}{\sum_{x=1}^{d}} N_{x, \, \nu} = \id \qquad \textnormal{for all } \;\; \nu \in \{0, \, 1, \, \hdots, \, d\}\,.
\end{equation}
For future convenience, note too that
\begin{equation}
   \underset{\nu \neq x}{\sum_{\nu=1}^{d}} N_{x, \, \nu} = \frac{1}{d} \left( \left( d-1 \right) A_\gamma \ketbra{m_x}{m_x} - \sqrt{A_\gamma \, \gamma \left( d-1\right)} \left( \ketbra{m_x}{\psi_{\neq x}} + \ketbra{\psi_{\neq x}}{m_x} \right) + \gamma \id_{\neq x}\right)\,,  
\end{equation}
such that
\begin{equation}
  \sum_{\nu=1}^{d} N_{x, \, \nu} = N_{x, \, x} +  \underset{\nu \neq x}{\sum_{\nu=1}^{d}} N_{x, \, \nu} = d M_{x, \, \gamma} \qquad \textnormal{for all} \;\; x \in \{1, \, \hdots, \, d\}\,. 
\end{equation}
We also have the useful properties
\begin{equation}
    \bra{\phi} \left( \id_{\neq x} - \ketbra{\psi_{\neq x}}{\psi_{\neq x}}\right) = \langle m_x | \sqrt{\rho_{\gamma}} \left( \id_{\neq x} - \ketbra{\psi_{\neq x}}{\psi_{\neq x}}\right) =0 \qquad \textnormal{for all} \;\; x \in \{1, \, \hdots, \, d\}\,,
\end{equation}
so we find that the projector $\id - \ketbra{\psi_{\neq x}}{\psi_{\neq x}}$ does not contribute to our relevant conditional probabilities. For any $x \in \{1, \, \hdots, \,d\}$, we have
\begin{equation}
\begin{aligned}
    \langle  \psi_x | N_{x, \, x} | \psi_x \rangle &= d^2 \abs{\langle m_x | \sqrt{\rho_{\gamma}} \sqrt{M_{x, \, \gamma}} | \phi  \rangle}^2 = \frac{1}{d^3} \left(   A_\gamma + 2 \left( d-1\right) \sqrt{A_\gamma \, \gamma} - \gamma \left( d-1\right)    \right)^2
    \\[2pt]
    &=  \frac{1}{d}\left( \left(\tr \sqrt{\rho_\gamma} \right)^2 - \gamma \left( d-1\right) \right)^2   \,
    \end{aligned}
\end{equation}
and, when $x \neq \mu$ and $\mu \in \{1, \, \hdots, \, d\}$, 
\begin{equation}
 \langle  \psi_\mu | N_{x, \, \mu} | \psi_\mu \rangle =  d \abs{ \langle m_\mu | \sqrt{\rho_{\gamma}} \sqrt{M_{x, \, \gamma}} \left( \ket{m_x} - \ket{m_\mu} \right)  \rangle   }^2 = \frac{1}{d^3} \left(A_\gamma - 2 \sqrt{A_\gamma \, \gamma} - \gamma \left( d-1\right)  \right)^2\,,
\end{equation}
such that
\begin{equation}\label{eqn: sum cond probs}
\sum_{\mu=1}^{d} \langle  \psi_\mu | N_{x, \, \mu} | \psi_\mu \rangle = \langle  \psi_x | N_{x, \, x} | \psi_x \rangle + \underset{\mu \neq x}{\sum_{\mu=1}^{d}} \langle  \psi_\mu | N_{x, \, \mu} | \psi_\mu \rangle =  \left( \tr \rho_\gamma \right)^2 =1\,.
\end{equation}
The relevant conditional probabilities are
\begin{equation}\label{eqn: state and meas}
    \begin{aligned}
        p \mleft( \mu  | \mu, \, 0\mright) &= \frac{1}{d} \left(  \tr \sqrt{\rho_{\gamma}}\right)^2 \,, \quad &&\textnormal{for all} \;\;  \mu \in \{1, \, \hdots, \, d\}\,,
        \\
        p \mleft( \nu  | 0, \, \nu \mright) &= \frac{1}{d} \left(  \tr \sqrt{M_{\nu, \, \gamma}}\right)^2 \,, \quad && \textnormal{for all} \;\;  \nu \in \{1, \, \hdots, \, d\}\,,
        \\
        p \mleft( \mu  | \mu, \, \mu \mright) &= \frac{1}{d}\left( \left(\tr \sqrt{\rho_\gamma} \right)^2 - \gamma \left( d-1\right) \right)^2 \quad && \textnormal{for all} \;\;  \mu \in \{1, \, \hdots, \, d\}\,.
    \end{aligned}
\end{equation}
We now split Eve's strategy into several cases, based on the error $\ve$ and the probability of error $T$, for which we also define a special value
\begin{equation}
    T_1 = 1- \frac{1}{d}\left( \left(\tr \sqrt{\rho_\ve} \right)^2 - \ve \left( d-1\right) \right)^2\,.
\end{equation}

\subsection{When $\ve \geq \ecrit$}
When $\ve=  \ecrit$, Eve chooses the states and measurements \eqref{eqn: decomp}, with parameter $Q=0$ (such that $\gamma=\ve$), along with
\begin{equation}\label{eqn: prob and f}
p_{\mu, \, \nu} = 
     \frac{1}{d} \, \delta_{\mu,\, \nu} \,, \qquad f \mleft( \mu, \, \mu \mright) = \mu\,, \qquad \textnormal{for all} \;\; \mu, \, \nu \in \{1, \, \hdots d\}\,. 
\end{equation}
This decomposition satisfies the constraints in \eqref{eqn: joint constraints}, as, for all $x \in \{1, \, \hdots\ d\}$,
\begin{equation}
    \begin{aligned}
     &\sum_{\mu, \, \nu=1}^{d} p_{\mu, \, \nu} \ketbra{\psi_\mu}{\psi_\mu} = \frac{1}{d} \sum_{\mu =1}^d \sqrt{d \rho_\ve} \ketbra{m_\mu}{m_\mu} \sqrt{d \rho_\ve} = \rho_{\ve}\,,
    \\
    &\sum_{\mu, \, \nu =1}^{d} p_{\mu, \, \nu} N_{x, \, \nu} = \frac{1}{d}  \sum_{\nu=1}^{d} N_{x, \, \nu} = M_{x, \, \ve}\,,
    \\
    & \sum_{\mu, \, \nu =1}^{d} p_{\mu, \, \nu} \langle \psi_\mu | N_{x, \, \nu} | \psi_\mu \rangle = \frac{1}{d}\sum_{\mu=1}^{d} \langle \psi_\mu | N_{x, \, \mu } | \psi_{\mu} \rangle    = \frac{1}{d} = \tr \mleft( \rho_\ve \, M_{x, \, \ve}\mright)\,, 
    \end{aligned}
\end{equation}
where in the last line we use \eqref{eqn: sum cond probs}. When $\ve= \ecrit$, Eve's guessing probability is perfect because
\begin{equation}
    \sqrt{A_{\ecrit}} = \left( \sqrt{d} +1\right) \sqrt{\ecrit}\,,
\end{equation}
such that her state and measurement vectors align,
\begin{equation}
    \sqrt{d \rho_{\ecrit}} \ket{m_x} =  \sqrt{d M_{x, \, \ecrit}} \ket{\phi}\,,
\end{equation}
so 
\begin{equation}
    \langle \psi_\mu | N_{\mu, \, \mu} | \psi_{\mu} \rangle = d^2 \langle m_\mu | \rho_{\ecrit} | m_\mu \rangle^2 = 1 \qquad \textnormal{for all} \;\; \mu \in \{1, \, \hdots, \, d\}\,.
\end{equation}
Note that no inconclusive outcome is necessary here. When $\ve > \ecrit$, Eve can maintain her perfect guessing probability by mixing this strategy with another one, according to a classical variable $\lambda \in \{0, \, 1, \, \hdots, \, d\}$. Her decomposition is then conditioned on $\lambda$, $\mathcal{D}_\lambda = \{\{p_{\mu, \, \nu | \lambda}\}, \, \{\ket{\psi_{\mu | \lambda }} \}, \, \{ \mathcal{N_{\nu | \lambda} } \}   \}$ and her mixing probability distribution $\{p_{\mu, \, \nu, \, \lambda}\}$ is given by
\begin{equation}
    p_{\mu, \, \nu, \, \lambda} = p_{\lambda} \, p_{\mu, \, \nu | \lambda}\,.
\end{equation}
When $\lambda=0$, Eve's will use the decomposition in \eqref{eqn: decomp}, while when $\lambda=1$, Eve uses the decomposition
\begin{equation}
   \begin{aligned}
       \ket{\psi_{\mu | \lambda=1}} &= \ket{m_\mu}\,, \qquad && \mu \in \{1, \, \hdots, \, d\}\,,
       \\[2pt]
       N_{x, \, \nu | \lambda=1} &= \ketbra{m_{x+\nu-1}}{m_{x+\nu-1}}\,, \qquad && x, \, \nu \in \{1, \, \hdots, \, d\}\,,
   \end{aligned} 
\end{equation}
with the probability distribution 
\begin{equation}
    p_{\mu, \, \nu | \lambda=1} = \frac{1}{d^2}\,.
\end{equation}
We find 
\begin{equation}
\begin{aligned}
    &\sum_{\mu, \, \nu=0}^{d} p_{\mu, \, \nu | \lambda=1} \, \ketbra{\psi_{\mu | \lambda=1}}{\psi_{\mu | \lambda=1}} = \frac{\id}{d}\,,
    \\
    &\sum_{\mu, \, \nu =0}^{d} p_{\mu, \, \nu | \lambda=1} \, N_{x, \, \nu | \lambda=1} = \frac{\id}{d} \,,
    \\
    & \sum_{\mu, \, \nu =0}^{d} \, p_{\mu, \, \nu | \lambda=1} \, \langle \psi_{\mu | \lambda=1} \, | \, N_{x, \, \nu | \lambda=1} \, | \, \psi_{\mu | \lambda=1} \rangle = \frac{1}{d^2} \sum_{\mu, \, \nu=1}^{d} \delta_{\mu, \, x + \nu -1}  = \frac{1}{d}\,.
\end{aligned}    
\end{equation}
Since we have proven before that 
\begin{equation}
\begin{aligned}
    &\sum_{\mu, \, \nu=0}^{d} p_{\mu, \, \nu | \lambda=0} \, \ketbra{\psi_{\mu | \lambda=0}}{\psi_{\mu | \lambda=0}} = \rho_{\ecrit}\,,
    \\
    &\sum_{\mu, \, \nu =0}^{d} p_{\mu, \, \nu | \lambda=0} \, N_{x, \, \nu | \lambda=0} = M_{x, \, \ecrit} \,,
    \\
    & \sum_{\mu, \, \nu =0}^{d} \, p_{\mu, \, \nu | \lambda=0} \, \langle \psi_{\mu | \lambda=0} \, | \, N_{x, \, \nu | \lambda=0} \, | \, \psi_{\mu | \lambda=1} \rangle = \frac{1}{d}\,,
\end{aligned}    
\end{equation}
the conditions \eqref{eqn: joint constraints} hold if we choose 
\begin{equation}
    p_\lambda = \begin{cases}
        \beta\,, \qquad & \lambda=0\,,
        \\
        1-\beta\,, \qquad & \lambda =1\,,
    \end{cases} \qquad  \qquad \beta= \frac{1- \ve}{1- \ecrit}\,.
\end{equation}
Eve maintains perfect guessing probability if she uses the guessing function
\begin{equation}
    f \mleft( \mu, \, \nu | \lambda \mright) = \begin{cases}
\mu\,, \qquad   & \lambda=0\,, \;\; \mu \in \{1, \, \hdots, \, d\}\,,  
\\
\mu-\nu+1\,, \qquad &    \lambda=1\,, \;\; \mu, \, \nu \in \{1, \, \hdots, \, d\} \,. 
    \end{cases}
\end{equation}

\subsection{When $\ve < \ecrit$}

\paragraph{When $T \geq T_1\,.$}
Eve will again use the strategy described in \eqref{eqn: decomp} and \eqref{eqn: prob and f}, with $Q=0$ such that $\gamma=\ve$, and her guessing probability will be 
\begin{equation}\label{eqn: pjoin bigT}
    \pj \mleft( \rho_\ve, \, \m_\ve | \mathcal{D}, \, f\mright) = \frac{1}{d}\left( \left(\tr \sqrt{\rho_\ve} \right)^2 - \ve \left( d-1\right) \right)^2\,.
\end{equation}
As an intermediate step to proving that this value increases continuously with $\ve$, note that
\begin{equation}
    \frac{\partial}{\partial \ve}  \left( \sqrt{\aep} + \left( d-1\right) \sqrt{\ve} \right)  = \frac{d-1}{2 \sqrt{A_\ve \,\ve}} \left( \sqrt{A_\ve} - \sqrt{\ve} \right) >0\,,
\end{equation}
so 
\begin{equation}
\sqrt{\ve} \left( \sqrt{\aep} + \left( d-1\right) \sqrt{\ve} \right) < \sqrt{\ecrit} \left( \sqrt{A_{\ecrit}} + \left( d-1\right) \sqrt{\ecrit} \right) = \frac{d}{2}   
\end{equation}
in the range $0 < \ve < \ecrit $.
Then, in this same range, we have 
\begin{equation}
    \frac{\partial }{\partial \ve} \left( \left(\tr \sqrt{\rho_\ve} \right)^2 - \ve \left( d-1\right) \right) = \frac{d-1}{d \sqrt{\aep \, \ve}} \left( d - 2 \sqrt{\ve} \left( \sqrt{\aep} + \left( d-1\right) \sqrt{\ve} \right) \right) >0\,.
\end{equation}
This guessing probability then increases continuously from $\frac{1}{d}$ (random guessing) to 1 as $\ve$ increases in our range. Since there is no inconclusive outcome, the probability of error is 
\begin{equation}
    \perr = 1 - \pj \mleft( \rho_\ve, \, \m_\ve \mright) = T_1\,,
\end{equation}
which is why this strategy is only suitable when $T \geq T_1$.

As a quick detour, remember that Eve's guessing probability in the single-noise case, with an amount $\kappa$ of noise and no inconclusive outcome, is 
\begin{equation}
    \psin^{*} \mleft( \rho_\kappa \mright) = \frac{1}{d} \left( \tr \sqrt{\rho_\kappa} \right)^2\,.
\end{equation}
This guessing probability is strictly increasing in $\kappa$, and it reduces to random guessing when $\kappa=0$. Fixing $0 < \ve < \ecrit$, there is at most one value of $\kappa$, $\kappa_{\ve}$, for which $\psin^{*} \mleft( \rho_\kappa \mright)$ is equal to the joint guessing probability in \eqref{eqn: pjoin bigT}. It is
\begin{equation}
\kappa_{\ve} = \frac{4 \ve \left( d - \left( d-1\right) \ve \right)}{d}\,,
\end{equation}
which satisfies 
\begin{equation}
    \ve = \frac{ d -\sqrt{d A_{\kappa_{\ve}}}  }{2 \left(d-1 \right)}\,.
\end{equation}
Note that $\kappa_\ve=0$ when $\ve=0$ and $\kappa_\ve=1$ when $\ve= \ecrit$.
We have 
\begin{equation}
    \kappa_{\ve} - 2 \ve = \frac{2 \ve}{d} \left( d - 2 \left( d-1\right) \ve \right) = \frac{2 \ve \sqrt{ A_{\kappa_\ve}}}{\sqrt{d}} > 0
\end{equation}
for all $0 < \ve \leq \ecrit$. Subbing in $\kappa=2\ve$, we conclude, finally, that in this range,
\begin{equation}\label{eqn: ineq}
    \frac{1}{d}\left( \left(\tr \sqrt{\rho_\ve} \right)^2 - \ve \left( d-1\right) \right)^2 - \frac{1}{d} \left( \tr \sqrt{\rho_{2 \ve} } \right)^2 > 0\,.
\end{equation}
We will use this fact in the following section.

\paragraph{When $T < T_1\,$.} Eve will now use a different strategy. She still uses the states and measurements in \eqref{eqn: decomp}, but her probability distribution and guessing function are now given by 
\begin{equation}\label{eqn: probs 2}
    p_{\mu, \, \nu} = \begin{cases}
        2 Q -1\,, \quad  & \mu=\nu=0\,,
        \vspace{0.1 cm}
        \\
        \frac{1-Q}{d}\,, \quad & \mu=0\,, \; \nu \in \{1, \, \hdots d\}\,,
        \vspace{0.1 cm}
        \\
        \frac{1-Q}{d}\,, \quad & \mu \in \{1, \, \hdots d\}\,, \; \nu =0\, 
        \vspace{0.1 cm}
        \\
        0 \,, \quad & \mu\,, \nu \in \{1, \, \hdots d\}\,,
    \end{cases}
\end{equation}
which is valid when she chooses $Q \geq \frac{1}{2}$, and 
\begin{equation}
    f \mleft( \mu, \, \nu \mright) = \begin{cases}
        0 \,, \quad & \mu = \nu =0\,,
        \\
        \mu\,, \quad & \mu \in \{1, \, \hdots, \, d\}\,, \;\; \nu =0\,,
         \\
         \nu\,, \quad & \mu = 0\,, \;\; \nu \in \{1, \, \hdots, \, d\}\,.
    \end{cases}
\end{equation}
This decomposition satisfies the constraints of \eqref{eqn: joint constraints}, because
\begin{equation}
    \begin{aligned}
    &\sum_{\mu, \, \nu=0}^{d} p_{\mu, \, \nu} \ketbra{\psi_\mu}{\psi_\mu} = Q \ketbra{\psi_0}{\psi_0} + \frac{1-Q}{d} \sum_{\mu=1}^{d} \ketbra{\psi_\mu}{\psi_\mu} = Q \ketbra{\phi}{\phi} + \left( 1- Q\right) \rho_{\gamma}  = \rho_{\ve}\,,
    \\
    &\sum_{\mu, \, \nu =0}^{d} p_{\mu, \, \nu} \, N_{x, \, \nu} = Q N_{x, \, 0} + \frac{1-Q}{d} \sum_{\nu=1}^{d} N_{x, \, \nu }  = Q \ketbra{m_x}{m_x} + \left( 1-Q\right) M_{x, \, \gamma}    =M_{x, \, \ve}\,,
    \\
    & \sum_{\mu, \, \nu =0}^{d} p_{\mu, \, \nu} \langle \psi_\mu | N_{x, \, \nu} | \psi_\mu \rangle = \left( 2Q-1\right) \abs{\langle \phi | m_x \rangle }^2 + \left( 1-Q\right) \left( \langle \phi | M_{x, \, \gamma} | \phi \rangle +  \langle m_x | \rho_\gamma | m_x \rangle \right)= \frac{1}{d}\,.
\end{aligned}    
\end{equation}
Remember that here, $Q$ is a parameter chosen by Eve, which we have not yet defined, and it is not in general equal to the probability of returning an inconclusive outcome. Using \eqref{eqn: state and meas}, her guessing probability is then
\begin{equation}\label{eqn: pguess small T}
    \pj \mleft( \rho_\ve, \, \m_\ve | \mathcal{D}, \, f\mright) = \frac{1-Q}{d} \left( \, \sum_{\mu=1}^{d} p \mleft( \mu | \mu, \, 0\mright)  + \sum_{\nu=1}^{d} p \mleft( \nu | 0 , \, \nu \mright) \right)= \frac{2  \left( 1-Q\right)}{d} \left( \tr \sqrt{\rho_\gamma} \right)^2\,. 
\end{equation}
The probability of giving an inconclusive outcome is
\begin{equation}
  P_{\textnormal{inc}} =   \sum_{\mu, \, \nu=0}^{d} p_{\mu, \, \nu} \,\delta_{f \left(\mu, \, \nu \right), \, 0} = 2Q-1\,,
\end{equation}
so the probability of error is 
\begin{equation}
  \perr =   1- P_{\textnormal{inc}} - \pj \mleft( \rho_\ve, \, \m_\ve | \mathcal{D}, \, f\mright) = 2 \left( 1- Q \right) \left( 1 - \frac{1}{d} \left( \tr \sqrt{\rho_\gamma}\right)^2 \right)\,.
\end{equation}
We should then choose the parameter $Q$ such that this error probability matches the chosen parameter $T$. The correct value for $Q$ is
\begin{equation}\label{eqn: Q T}
    Q \mleft( T\mright) = \frac{2 \left( d-1\right) \left( 1 - \ve\right) -dT -2 \sqrt{2 T \ve \left( d-1\right)}}{2 \left( d-1\right)}\,,
\end{equation}
which decreases strictly with $T$ (note that $Q \mleft( 0 \mright)= 1- \ve$).
This leaves us with the guessing probability 
\begin{equation}\label{eqn: pguess smaller T}
\begin{aligned}
    \pj \mleft( \rho_\ve, \, \m_\ve | \mathcal{D}, \, f \mright) &= 1 - P_{\textnormal{inc}} - P_{\textnormal{err}} = 2 - 2Q \mleft( T \mright) - T
    \\
    &= \frac{2 \ve \left( d-1\right)+ T + 2 \sqrt{2 T \ve \left( d-1\right)} }{d-1} = \frac{\left( \sqrt{T} + \sqrt{2 \ve \left( d-1\right)} \right)^2}{d-1}\,.
    \end{aligned}
\end{equation}
As a final step to proving that this decomposition is valid, we still have to show that $Q \mleft( T\mright) \geq \frac{1}{2}$ when $\ve < \ecrit$ and $T \geq T_1$, so that the mixing probability distribution $\{p_{\mu, \, \nu}\}$ is well-defined. Let's define another parameter,
\begin{equation}
  T_2 = \frac{d-1}{d^2} \left( \sqrt{A_{2 \ve}} - \sqrt{2 \ve}\right)^2 = 1- \frac{1}{d} \left( \tr \sqrt{\rho_{2 \ve}}\right)^2\,,  \end{equation}
which satisfies
\begin{equation}
    Q \mleft( T_2 \mright) = \frac{1}{2}\,.
\end{equation}
Since the function $Q \mleft(T \mright)$ in \eqref{eqn: Q T} decreases strictly with $T$, we have
\begin{equation}
    Q \mleft( T \mright) \geq \frac{1}{2} \qquad \textnormal{for all} \;\; 0 \leq  T \leq T_2\,,
\end{equation}
so this decomposition is valid for all $T \leq T_2$. If we can show that $T_1 \leq T_2$ then we can show that the decomposition is valid for all $T < T_1$, as desired. 
We have 
\begin{equation}
    T_2 - T_1 = \frac{1}{d}\left( \left(\tr \sqrt{\rho_\ve} \right)^2 - \ve \left( d-1\right) \right)^2 - \frac{1}{d} \left( \tr \sqrt{\rho_{2 \ve}}\right)^2 >0\,,
\end{equation}
where we use the inequality \eqref{eqn: ineq}. The guessing probability \eqref{eqn: pguess small T} is then achievable in this region. 

\subsection{Single versus joint noise}
\begin{figure}[h]
    \centering
\begin{subfigure}{0.45\textwidth}
    \begin{tikzpicture}
\begin{scope}[scale=0.75]
    \begin{axis}[
        axis lines=middle,
        xlabel={$\delta$},
        ylabel={$P$},
        xlabel style={at={(ticklabel* cs:1.02)}, anchor=west},
        ylabel style={at={(ticklabel* cs:1.02)}, anchor=south},
        ymin=0, ymax=1,
        xmin=0, xmax=1,
        xtick={0,0.5,1},
        ytick={0,0.5,1},
        domain=0:1,
        samples=200,
        legend style={
    at={(0.95,0.05)},
    anchor=south east,
    font=\small},
        legend style={font=\small}]
        \addplot[
            thick,
            black
        ] {x < 0.5 ? 0.5*( 1+ 2*sqrt(x*(1-x))  )  : 1};
        \addlegendentry{shared noise};
        \addplot[
            dashed,
            black
        ] { pow( sqrt(0.5*(1 - sqrt(x*(2 - x)))) + sqrt(x) , 2 ) };
        \addlegendentry{single noise};
    \end{axis}
\end{scope}
\end{tikzpicture}
\vspace{0.5 cm}
    \caption{$T= \tmax\mleft( \delta \mright)$.}
    \label{fig: tmax}
\end{subfigure}
\hfill
\begin{subfigure}{0.45\textwidth}
    \begin{tikzpicture}
\begin{scope}[scale=0.75]
    \begin{axis}[
        axis lines=middle,
        xlabel={$\delta$},
        ylabel={$P$},
        xlabel style={at={(ticklabel* cs:1.02)}, anchor=west},
        ylabel style={at={(ticklabel* cs:1.02)}, anchor=south},
        ymin=0, ymax=1,
        xmin=0, xmax=1,
        xtick={0,0.5,1},
        ytick={0,0.5,1},
        domain=0:1,
        samples=200,
        legend style={
    at={(0.95,0.05)},
    anchor=south east,
    font=\small},
        legend style={font=\small}]
        \addplot[black, thick, domain=0:0.113] {pow( sqrt(0.33*(1 - sqrt(x*(2 - x)))) + sqrt(2*(1 - sqrt(1-x)) ) , 2 )};
        \addlegendentry{shared noise};
        \addplot[
            dashed,
            black
        ] {pow( sqrt(0.33*(1 - sqrt(x*(2 - x)))) + sqrt(x) , 2 )};
        \addlegendentry{single noise};
        \addplot[thick, black, domain=0.12:1] {x < 0.5 ? 0.5*( 1+ 2*sqrt(x*(1-x))  )  : 1};
    \addplot[only marks, mark=o, mark size=2pt, black] coordinates {(0.12, 0.59)};
    \addplot[only marks, mark=*, mark size=2pt, black] coordinates {(0.12,0.82)};
    \end{axis}
\end{scope}
\end{tikzpicture}
\vspace{0.5 cm}
    \caption{$T= \frac{2}{3} \, \tmax \mleft( \delta \mright)$.}
    \label{fig: 2/3 tmax}
\end{subfigure}
\vspace{0.5 cm}
\begin{subfigure}{0.45\textwidth}
    \begin{tikzpicture}
\begin{scope}[scale=0.75]
    \begin{axis}[
        axis lines=middle,
        xlabel={$\delta$},
        ylabel={$P$},
        xlabel style={at={(ticklabel* cs:1.02)}, anchor=west},
        ylabel style={at={(ticklabel* cs:1.02)}, anchor=south},
        ymin=0, ymax=1,
        xmin=0, xmax=1,
        xtick={0,0.5,1},
        ytick={0,0.5,1},
        domain=0:1,
        samples=200,
        legend style={
    at={(0.95,0.05)},
    anchor=south east,
    font=\small},
        legend style={font=\small}]
        \addplot[black, thick, domain=0:0.272] {pow( sqrt(0.166*(1 - sqrt(x*(2 - x)))) + sqrt(2*(1 - sqrt(1-x)) ) , 2 )};
        \addlegendentry{shared noise};
        \addplot[
            dashed,
            black
        ] {pow( sqrt(0.166*(1 - sqrt(x*(2 - x)))) + sqrt(x) , 2 )};
        \addlegendentry{single noise};
        \addplot[thick, black, domain=0.28:1] {x < 0.5 ? 0.5*( 1+ 2*sqrt(x*(1-x))  )  : 1};
    \addplot[only marks, mark=o, mark size=2pt, black] coordinates {(0.28, 0.6)};
    \addplot[only marks, mark=*, mark size=2pt, black] coordinates {(0.28,0.95)};
    \end{axis}
\end{scope}
\end{tikzpicture}
\vspace{0.5 cm}
    \caption{$T= \frac{1}{3} \, \tmax \mleft( \delta \mright)$.}
    \label{fig: 1/3 tmax}
\end{subfigure}
\hfill
\begin{subfigure}{0.45\textwidth}
    \begin{tikzpicture}
\begin{scope}[scale=0.75]
    \begin{axis}[
        axis lines=middle,
        xlabel={$\delta$},
        ylabel={$P$},
        xlabel style={at={(ticklabel* cs:1.02)}, anchor=west},
        ylabel style={at={(ticklabel* cs:1.02)}, anchor=south},
        ymin=0, ymax=1,
        xmin=0, xmax=1,
        xtick={0,0.5,1},
        ytick={0,0.5,1},
        domain=0:1,
        samples=200,
        legend style={
    at={(0.95,0.05)},
    anchor=south east,
    font=\small},
        legend style={font=\small}]
        \addplot[black, thick, domain=0:0.491] { 2*(1 - sqrt(1-x))  };
        \addlegendentry{shared noise};
        \addplot[
            dashed,
            black
        ] {x};
        \addlegendentry{single noise};
        \addplot[thick, black, domain=0.5:1] {1};
    \addplot[only marks, mark=o, mark size=2pt, black] coordinates {(0.5, 0.58)};
    \addplot[only marks, mark=*, mark size=2pt, black] coordinates {(0.5,1)};
    \end{axis}
\end{scope}
\end{tikzpicture}
\vspace{0.5 cm}
    \caption{$T= 0$.}
    \label{fig: T=0}
\end{subfigure}
    \caption{\emph{Shared versus single noise}. Eve's optimal guessing probability $P$ for the single-noise case (dashed) and a lower bound on her guessing probability in the shared-noise case (undashed), both as a function of the total noise $\delta$ and given a maximum probability $T$ of error.}
    \label{fig: joint noise app}
\end{figure}
We have found a lower bound on Eve's joint guessing probability for all $\ve$ and $T$, which we summarize as
\begin{equation}
    \pj \mleft( \rho_\ve, \, \m_\ve\mright) \geq 
    \begin{cases}
    1\,, \quad & \ve \geq \ecrit\,, \vspace{0.2 cm}
    \\
    \frac{1}{d}\left( \left(\tr \sqrt{\rho_\ve} \right)^2 - \ve \left( d-1\right) \right)^2\,, \quad & \ve < \ecrit\,, \;\; T \geq T_1\,, \vspace{0.2 cm}
    \\
    \frac{\left( \sqrt{T} + \sqrt{2 \ve \left( d-1\right)} \right)^2}{d-1}\,, \quad & \ve < \ecrit\,, \;\; T < T_1\,.
    \end{cases}
\end{equation}
We will now compare this performance to that in the single-noise case, where only the state is noisy. In order to compare these two cases, we will assume throughout that a total amount of noise $\delta$ is shared between the devices. In the single-noise case, this is all contained in the state $\rho_\delta$, which is measured in $\m= \{M_x\}_x$, $M_x = \ketbra{m_x}{m_x}$,  while in the joint-noise case, there is an amount $\ve$ of noise in each device, where 
\begin{equation}
    \ve = 1 - \sqrt{1 - \delta}\,, \qquad \delta = \ve \left( 2 - \ve \right)\,.
\end{equation}
The value $\ve$ is chosen such that 
\begin{equation}
  \tr \mleft(   \Delta_\delta \mleft( \ketbra{\phi}{\phi} \mright) \, M_x \mright) = \tr \mleft( \Delta_\ve \mleft( \ketbra{\phi}{\phi} \mright)\, \Delta_{\ve} \mleft( M_x \mright) \mright)  \qquad \textnormal{for all} \;\; x
\end{equation}
for any state $\ket{\phi}$ and any measurement $\m=\{M_x\}_x$ for which $\tr M_x=1$ for all $x$. Remember that from \eqref{eqn: noisy Q T}, for a fixed bound $T$ on the error probability, the optimal guessing probability in the single-noise case is
\begin{equation}
    \psin \mleft( \rho_\delta, \, \m \mright) = \frac{1}{d-1} \left( \sqrt{T} + \sqrt{\delta \left(d-1 \right)} \right)^2\,, \qquad 0 \leq T < \tmax \mleft( \delta \mright)\,,
\end{equation}
where 
\begin{equation}
    \tmax \mleft( \delta \mright) = 1 -\frac{1}{d} \left( \tr \sqrt{\rho_\delta}\right)^2\,
\end{equation}
(note that we now include the extreme points $T=0$ and $T=\tmax \mleft( \delta \mright)$).
This guessing probability is only perfect when $\delta=1$, so when $\ve \geq \ecrit$, the joint guessing probability is clearly higher. 
Let's define the difference
\begin{equation}
    \left( *\right) = \pj \mleft( \rho_\ve, \, \m_\ve | \mathcal{D}, \, f\mright) - \psin \mleft( \rho_\delta, \, \m \mright) \,.
\end{equation}
When $\ve < \ecrit$ and when $ T_1 \leq T \leq \tmax \mleft( \delta \mright)$, using \eqref{eqn: pjoin bigT}, we have 
\begin{equation}
\begin{aligned}
    \left( *\right)  &= \frac{1}{d}\left( \left(\tr \sqrt{\rho_\ve} \right)^2 - \ve \left( d-1\right) \right)^2 - \frac{1}{d-1} \left( \sqrt{T} + \sqrt{\delta \left(d-1 \right)} \right)^2
    \\
    &\geq \frac{1}{d}\left( \left(\tr \sqrt{\rho_\ve} \right)^2 - \ve \left( d-1\right) \right)^2 - \frac{1}{d} \left( \tr \sqrt{\rho_{\delta}}\right)^2 > 0\,,
    \end{aligned}
\end{equation}
where in the final inequality we use $T < \tmax$, the inequality \eqref{eqn: ineq} and
\begin{equation}
2 \ve - \delta = 2 \ve - \ve \left( 2 - \ve\right) = \ve^2 >0\,.
\end{equation}
When $0 \leq T < T_1$, using \eqref{eqn: pguess smaller T}, we have
\begin{equation}
\begin{aligned}
    \left( * \right) &= \frac{\left( \sqrt{T} + \sqrt{2 \ve \left( d-1\right)} \right)^2}{d-1} -  \frac{\left( \sqrt{T} + \sqrt{\delta \left(d-1 \right)} \right)^2}{d-1}     
\\
&= \frac{1}{\sqrt{2 \ve} + \sqrt{\delta}} \left( 2 \sqrt{\frac{T}{d-1} } + \sqrt{2 \ve} + \sqrt{\delta} \right) \left( 2 \ve - \delta\right) >0\,.
\end{aligned}
\end{equation}
We have shown, then, that in the case of an unbiased state and measurement pair, for any bound $T$ on the error probability, any dimension $d$ and any amount $\delta$ of total noise, the guessing probability of a correlated eavesdropper is strictly higher when the state and measurement are \emph{both} noisy, rather than the state alone. Figure \ref{fig: joint noise app} shows a comparison of the eavesdropper's strategies in the qubit case for various values of $T$.

\end{document}